\documentclass{aa}
\usepackage{natbib}
\bibpunct{(}{)}{;}{a}{}{,} 

\usepackage[english]{babel}           
\usepackage{graphicx,epsfig}
\usepackage{txfonts}
\usepackage{mathrsfs}
\usepackage{lscape}
\usepackage{eucal}
\usepackage{pifont}
\usepackage{rotating}
\usepackage{multirow}
\usepackage{tikz}
\usepackage[colorlinks=true,citecolor=blue,linkcolor=blue]{hyperref}
\usepackage{longtable}
\usepackage[fleqn]{amsmath}
\usepackage{breqn}

\def \cc    {\ifmmode{{\rm cm}^{-3}}\else{${\rm cm}^{-3}$}\fi}
\def \cq    {\ifmmode{{\rm cm}^{-2}}\else{${\rm cm}^{-2}$}\fi}
\def \mic   {\ifmmode{\mu{\rm m}}\else{$\mu$m}\fi}
\def \eccs  {\ifmmode{{\rm erg}\,\,{\rm cm}^{-3}\,\,{\rm s}^{-1}}\else{${\rm erg}\,\,{\rm cm}^{-3}\,\,{\rm s}^{-1}$}\fi}
\def \ecc   {\ifmmode{\,{\rm erg}\,{\rm cm}^{-3}}\else{$\,{\rm erg}\,{\rm cm}^{-3}$}\fi}
\def \ecqs  {\ifmmode{\,{\rm erg}\,{\rm cm}^{-2}\,{\rm s}^{-1}\,{\rm 
             sr}^{-1}}\else{$\,{\rm erg}\,{\rm cm}^{-2}\,{\rm s}^{-1}\,{\rm sr}^{-1}$}\fi}
\def \ecss  {\ifmmode{\,{\rm erg}\,{\rm cm}^{-2}\,{\rm s}^{-1}}\else{$\,{\rm erg}\,{\rm cm}^{-2}\,{\rm s}^{-1}$}\fi}
\def \deg   {\ifmmode{^{\circ}}\else{$^{\circ}$}\fi} 
\def \pc    {\ifmmode{\,{\rm pc}}\else{$\,{\rm pc}$}\fi} 
\def \kms   {\ifmmode{\,{\rm km}\,{\rm s}^{-1}}\else{km~s$^{-1}$}\fi} 
\def \kmspc {\ifmmode{\,{\rm km}\,{\rm s}^{-1}\,{\rm pc}^{-1}}\else{km s$^{-1}$ pc$^{-1}$}\fi} 
\def \MJysr {\ifmmode{\,{\rm MJy\,sr}^{-1}}\else{$\,{\rm MJy\,sr}^{-1}$}\fi} 
\def \Kkms  {\ifmmode{\,{\rm K\,km\,s}^{-1}}\else{$\,{\rm K\,km\,s}^{-1}$}\fi}
\def \epso{\ifmmode{\overline{\varepsilon}_{\rm obs}}\else{$\overline{\varepsilon}_{\rm obs}$}\fi}
\def \utM{\ifmmode{u_{\theta,{\rm M}}}\else{$u_{\theta,{\rm M}}$}\fi}
\def \urM{\ifmmode{u_{r,{\rm M}}}\else{$u_{r,{\rm M}}$}\fi}
\def \twCO{\ifmmode{\rm ^{12}CO}\else{$\rm^{12}CO$}\fi} 
\def \thCO{\ifmmode{\rm ^{13}CO}\else{$\rm^{13}CO$}\fi} 
\def \CeiO{\ifmmode{\rm C^{18}O}\else{$\rm C^{18}O$}\fi} 
\def \twCN{\ifmmode{\rm ^{12}CN}\else{$\rm^{12}CN$}\fi} 
\def \thCN{\ifmmode{\rm ^{13}CN}\else{$\rm^{13}CN$}\fi} 
\def \HdCO{\ifmmode{\rm H_{2}CO}\else{$\rm H_{2}CO$}\fi} 
\def \twHdCO{\ifmmode{\rm ^{12}H_{2}CO}\else{$\rm^{12}H_{2}CO$}\fi} 
\def \thHdCO{\ifmmode{\rm ^{13}H_{2}CO}\else{$\rm^{13}H_{2}CO$}\fi} 
\def \twC{\ifmmode{\rm ^{12}C}\else{$\rm^{12}C$}\fi} 
\def \thC{\ifmmode{\rm ^{13}C}\else{$\rm^{13}C$}\fi} 
\def \Hp{\ifmmode{\rm H^+}\else{$\rm H^+$}\fi} 
\def \Hep{\ifmmode{\rm He^+}\else{$\rm He^+$}\fi} 
\def \Hepp{\ifmmode{\rm He^{++}}\else{$\rm He^{++}$}\fi} 
\def \Cp{\ifmmode{\rm C^+}\else{$\rm C^+$}\fi} 
\def \Sp{\ifmmode{\rm S^+}\else{$\rm S^+$}\fi} 
\def \Sip{\ifmmode{\rm Si^+}\else{$\rm Si^+$}\fi} 
\def \Fep{\ifmmode{\rm Fe^+}\else{$\rm Fe^+$}\fi} 
\def \Op{\ifmmode{\rm O^+}\else{$\rm O^+$}\fi} 
\def \CFp{\ifmmode{\rm CF^+}\else{$\rm CF^+$}\fi}
\def \CHp{\ifmmode{\rm CH^+}\else{$\rm CH^+$}\fi}
\def \CHdp{\ifmmode{\rm CH_2^+}\else{$\rm CH_2^+$}\fi}
\def \CHtp{\ifmmode{\rm CH_3^+}\else{$\rm CH_3^+$}\fi} 
\def \SHp{\ifmmode{\rm SH^+}\else{$\rm SH^+$}\fi}
\def \SHdp{\ifmmode{\rm SH_2^+}\else{$\rm SH_2^+$}\fi}
\def \SHtp{\ifmmode{\rm SH_3^+}\else{$\rm SH_3^+$}\fi}
\def \twCHp{\ifmmode{\rm ^{12}CH^+}\else{$\rm^{12}CH^+$}\fi}
\def \thCHp{\ifmmode{\rm ^{13}CH^+}\else{$\rm^{13}CH^+$}\fi}
\def \CtH{\ifmmode{\rm C_2H}\else{$\rm C_2H$}\fi} 
\def \CthHt{\ifmmode{\rm C_3H_2}\else{$\rm C_3H_2$}\fi} 
\def \Htp{\ifmmode{\rm H_3^+}\else{$\rm H_3^+$}\fi} 
\def \COp{\ifmmode{\rm CO^+}\else{$\rm CO^+$}\fi} 
\def \HCOp{\ifmmode{\rm HCO^+}\else{$\rm HCO^+$}\fi} 
\def \HtOp{\ifmmode{\rm H_3O^+}\else{$\rm H_3O^+$}\fi} 
\def \HCfiN{\ifmmode{\rm HC_5N}\else{$\rm HC_5N$}\fi} 
\def \wat{\ifmmode{\rm H_2O}\else{$\rm H_2O$}\fi} 
\def \HdO{\ifmmode{\rm H_2O}\else{$\rm H_2O$}\fi} 
\def \OHp{\ifmmode{\rm OH^+}\else{$\rm OH^+$}\fi} 
\def \HdOp{\ifmmode{\rm H_2O^+}\else{$\rm H_2O^+$}\fi} 
\def \HtOp{\ifmmode{\rm H_3O^+}\else{$\rm H_3O^+$}\fi} 
\def \NHd{\ifmmode{\rm NH_2}\else{$\rm NH_2$}\fi} 
\def \NHtrois{\ifmmode{\rm NH_3}\else{$\rm NH_3$}\fi} 
\def \oxy{\ifmmode{\rm O_2}\else{$\rm O_2$}\fi} 
\def \HH{\ifmmode{\rm H_2}\else{$\rm H_2$}\fi}
\def \Jone{\ifmmode{\rm {(J=1--0)}}\else{{(J=1--0)}}\fi} 
\def \Jtwo{\ifmmode{\rm {(J=2--1)}}\else{{(J=2--1)}}\fi} 
\def \Jthr{\ifmmode{\rm {(J=3--2)}}\else{{(J=3--2)}}\fi} 
\def \Jfou{\ifmmode{\rm {(J=4--3)}}\else{{(J=4--3)}}\fi} 
\def \Jfiv{\ifmmode{\rm {J=4--3}}\else{{J=4--3}}\fi} 
\def \Ta{\ifmmode{\rm T_A}\else{$\rm T_A$}\fi} 
\def \Tas{\ifmmode{\rm T_A^*}\else{$\rm T_A^*$}\fi} 
\def \Tmb{\ifmmode{\rm T_{mb}}\else{$\rm T_{mb}$}\fi} 
\def \Tr{\ifmmode{\rm T_r}\else{$\rm T_r$}\fi} 
\def \Trs{\ifmmode{\rm T_r^*}\else{$\rm T_r^*$}\fi}
\def \NHt{\ifmmode{N_{\rm H}}\else{$N_{\rm H}$}\fi}
\def \NH{\ifmmode{N({\rm H})}\else{$N({\rm H})$}\fi}
\def \NH2{\ifmmode{N({\rm H}_2)}\else{$N({\rm H}_2)$}\fi}
\def \NCH{\ifmmode{N({\rm CH})}\else{$N({\rm CH})$}\fi}
\def \NHF{\ifmmode{N({\rm HF})}\else{$N({\rm HF})$}\fi}
\def \dens {\ifmmode{n_{\rm H}}\else{$n_{\rm H}$}\fi}
\def \densini{\ifmmode{n_{\rm H}^0}\else{$n_{\rm H}^0$}\fi}
\def \densfin{\ifmmode{n_{\rm H}^{\rm f}}\else{$n_{\rm H}^{\rm f}$}\fi}
\def \nCO{\ifmmode{n({\rm CO})}\else{$n({\rm CO})$}\fi}
\def \nHF{\ifmmode{n({\rm HF})}\else{$n({\rm HF})$}\fi}
\def \nH2{\ifmmode{n({\rm H}_2)}\else{$n({\rm H}_2)$}\fi}

\usepackage{color}

\begin{document}

\title{Shocks in the warm neutral medium}
\subtitle{I. Theoretical model}

\author{
  B. Godard            \inst{\ref{lerma}, \ref{ENS}}, 
  G. Pineau des Forêts \inst{\ref{IAS},   \ref{lerma}}, \and
  S. Bialy             \inst{\ref{Technion}}
}

\institute{
Observatoire de Paris, Université PSL, Sorbonne Université, LERMA, 75014 Paris, France
\label{lerma}
\and
Laboratoire de Physique de l’Ecole Normale Supérieure, ENS, Université PSL, CNRS, Sorbonne Université, Université de Paris, F-75005 Paris, France
\label{ENS}
\and
Université Paris-Saclay, CNRS, Institut d’Astrophysique Spatiale, 91405, Orsay, France
\label{IAS}
\and
Physics Department, Technion, Haifa, 3200003, Israel
\label{Technion}
}

 \date{Received 21 December 2023 / Accepted 25 April 2024}

\abstract{
Atomic and molecular line emissions from shocks may provide valuable information on the injection of mechanical energy into the interstellar medium (ISM), the generation of turbulence, and the processes of phase transition between the warm neutral medium (WNM) and the cold neutral medium (CNM).}
{In this series of papers, we investigate the properties of shocks propagating in the WNM. 
Our objective is to identify the tracers of these shocks, use them to interpret ancillary observations of the local diffuse matter, and provide predictions for future observations.}
{Shocks propagating in the WNM are studied using the Paris-Durham shock code, a multi-fluid model built to follow the thermodynamical and chemical structures of shock waves at steady-state in a plane-parallel geometry. The code, designed to take into account the impact of an external radiation field, is updated to treat self-irradiated shocks at intermediate ($30 < V_S < 100$ \kms) and high velocity ($V_S \geqslant 100$ \kms), which emit ultraviolet (UV), extreme-ultraviolet (EUV), and X-ray photons. The couplings between the photons generated by the shock, the radiative precursor, and the shock structure are computed self-consistently using an exact radiative-transfer algorithm for line emission. The resulting code is explored over a wide range of parameters ($0.1 \leqslant \dens \leqslant 2$ \cc, $10 \leqslant V_S \leqslant 500$ \kms, and $0.1 \leqslant B \leqslant 10$ $\mu$G), which covers the typical conditions of the WNM in the solar neighborhood.}
{The explored physical conditions lead to the existence of a diversity of stationary magnetohydrodynamic solutions, including J-type, CJ-type, and C-type shocks. These shocks are found to naturally induce phase transition between the WNM and the CNM, provided that the postshock thermal pressure is higher than the maximum pressure of the WNM and that the maximum density allowed by magnetic compression is greater than the minimum density of the CNM. The input flux of mechanical energy is primarily reprocessed into line emissions from the X-ray to the submillimeter domain. Intermediate- and high-velocity shocks are found to generate a UV radiation field that scales as $V_S^3$ for $V_S < 100$~\kms\ and as $V_S^2$ at higher velocities, and an X-ray radiation field that scales as $V_S^3$ for $V_S \geqslant 100$~\kms. Both radiation fields may extend over large distances in the preshock depending on the density of the surrounding medium and the hardness of the X-ray field, which is solely driven by the shock velocity.}
{This first paper presents the thermochemical trajectories of shocks in the WNM and their associated spectra. It corresponds to a new milestone in the development of the Paris-Durham shock code and a stepping stone for the analysis of observations that will be carried out in forthcoming works.}


\authorrunning{B. Godard et al.}
\titlerunning{Phase transition shocks in the Warm Neutral Medium} 
\maketitle

\section{Introduction}

\begin{figure}[!h]
\begin{center}
\includegraphics[width=9.2cm,trim = 1.5cm 2.8cm 0.5cm 1.0cm, clip,angle=0]{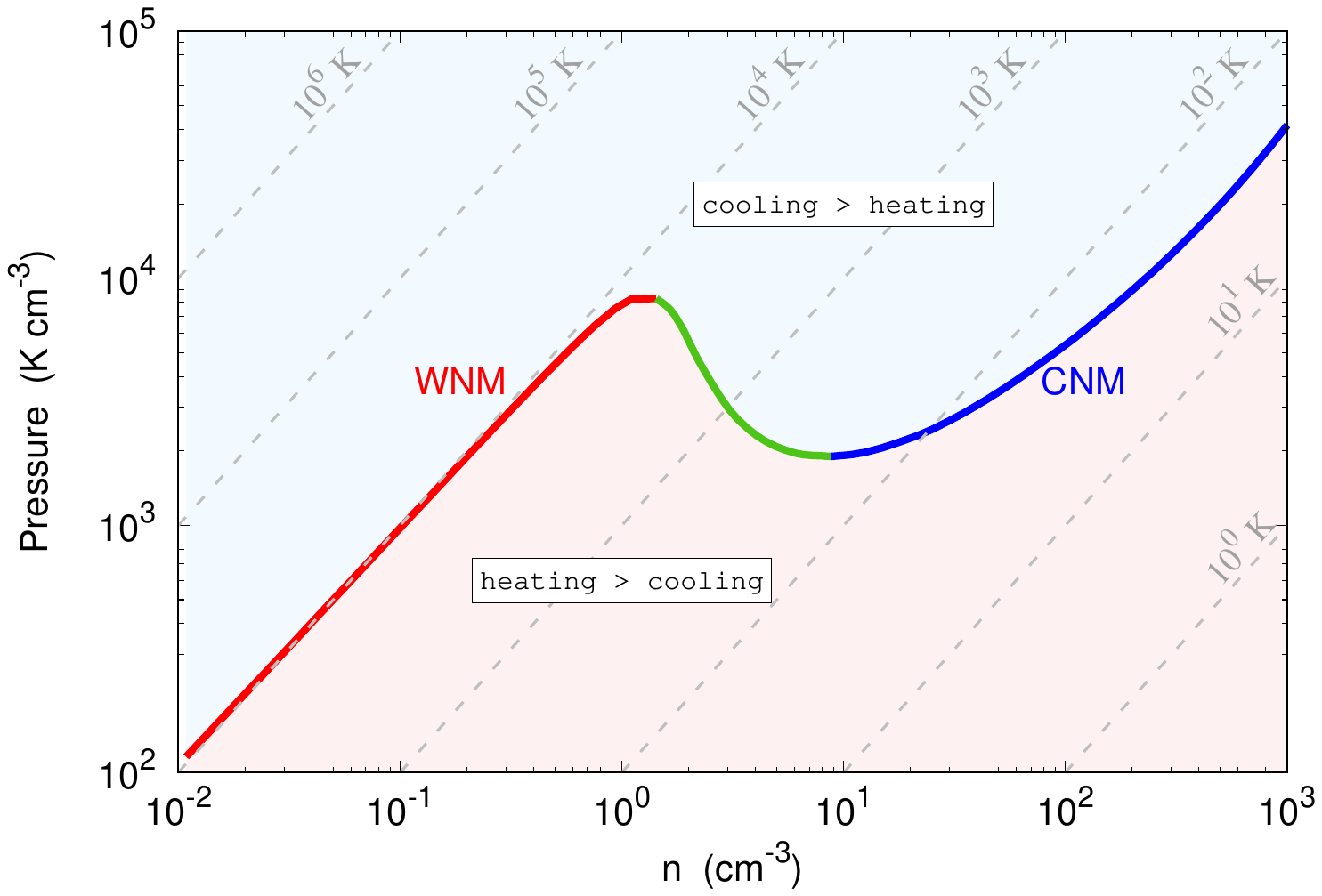}
\caption{Thermal equilibrium state of the diffuse interstellar gas obtained with the Paris-Durham code and displayed in a particle density versus thermal pressure diagram. The color code highlights the three branches of the equilibrium state: the dynamically stable WNM (red) and CNM (blue) phases at $\sim 10^4$\,\,K and $\sim 80$\,\,K, respectively, and the dynamically unstable branch (green) at intermediate temperatures. The light blue and red zones show the regions of the diagram where the cooling rate is larger or smaller, respectively, than the heating rate. Light gray lines are isothermal contours from 1 to $10^6$ K (from bottom right to top left).}
\label{Fig-equilibrium}
\end{center}
\end{figure}

The dynamical evolution of the diffuse matter toward the formation of dense structures is a fascinating problem of physics. Because of thermal instability (see Fig.~\ref{Fig-equilibrium}), the diffuse neutral medium is known to be separated into two stable thermal states, a warm phase at $\sim 10\,000$~K, the warm neutral medium (WNM), and a cold phase at $\sim 80$~K, the cold neutral medium (CNM) (e.g., \citealt{Field1965a, Heiles2003, Wolfire2003, Murray2018a, Marchal2019a}). Although these two phases are found to be at thermal pressure equilibrium \citep{Jenkins2011}, they do not peacefully coexist. By inducing pressure variations and shear motions at all scales, magnetized turbulence drives transfers of mass and energy between the WNM and the CNM (e.g., \citealt{Audit2005a,Seifried2011}). Under the combined actions of turbulence and thermal instability, the diffuse gas therefore constantly flows from one phase to the other, eventually leading to the formation of structures massive enough to trigger gravitational collapse.

One remarkable aspect is the role of interstellar shocks at each stage of this dynamical evolution. The driving of interstellar turbulence in galaxies occurs through a wealth of supersonic events, which include outflows from young stellar objects, infall of material onto the galactic disk, supernova explosions, outflows from active galactic nuclei, and galaxy collisions. All these events inevitably form shocks, which not only dissipate a fraction of the initial kinetic energy but also generate a turbulent cascade in their wake; for example, through the production of baroclynic vorticity \citep{Elmegreen2004, Vazquez-Semadeni1996a}. On the theoretical side, this process is particularly well illustrated by simulations of supernova-driven turbulence 
(e.g., \citealt{Pan2016a}). On the observational side, one of the most iconic examples of this process is the Stephan’s Quintet galaxy collision, where colliding galaxies at relative velocities of $\sim 1000$~\kms\ generate a turbulent flow whose mechanical energy is traced by the rovibrational lines of \HH\ \citep{Guillard2009, Appleton2023a}. Interstellar shocks are therefore the fingerprints of the mechanisms by which mechanical energy is injected into the ISM. 

By altering the physical state of matter, shocks propagating in the WNM are known to bring the gas in an unstable cooling state (red region in Fig.~6 of \citealt{Falle2020a}). Hydrodynamical simulations show that if the postshock thermal pressure is higher than the maximum pressure of the WNM and the cooling time is shorter than the dynamical time of the perturbation, shocks inevitably induce phase transition between the WNM and the CNM (e.g., \citealt{Hennebelle1999a,Kupilas2021a}). This process is mitigated by the presence of a magnetic field. Magnetohydrodynamic simulations of colliding flows \citep{Hennebelle2000a, Inoue2008a} and of interstellar shocks \citep{Falle2020a} show that the magnetic field efficiently prevents the formation of very dense gas and may force the final state to lie on the unstable part of the thermal equilibrium (green curve in Fig.~\ref{Fig-equilibrium}). Nevertheless, shocks still induce phase transition in the sense that this unstable state subsequently splits into warm and cold environments \citep{van-Loo2010a}. Shocks may therefore play a substantial role in the mass transfer between the WNM and the CNM.

Finally, as the turbulence develops, the kinetic energy injected at large scale is transferred toward small scales. This turbulent cascade, which operates in a multiphase environment, involves a complex interplay between compressive and solenoidal modes (e.g., \citealt{Porter2002,Vestuto2003,Pan2016a}) that leads to the formation of low-velocity shocks, which propagate in the WNM and the CNM and dissipate a substantial fraction of the kinetic energy \citep{Stone1998a, Lehmann2016, Lesaffre2020, Richard2022a}. Shocks could therefore be the fingerprints of the dissipation of the mechanical energy stored in interstellar turbulence.

While extensive studies have been performed to identify and interpret atomic and molecular tracers of low-velocity shocks ($V_S \leqslant 30$~\kms) propagating in the CNM (e.g., \citealt{Draine1986a,Monteiro1988,Flower1998c,Lesaffre2013}), little has been done so far regarding the observational tracers of shocks propagating in the WNM at low ($V_S \leqslant 30$~\kms), intermediate ($30 < V_S < 100$~\kms), and high velocities ($V_S \geqslant 100$~\kms) that induce phase transition between the WNM and the CNM. This oversight may be attributed to the numerical challenges associated with the problem. Indeed, one major difficulty is that intermediate- and high-velocity shocks produce high-energy photons that modify the state of the preshock gas and interact with the shock itself \citep{Raymond1979,Hollenbach1989}. Numerical codes capable of describing all the intricacies of self-irradiated shocks are rare. 

The impact of ionizing radiation has long been included in numerical simulations to study the thermal instability induced by interstellar shocks (e.g., \citealt{Sutherland2003a}) and more recently in simulations of supernova remnants (SNRs) \citep{Sarkar2021b, Sarkar2021a} and Galactic winds \citep{Sarkar2022a}. However, as detailed simulations are numerically expensive, they are usually limited to specific cases. It follows that systematic studies involving the complete exploration of a large parameter domain can only be achieved with 1D steady-state models. To the best of our knowledge, the only existing public model that follows the physics of self-irradiated shocks is the MAPPINGS V code. Developed over the past 40 years \citep{Dopita1978a, Dopita1995, Dopita1996, Allen2008, Dopita2013a, Sutherland2017}, MAPPINGS is a self-consistent model designed to compute the thermochemical structure and emission of photoionized regions and fast atomic shocks. For instance, MAPPINGS has been successfully applied to study narrow line regions and active galactic nuclei (e.g., \citealt{Best2000a,Zhu2023a}), Herbig-Haro objects (e.g., \citealt{Dopita2017a}), and SNRs (e.g., \citealt{Kopsacheili2020a, Priestley2022a}).

Given the existence of a highly sophisticated and regularly updated model such as MAPPINGS, one might wonder why we would develop another code. In addition to the fact that it is always advisable to have several models at hand to describe the same physical problem, the Paris-Durham shock code has a few advantages for the study of shocks propagating in the WNM. First, the Paris-Durham shock code is a multifluid model capable of computing various types of magnetohydrodynamic stationary solutions with distinctive physical properties, including C-type, C*-type, and CJ-type shocks \citep{Chernoff1987, Roberge1990, Godard2019}. Second, the code was recently improved to take into account the impact of an external UV radiation field \citep{Lesaffre2013, Godard2019}, and therefore contains all the ingredients required  to treat the phase transition process between the WNM and the CNM, which results from the thermal instability that naturally occurs in diffuse irradiated environments. Last, this code is designed to study the formation and excitation of molecules in shocks. Although this aspect is not addressed here, developing the model for higher velocities allows to identify molecular tracers of high-velocity shocks, which are naturally produced during the phase transition and the formation of CNM structures.

In the first paper of this series, we therefore introduce a new version of the Paris-Durham shock code capable of treating the physics of self-irradiated shocks. As a first application, we present a study of shocks propagating at velocities between 10 and 500 \kms\ in the diffuse WNM irradiated by the standard interstellar radiation field. This work is a stepping stone toward two follow-up papers dedicated to specific observational tracers of these shocks (e.g. \citealt{Godard2024b}).

The main physical ingredients of the code and the recent updates performed to model self-irradiated shocks are presented in Sect. \ref{Sect-method} and Appendix \ref{Append-radtrans}. The dynamical, thermal, and chemical structures of a prototypical shock and the induced phase transition between the WNM and the CNM are described in Sect.~\ref{Sect-thermochem}. The reprocessing of the input mechanical energy flux into line and continuum radiation and the corresponding heating and ionizing rates of the interstellar gas are shown and discussed in Sect.~\ref{Sect-grid}.

\section{Physical ingredients and numerical method} \label{Sect-method}

The theoretical model presented in this work is based on the Paris-Durham shock code, a public numerical tool\footnote{Available on the ISM plateform \url{https://ism.obspm.fr}.} built to follow the dynamical, thermal, and chemical structures of a multi-fluid plane-parallel interstellar shock at steady-state. Initially designed to  study low-velocity shocks ($V_S \leqslant 30$ \kms) propagating in molecular environments, the code was recently improved to treat shocks irradiated by an external UV radiation field \citep{Godard2019} and intermediate-velocity shocks at velocities up to $60$~\kms\ self-irradiated by UV photons produced through the collisional excitation of atomic hydrogen \citep{Lehmann2020,Lehmann2022}. The model used here is built upon these recent improvements with additional updates on the chemistry and the radiative transfer, presented below, that allow the computation of shocks propagating at velocities larger than $60$~\kms\ in ionized or partially ionized atomic regions such as the WNM.

The physics of photodissociating and photoionizing shocks is well understood \citep{Dopita2003} and has been the subject of numerous studies over the past 60 years (e.g., \citealt{Shull1979, Raymond1979, Hollenbach1989, Dopita1995, Dopita1996, Allen2008, Sutherland2017}). In a nutshell, high-velocity shocks produce UV, EUV, and soft X-ray photons that not only propagate downstream where they interact with the postshock gas but also propagate upstream where they interact with the preshock medium and induce the formation of a radiative precursor. Depending on the shock speed, the flux of ionizing photons escaping upstream may become large enough to fully ionize the preshock, effectively leading to the formation of an HII region. As it modifies the entrance conditions of the gas, the radiation field generated by the shock itself therefore alters the shock own thermodynamical structure. This brief summary shows that modeling high-velocity shocks requires to take into account all the chemical and cooling processes that occur in HII regions and to implement a numerical scheme that self-consistently computes the physical state of the radiative precursor and its feedback on the shock structure. Following \citet{Sutherland2017} and \citet{Lehmann2020}, we apply here a simple iterative scheme. The thermochemical structure of a shock is first computed using the Paris-Durham shock code. The propagation of the photons generated by the shock is then solved using a post-processing radiative-transfer algorithm. Both steps are repeated until convergence.

\subsection{Parameters and initial conditions} \label{Sect-param}

\begin{table*}
\begin{center}
\caption{Main parameters of the shock code, standard model, and range of values explored in this work.}
\label{Tab-main}
\begin{tabular}{l r r r l}
\hline
name & standard & range & unit & definition \\
\hline
$n_{\rm H}^0$      & 1.0                 & $0.1$ $-$ $2$     & cm$^{-3}$           & preshock proton density$^a$ \\
$G_0$              & 1.0                 & $-$               &                     & radiation field scaling factor$^b$ \\
$\zeta_{\HH}$      & $3 \times 10^{-16}$ & $-$               & s$^{-1}$            & \HH\ cosmic ray ionization rate \\
$V_{\rm s}$        & 200                 & 10 $-$ 500        & km s$^{-1}$         & shock velocity \\
$B_0$              & 1.0                 & 0.1 $-$ 10        & $\mu$G              & initial transverse magnetic field \\
\hline
\end{tabular}
\begin{list}{}{}
(a) defined as $n_{\rm H}^0=n^0({\rm H}) + 2n^0(\HH) + n^0(\Hp)$, where $n^0({\rm H})$, $n^0(\HH)$, and $n^0(\Hp)$ are the initial densities of H, \HH, and \Hp.\\
(b) the scaling factor $G_0$ is applied to the standard ultraviolet radiation field of \citet{Mathis1983}.
\end{list}
\end{center}
\end{table*}

\begin{figure}[!h]
\begin{center}
\includegraphics[width=9.0cm,trim = 2cm 2.8cm 1cm 2.0cm, clip,angle=0]{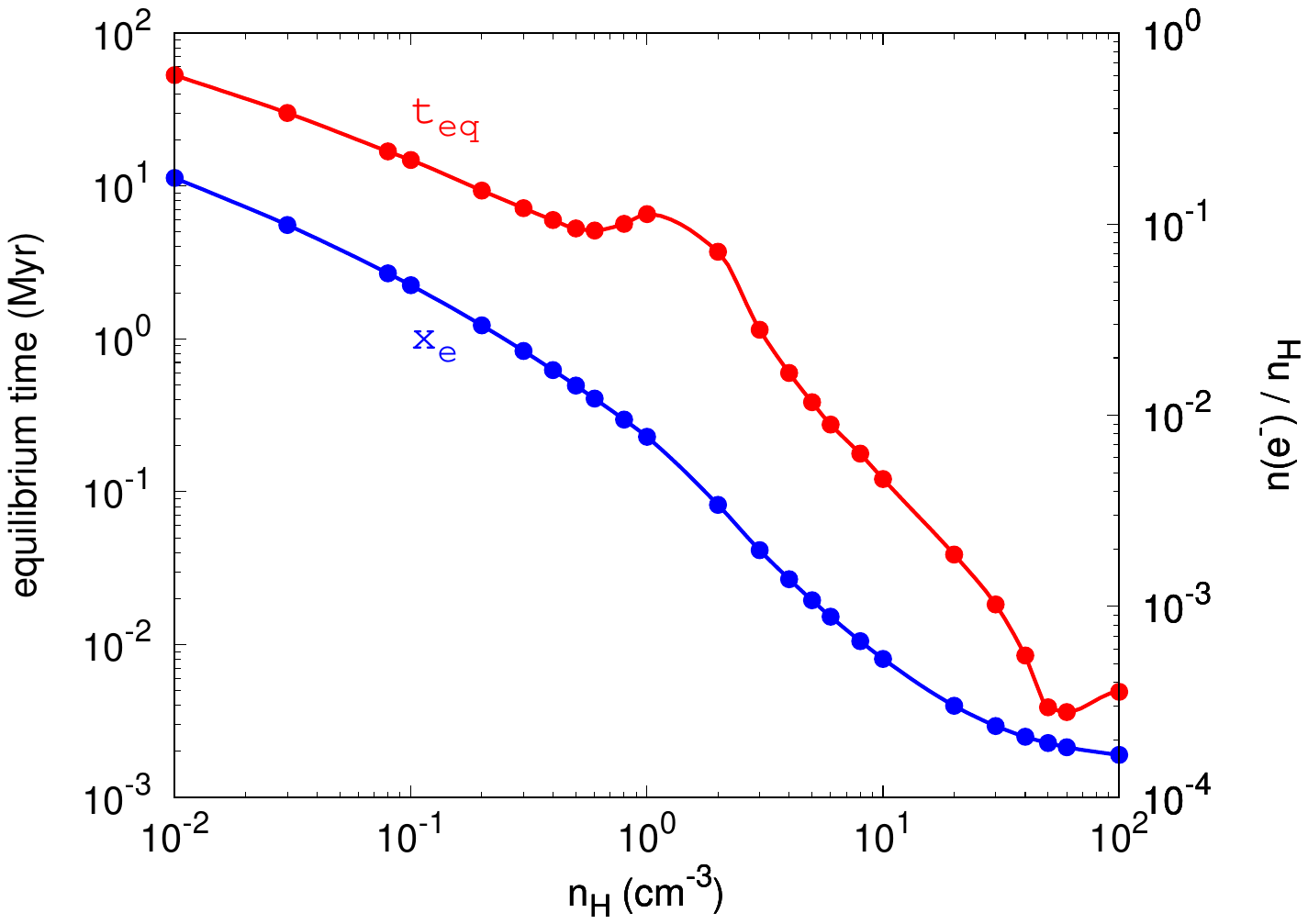}
\caption{Timescale required to reach the ionization, thermal, and chemical equilibrium states in the diffuse medium as a function of the proton density $n_{\rm H}$ (red curve), and corresponding equilibrium values of the ionization fraction, $x_e = n({\rm e}^-)/n_{\rm H}$ (blue curve). The equilibrium timescale is estimated as the time at which the kinetic temperature, the electronic fraction, and the chemical abundances of all species are within 10 \% of their equilibrium values, starting from random initial conditions.}
\label{Fig-timeequil}
\end{center}
\end{figure}

We consider a plane-parallel shock propagating at a velocity $V_S$ in a magnetized medium with a constant proton density $n_{\rm H}^0$ and a magnetic field $B_0$ set in the direction perpendicular to the direction of propagation\footnote{As shown by \citet{Hollenbach1979}, transverse shocks accurately describe oblique J-type shocks if the Alfv\'enic Mach number is sufficiently large such that the magnetic pressure behind the shock front is negligible compared to the thermal pressure. This is the case in most of the models explored in this work (see Sect. \ref{Sect-grid}).}. In addition to the radiation produced by the shock itself, we assume that the shock is externally irradiated by an isotropic UV and optical photon flux. This flux is set to the standard interstellar radiation field (ISRF, \citealt{Mathis1983}) and scaled with a parameter $G_0$. Given the typical distances between OB star associations in the solar neighborhood ($\sim 100$ pc, e.g., \citealt{Zari2018}) and the typical density of the WNM ($\sim 0.5$ \cc), we neglect the absorption of the external radiation field by the surrounding medium in which the shock propagates \citep{Bellomi2020}. Neglecting also the absorption by the shock itself (see Sect. \ref{Sect-absorption}), the external UV and optical photon flux is therefore assumed constant throughout the entire model. The preshock and shock gas are finally assumed to be pervaded by cosmic ray particles with an equivalent \HH\ ionization rate $\zeta_{\HH}$. The main parameters of the model, the standard values adopted, and the range of values explored in this work are given in Table~\ref{Tab-main}.

Within this geometry and under the steady-state approximation, the magnetohydrodynamics equations and the chemical network adopted in the model lead to a set of coupled first order differential equations that describe the dynamical, thermal, and chemical evolutions of a fluid particle during its trajectory from the preshock medium to the postshock gas \citep{Flower2010}. This system is solved using the DVODE forward integration algorithm \citep{Hindmarsh1983} starting from a given position in the preshock with known initial conditions. The initial position of the fluid particle (i.e., its distance from the shock front) is chosen so that the dissociating and ionizing radiation field generated by the shock is entirely absorbed by the radiative precursor. To limit the number of free parameters, we finally assume that the gas at this position is at thermochemical equilibrium. While it entirely fixes the initial conditions of the problem, this choice is highly debatable.  As an illustration, we display in Fig.~\ref{Fig-timeequil} the equilibrium value of the electronic fraction, $n({\rm e}^-) / \dens$, and the timescale required to reach thermochemical equilibrium in the diffuse interstellar medium as functions of the density of the gas. This figure shows that the WNM ($0.1\,\cc \leqslant n_{\rm H}^0 \leqslant 2$ \cc) requires $\sim 10$ Myr to reach thermochemical equilibrium, a value larger than the dynamical evolution timescale of the diffuse matter (turnover timescale and evaporation and condensation timescales of the turbulent ISM) or the typical timescale separating the arrival of two successive SNRs at a given point in the Galaxy \citep{Draine2011}.

\subsection{Dust grains and chemical network}  \label{Sect-chem}

The chemical composition of a fluid particle across an interstellar shock is computed taking into account gas, dust, and PAHs. The dust-to-gas mass ratio is set to 0.01 and the dust size distribution is assumed to follow the classical MRN distribution \citep{Mathis1977} with a power law index of -3.5. The fractional abundance of PAHs, $n_{\rm PAH}/n_{\rm H}$, is set to $10^{-6}$, which corresponds to standard conditions for unshielded diffuse environments \citep{Draine2007}. Oppositely to our previous work \citep{Godard2019} we neglect here the formation and destruction of grain mantles. For simplification, we also neglect shattering, vaporization, and coagulation of grain cores \citep{Jones1996}. The grain size distribution is therefore solely affected by erosion through thermal sputtering (Sect. 2.2 and Appendix A of \citealt{Godard2019}). This is a strong limitation of the model as shattering and vaporization of large grains are expected to occur in high-velocity shocks (see for instance Fig.~7 of \citealt{Guillet2009}) and therefore modify the surface available for the recombination of ionized species.

The gas-phase composition of the fluid particle results from the setup of an extensive chemical network. Throughout this work, the gas-phase elemental abundances and the depletion in grain cores are set to the standard values derived in the solar neighborhood (see Table 1 of \citealt{Kristensen2023a}). Initially designed to study the formation of molecules in dense environments, the chemical network implemented in the online version of the Paris-Durham shock code includes nine different elements (H, He, C, N, O, S, Mg, Si, and Fe) and 141 species that interact with one another through an ensemble of $\sim 2800$ chemical reactions whose rates were updated in 2022. These include cosmic ray ionizations, photodissociation and photoionization processes, radiative and dissociative recombinations, collisional ionizations and dissociations, and ion-neutral and neutral-neutral reactions. 

For the purposes of the present paper and the follow-up papers, this chemical network has been further extended to take into account the evolution of the abundances of multi-ionized species. The ionization stages of all the elements listed above with a formation enthalpy below $10^8$ K have been included, along with photoionization processes, electron impact collisional ionizations, radiative and dielectronic recombinations, and recombinations onto dust and PAHs. These modifications have increased the number of species treated in the code to 209. In practice, the photoionization rate of any ion $i$ at a position $z$ along the shock trajectory (see Fig.~1 of \citealt{Godard2019}) is computed from the integration of its frequency dependent ionization cross section, $\sigma_i(\nu)$, over the specific intensity of the local radiation field, $I_\nu$, as
\begin{equation}
k_\gamma^i (z) = \int \int \sigma_i (\nu) \frac{1}{h\nu} I_\nu(z) \, d\nu \, d\Omega,
\end{equation}
where $I_\nu = I_\nu^{\rm ext} + I_\nu^{\rm int}$ includes the contribution of both the external ISRF (see Sect. \ref{Sect-param}) and the radiation field generated by the shock itself (see Sect. \ref{Sect-radtrans} and Appendix \ref{Append-radtrans}). As illustrated in Appendix \ref{Append-photo}, the photoionization cross sections are taken from \citet{Verner1995}. The product of any photoionization process and the probability to produce Auger's electrons are taken from \citet{Kaastra1993}. The collisional ionization rates of all atoms and atomic ions via electron impact are computed using the database of \citet{Lennon1988} (Tables 2 and 3 and Eqs. 6 and 7 in their paper). The radiative recombination rates are calculated from the data and prescriptions given by \citet{Badnell2006} (Table 1 and Eqs. 1 \& 2 in their paper), \citet{Altun2007} (Table 2 in their paper), and \citet{Abdel-Naby2012} (Table 4 in their paper). The dielectronic recombination rates are derived from the database\footnote{available on \url{http://amdpp.phys.strath.ac.uk/tamoc/DR/}} provided by \citet{Badnell2003}. In the few cases where this database is incomplete, the dielectronic recombination rates are alternatively taken from \citet{Landini1990} (Table 1 in their paper). The recombination rates on dust and PAHs are finally computed using the prescription of \citet{Draine1987}.

\subsection{Cooling processes} \label{Sect-cooling}

The cooling processes treated in the public version of the Paris-Durham shock code are described in \citet{Flower2003}, \citet{Lesaffre2013}, and \citet{Godard2019}. In particular, the code includes the excitation of the fine structure and metastable levels of all neutral and singly ionized atoms, the excitation of the rovibrational levels of several molecules (e.g., \HH, $^{12}$CO, and $^{13}$CO), and the energy lost through the ionization and dissociation of atoms and molecules by collisions with H, He, H$^+$, \HH, and ${\rm e}^-$. To accurately follow the thermal evolution of a plasma shocked at high temperature ($T>10^5$ K), three additional processes have been implemented in the shock model: the continuum Bremsstrahlung emission, the loss of energy due to the ionization of ionized species by electron impact, and the collisional excitation of multi-ionized species.

The total Bremsstrahlung cooling rate per unit volume (in erg cm$^{-3}$ s$^{-1}$) is modeled with Eq. 6.17 of \citet{Dopita2003}
\begin{equation}
\Lambda_{\rm ff} = \frac{16}{3\sqrt{3}} \left(\frac{(2\pi)^3 k T_{\rm e}}{h^2 m_{\rm e}^3} \right)^{1/2} \left(\frac{q_e^2}{c}\right)^3 n_{\rm e} \sum_i n_i Z_i^2 \langle g_{\rm ff}(\gamma^2)\rangle,
\end{equation}
where $k$ and $h$ are the Boltzmann and the Planck constants, $c$ is the speed of light, $q_{\rm e}$, $m_{\rm e}$, $T_{\rm e}$, and $n_{\rm e}$ are the charge, mass, temperature, and density of the electrons, and $n_i$ and $Z_i$ are the density and the ionization stage of each ion $i$. The total free-free Gaunt factors, $\langle g_{\rm ff}(\gamma^2)\rangle$, are calculated using a logarithmic interpolation of Table 4 of \citet{van-Hoof2014a} who computed these factors over a wide range of values of the scaled quantity $\gamma^2 = Z_i^2 E_{\rm Ry} / k T_e$, where $E_{\rm Ry}$ is the energy associated with one Rydberg ($2.17987 \times 10^{-11}$ erg).

The cooling induced by electron impact ionization is treated in the same way than the thermal energy lost during any endothermic chemical reaction \citep{Flower1985}. The cooling rate following the ionization of an ion $i$ writes
\begin{equation}
\Lambda_{\rm ion}^i = n_{\rm e} n_i I_i k_{\rm ion}^i (T_{\rm e}),
\end{equation}
where $I_i$ is the ionization potential of the ion and $k_{\rm ion}^i(T_{\rm e})$ is the temperature dependent rate coefficient for collisional ionization computed by \citet{Lennon1988} and used in the chemical network (see Sect \ref{Sect-chem}).

Finally, the cooling induced by the excitation of the electronic levels of any atom in any ionization stage by collision with electrons is computed using CHIANTI\footnote{version 10.0.2 available on \url{https://www.chiantidatabase.org}} \citep{Dere1997,Dere2019,Del-Zanna2021}. Originally developed to study solar and stellar atmospheres, the CHIANTI database contains an up-to-date set of atomic data (energy levels, radiative transition probabilities, and collisional excitation rates) for a large number of ions. For the species treated in this work, the database includes 22\,948 energy levels and 763\,970 radiative transitions. As described in Sect. \ref{Sect-radtrans}, the populations of all the levels and the emissivities of all the transitions are computed self-consistently in the postprocessing radiative transfer. Such a detailed treatment is, however, numerically prohibitive for the Paris-Durham shock code. To simplify, the time-dependent cooling is computed with precalculated functions. Assuming that each species is in its ground state, the cooling induced by any ion $i$ writes
\begin{equation} \label{Eq-cool-exc}
\Lambda_{\rm exc}^i = n_e n_i \sum_j \Delta E_j C_j(T_{\rm e}),
\end{equation}
where $\Delta E_j$ is the energy difference between the level $j$ and the ground state, $C_j(T_{\rm e})$ are the temperature dependent excitation rate coefficient between the ground state and the level $j$ provided by the CHIANTI database, and the sum is performed over all excited electronic levels. The sum in Eq. \ref{Eq-cool-exc} is fitted, with an accuracy of 10\% between 10 and $10^8$ K, with a sum of modified Arrhenius components using the curve fit optimization algorithm provided by the SciPy python library, and is used as a cooling function in the Paris-Durham shock code. The validity of this approach is checked, at each iteration, by comparing the cooling computed in the Paris-Durham shock code to the total emission of each species computed by the postprocessing radiative transfer tool.

\subsection{Postprocessing of the radiative transfer} \label{Sect-radtrans}

The temperature reached in high-velocity shocks, the productions of multi-ionized species and free electrons, and the collisional excitation of atoms and ions, lead to the emission of continuum and line radiations. By heating and ionizing the shocked and preshocked materials, the UV, EUV, and X-ray photons exert a negative feedback that acts against their own production. An accurate description of the propagation of these photons is therefore essential to compute the thermodynamical structure of the shock. The difficulty of the task lies in the treatment of photon trapping, that is the resonant absorption and reemission of line photons, which can only escape through frequency scattering in the line wings.

In this paper, the propagation of photons is computed in postprocessing with the following methodology. (1) All the emission lines contained in the CHIANTI database for any atom in any ionization stage are calculated with an exact radiative-transfer algorithm outlined in Appendix \ref{Append-radtrans}. Following the principles of coupled escape probability \citep{Elitzur2006a}, absorption and induced emission processes are handled as non-local couplings within the evolution equations of the level populations. The resulting system of equations is then solved, at equilibrium, taking into account all available levels, collisional rates, and radiative transitions. (2) The emission of continuum radiation is calculated including Bremsstrahlung, recombination, and two-photon processes. For the sake of simplicity and for the purposes of the present paper, dust emission is not included. (3) The line fluxes and the flux density of continuum radiation are then computed over the shock profile and the preshock medium taking into account the absorption due to photoionization and the extinction induced by dust and PAHs (see Appendix \ref{Append-photo}).

The emissivity of free-free radiation (in erg cm$^{-3}$ s$^{-1}$ Hz$^{-1}$ sr$^{-1}$) at frequency $\nu$ is taken from Eq. 6.14 of \citet{Dopita2003},
\begin{equation}
\begin{split}
\epsilon_{\rm ff}(\nu) & = \frac{16}{3\sqrt{3}} \left(\frac{ \pi }{2 k m_{\rm e}^3 T_e} \right)^{1/2} \left(\frac{q_e^2}{c}\right)^3 n_{\rm e} {\rm exp}\left(-\frac{h\nu}{kT_e}\right) \\
& \times \sum_i n_i Z_i^2 \left\langle g_{\rm ff} \left(\gamma^2,\frac{h\nu}{kT_e}\right) \right\rangle,
\end{split}
\end{equation}
where $\langle g_{\rm ff}(\gamma^2,h\nu/kT_e) \rangle$ are the frequency dependent free-free Gaunt factors (taken from Table 3 of \citealt{van-Hoof2014a}) and the sum is performed over all ions $i$. The emissivity of the free-bound radiation induced by the recombination of an ion $i$ onto a bound state ${\rm b}$ of the recombined ion $i'$ is modeled using Eq. 10.25 of \citet{Draine2011} as
\begin{equation}
\epsilon_{\rm fb}^{i,{\rm b}}(\nu) = \frac{1}{2}\frac{g_{\rm b}}{g_i}\frac{h^4\nu^3}{(2\pi m_e k T_e)^{3/2}c^2} n_e n_i \sigma_{i'}^{\rm b}(\nu) {\rm exp} \left(-\frac{h\nu - I_{\rm b}}{kT_e} \right),
\end{equation}
where $g_i$ is the degeneracy of the ground state of the ion $i$, and $g_{\rm b}$, $I_{\rm b}$, and $\sigma_{i'}^{\rm b}$ are the degeneracy, the ionization potential, and the photoionization cross section of the recombined ion $i'$ in the bound state ${\rm b}$. The photoionization cross section of the ground state of the recombined ion is taken from \citet{Verner1995}. For all the other states, the photoionization cross sections are derived from the prescription of \citet{Karzas1961a}, setting the associated bound-free Gaunt factors to one. Finally, and as described in Appendix \ref{Append-radtrans}, any transition that occurs through the emission of two photons is treated by the radiative-transfer algorithm as an optically thin line with a radiative decay rate, $A_{\rm 2ph}$, provided by the CHIANTI database. The emissivity of the two-photon emission process is then computed with the prescription of \citet{Shull1979} as
\begin{equation}
\epsilon_{\rm 2ph}(\nu) = \frac{h \nu_{\rm 2ph}}{4 \pi} A_{\rm 2ph} n_{\rm 2ph} \frac{6 \nu}{\nu_{\rm 2ph}^2} \left( 1 - \frac{\nu}{\nu_{\rm 2ph}} \right),
\end{equation}
where $\nu_{\rm 2ph}$ is the frequency associated with the transition and $n_{\rm 2ph}$ is the density of the upper level obtained with the radiative-transfer algorithm.

As shown above and in Appendix \ref{Append-radtrans}, we made the choice to address the propagation of photons in interstellar shocks with a detailed treatment that involves an exact and convoluted radiative-transfer algorithm. Such a detailed treatment is paramount to predict the fluxes and the line profiles of specific resonant transitions like, for instance, the Lyman series of atomic hydrogen (see Fig.~\ref{Fig-H-profile} of Appendix \ref{Append-radtrans}). It should be noted, however, that all these intricacies have a weak influence on the couplings between the self-generated radiation field and the thermochemical state of the gas. Indeed, we find that treating all lines as optically thin transitions provides similar results regarding the thermochemical structure of the shock and the extent and physical state of the radiative precursor. This is not due to the velocity broadening or velocity gradients induced by shocks. It comes from the fact that if a specific line is optically thick, either the photons escape in the wings of the line, or the associated upper level cascade through the emission of other lines which, in total, produce a similar contribution to the ionizing and dissociating radiation field.

\section{Thermochemical structure of WNM shocks} \label{Sect-thermochem}

\subsection{Standard model}

The parameters of the standard model are given in Table~\ref{Tab-main}. As a prototypical case, we consider an interstellar shock at a velocity of 200 \kms\ propagating in a medium with an initial density $n_{\rm H}^0 = 1$ \cc\ and illuminated by the standard ISRF ($G_0 = 1$). The total \HH\ cosmic ray ionization rate $\zeta_{{\rm H}_2}$ is set to $3 \times 10^{-16}$ s$^{-1}$, as deduced from the observations of H$_3^+$, OH$^+$, H$_2$O$^+$, and ArH$^+$ in the local diffuse ISM \citep{McCall2003, Indriolo2007, Indriolo2015, Neufeld2017}. Finally, the transverse magnetic field strength $B_0$ is  set to 1 $\mu$G, in agreement with the magnetic field plateau obtained from Zeeman measurements for gas at low total column density \citep{Crutcher2010} and with the Faraday rotation measurements along diffuse lines of sight \citep{Frick2001a}. All the parameters are set to the typical conditions of the WNM observed in the local ISM, except for the shock velocity. Indeed,  shocks at $V_S=200$ \kms\ are not particularly frequent in the diffuse local ISM. This choice is mostly made here to test the validity of the model and the interplay of the physical processes implemented in the code. With these parameters, the standard model is a single fluid J-type shock with no decoupling between the ionized and the neutral flows.

\subsection{Thermal profiles and electronic fraction}

\begin{figure*}
\begin{center}
\includegraphics[width=17.0cm,trim = 1.5cm 1.5cm 1.0cm 1.0cm, clip,angle=0]{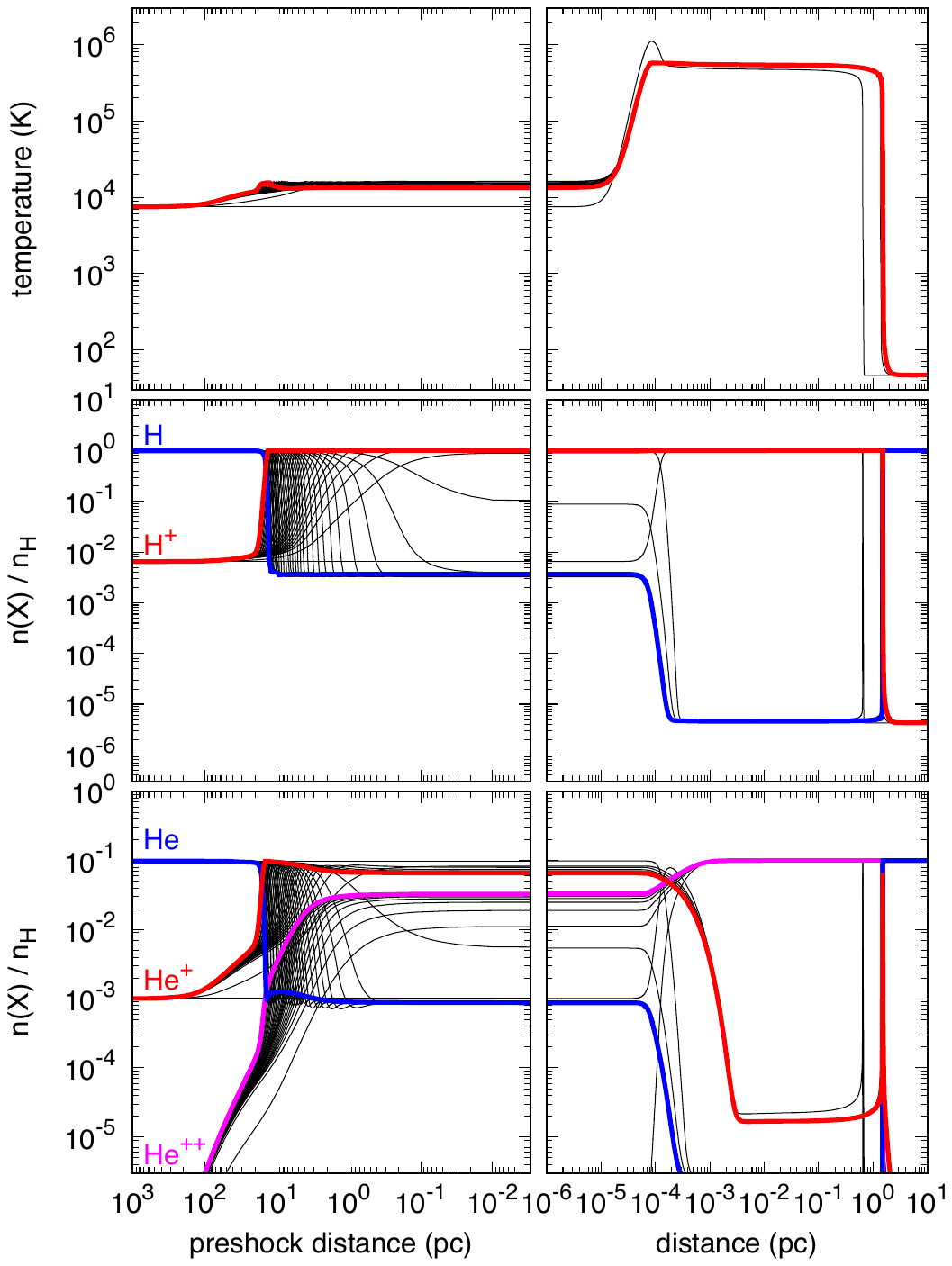}
\caption{Convergence of the standard model on the 
shock thermochemical profiles. Profiles of the kinetic temperature (top panels), the fractional abundances of H and \Hp\ (middle panels), and the fractional abundances of He, \Hep, and \Hepp\ (bottom panels) as functions of the distance in the preshock (left panels) and in the shock (right panels). The black curves show the intermediary profiles computed during the iterative procedure and the colored curves the profiles obtained in the final converged iteration.}
\label{Fig-convergence}
\end{center}
\end{figure*}

\begin{figure*}
\begin{center}
\includegraphics[width=17.0cm,trim = 1.5cm 1.5cm 1.0cm 1.0cm, clip,angle=0]{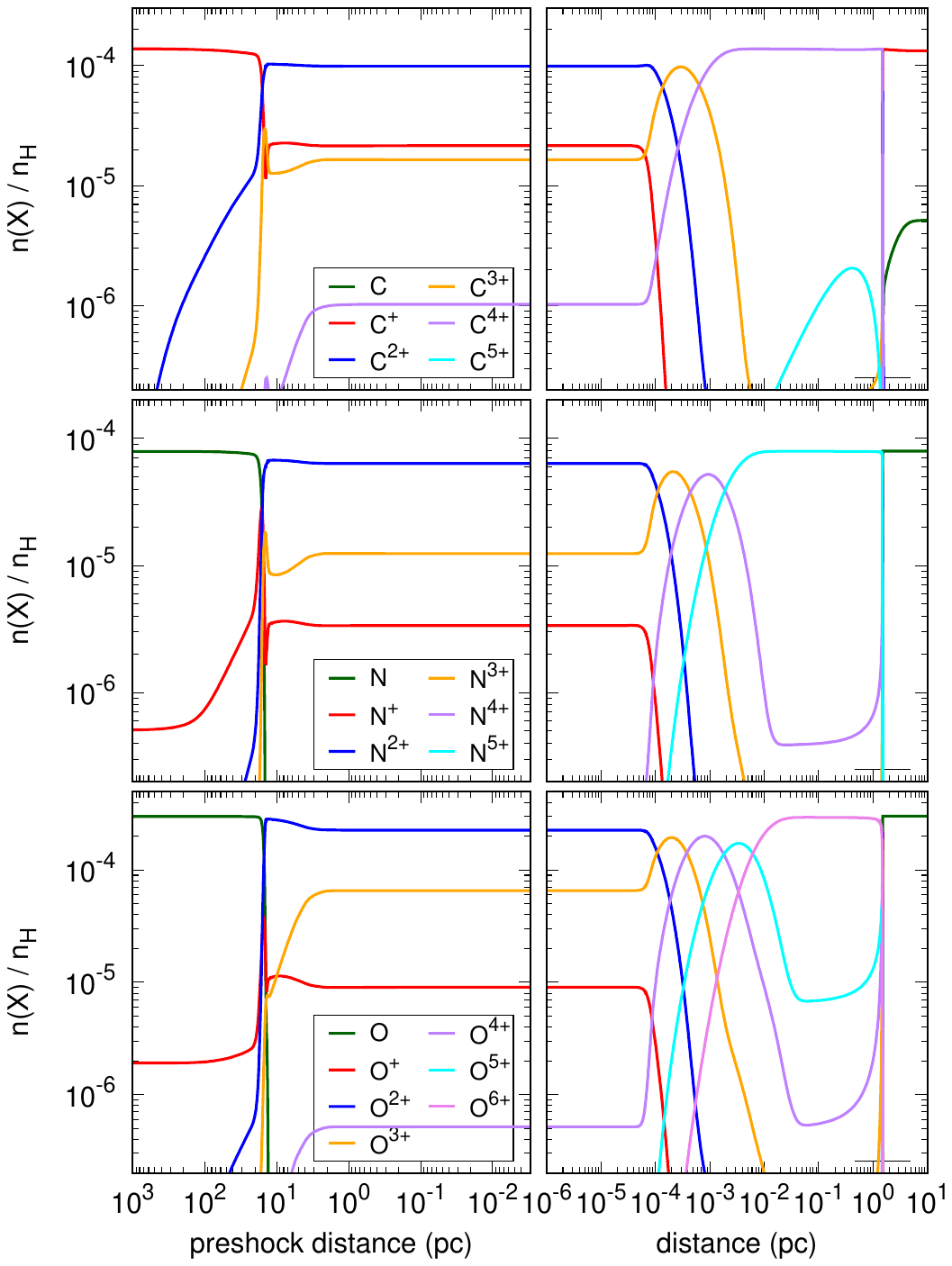}
\caption{Production of multi-ionized species in the standard model. Abundances profiles of carbon-bearing species (top panels), nitrogen-bearing species (middle panels), and oxygen-bearing species (bottom panels) as functions of the distance in the preshock (left panels) and in the shock (right panels). The high ionization stages of carbon, nitrogen, and oxygen (C$^{6+}$, N$^{6+}$, N$^{7+}$, O$^{7+}$, and O$^{8+}$) are not included because their relative abundances are all below the lowest value shown in this figure.}
\label{Fig-multiions}
\end{center}
\end{figure*}

The outcome of the iterative procedure is illustrated in Fig.~\ref{Fig-convergence}, which displays the profiles of the kinetic temperature (top panels), the fractional abundances of H and \Hp\ (middle panels), and the fractional abundances of He, \Hep, and \Hepp\ (bottom panels) obtained with the standard model in the preshock (left panels) and postshock medium (right panels). The left panels of this figure are similar to Fig.~2 of \citet{Sutherland2017}, which shows the ionization and thermal structures of the preshock for a shock at 150 \kms\ with no external UV and optical radiation field. As found by \citet{Sutherland2017}, we find that about $20$ iterations are required to reach convergence on the radiative transfer and therefore on the thermochemical profiles of the preshock gas.

The magnetohydrodynamics and thermochemical equations are solved in the Paris-Durham shock code using a forward integration technique. Figure~\ref{Fig-convergence} therefore shows the out-of-equilibrium evolution of a fluid particle traveling from the preshock gas to the postshock medium. The initial conditions of the gas are typical of the WNM, with a kinetic temperature of $\sim 8000$~K and an ionization fraction of $\sim 0.01$ (see Fig.~\ref{Fig-timeequil}). As it encounters the radiative precursor, the gas becomes almost fully ionized and is heated to $\sim 13\,400$ K. The flux of ionizing photons generated by a shock at 200 \kms\ is so strong that the gas is found to be at thermochemical equilibrium along its entire trajectory through the preshock. This is in line with the results of \citet{Sutherland2017}, who show that the preshock gas can be considered at thermochemical equilibrium for shock velocities $V_S \geqslant 120-150$ \kms.

As described in previous studies, the radiative precursor has mainly two feedbacks on the shock structure. First, the increase of the ionization state of the gas in the radiative precursor implies that the mean mass of particles drops from $\sim 1.27$ to $\sim 0.6$ a.m.u. As dictated by the jump conditions of adiabatic magnetohydrodynamic shocks \citep{Roberge1990}, the postshock maximal temperature is thus reduced by a factor of $\sim 2.1$ (see the initial and final trajectories in Fig.~\ref{Fig-convergence}). Second, the gradient of temperature induced in the preshocked gas exerts a force that slows down and compresses the fluid. The gas therefore enters the shock with a reduced effective velocity. This effect is mitigated here because the initial ram pressure is about three orders of magnitude larger than the maximum thermal pressure in the preshock.

With all these effects self-consistently computed, a shock propagating at $200$~\kms\ in the WNM is found to heat the gas to a postshock temperature of $\sim 6 \times 10^5$ K. The evolution of the postshock medium is then driven by the cooling mechanisms and the recombination processes. At these temperatures, we find that the cooling time is $\sim 4 \times 10^4$~yr, which corresponds to a cooling length of $2$ pc (see Fig.~\ref{Fig-convergence}). Both these results are in fair agreement (within a factor of two) with Eq. 36.33 of \citet{Draine2011} who estimated the cooling time of high-velocity shocks with a simplified cooling function for hot plasmas ($T > 10^5$ K).

\subsection{Chemical profiles}

Figure~\ref{Fig-convergence} (right panel) shows that the large temperatures produced by the standard model imply that both hydrogen and helium are fully ionized in the postshock medium. Interestingly, in both cases, an ionization equilibrium is reached between collisional ionization and radiative and dielectronic recombinations that lasts until the gas cools down to temperatures below $10^5$ K. As a comparison, and for future analyses of observational tracers, Fig.~\ref{Fig-multiions} displays the abundances of the different ionization stages of carbon, nitrogen, and oxygen obtained at the final iteration of the standard model. In contrast to hydrogen and helium, none of the elements shown in Fig.~\ref{Fig-multiions} (right panel) is fully ionized in the standard shock model and none of them reaches ionization equilibrium. This particular result highlights the need to follow the time dependent evolution of the chemistry even in stationary shock structures.

The analysis of the preshock (Fig.~\ref{Fig-multiions}, left panel) shows that the soft X-ray radiation field generated by the standard model is strong enough to doubly ionize carbon, nitrogen, and oxygen in the radiative precursor. This feature is particularly important for the interpretation of observations of atoms obtained in the diffuse local ISM.

\subsection{Phase transition} \label{Sect-phase-standard}

\begin{figure}[!h]
\begin{center}
\includegraphics[width=9.0cm,trim = 1.5cm 2.2cm 0.5cm 2.2cm, clip,angle=0]{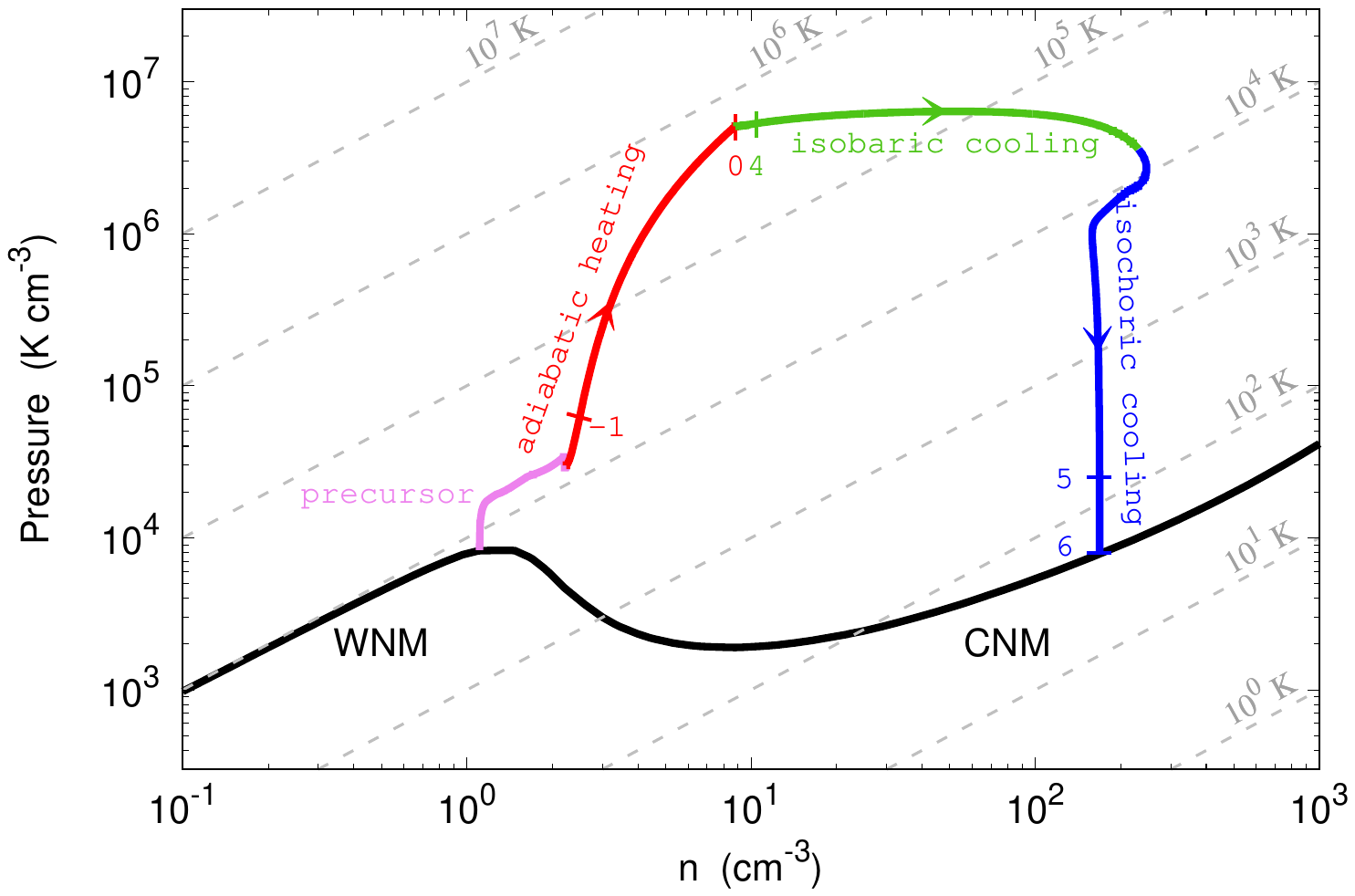}
\includegraphics[width=9.0cm,trim = 1.5cm 2.2cm 0.5cm 2.2cm, clip,angle=0]{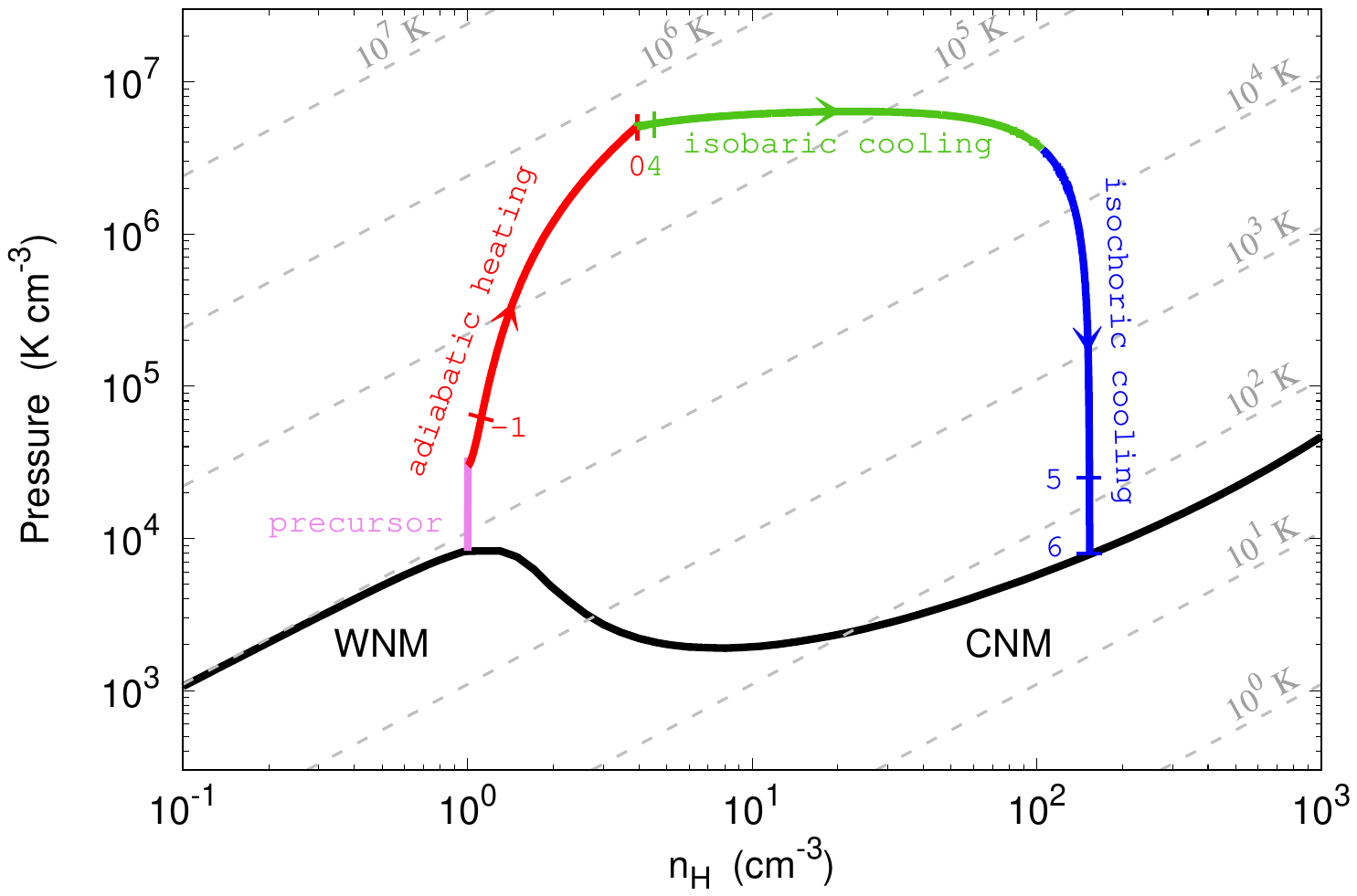}
\caption{Trajectory of a shocked fluid particle from the ambient medium to the postshock region obtained in the standard model. The trajectory is displayed in a particle density versus thermal pressure diagram (top panel) and in a proton density versus thermal pressure diagram (bottom panel) to highlight the phase transition from the WNM to the CNM induced by a shock at 200 \kms. A fluid particle initially evolves through the radiative precursor (pink curve). As it crosses a shock front, it undergoes three successive regimes: an adiabatic heating (red curve) followed by a quasi-isobaric cooling (green curve), and finally a quasi-isochoric cooling (blue curve) as the magnetic pressure becomes dominant in the postshock gas. The black curve indicates the thermal equilibrium state of the diffuse gas obtained for $G_0=1$ (see Fig.~\ref{Fig-equilibrium}). Light gray lines are isothermal contours from 1 to $10^7$ K (from bottom right to top left). We note that these isocontours are evenly spaced in the top panel but not in the bottom panel. Tics on the trajectories indicate the time, in log scale, starting from the adiabatic phase, ranging from $10^{-1}$~yr to $10^6$~yr.}
\label{Fig-Trajectory-Main}
\end{center}
\end{figure}

The impact of an interstellar shock on the thermodynamical state of the diffuse matter is shown in the top panel of Fig.~\ref{Fig-Trajectory-Main}, which displays the trajectory of the standard model in a particle density versus thermal pressure diagram. To bring out the physical properties of shocks, leaving aside the variation of the number of particles induced by the ionization in the preshock and the recombination in the postshock, we also display this trajectory in a proton density versus thermal pressure diagram in the bottom panel of Fig.~\ref{Fig-Trajectory-Main}. While this figure is drawn for a particular set of parameters, it depicts a very general modus operandi of stationary interstellar shocks. 
A fluid particle crossing a shock front basically undergoes three successive regimes.

First, the gas is heated adiabatically (red curve in Fig.~\ref{Fig-Trajectory-Main}). For strong magnetohydrodynamic shocks, that is shocks where the sonic and the Alfv\'enic Mach numbers are much higher than one, the Rankine-Hugoniot relations imply that the mass density $\rho$, velocity, $V$, and temperature, $T$, of the gas after the adiabatic jump are \citep{Roberge1990}
\begin{equation}
\rho = \frac{\gamma+1}{\gamma-1} \rho_0,
\end{equation}
\begin{equation}
V    = \frac{\gamma-1}{\gamma+1} V_S,
\end{equation}
and
\begin{equation}
T    = \frac{2(\gamma-1)}{(\gamma+1)^2} \frac{\mu V_S^2}{k},
\end{equation}
where $\rho_0$ is the mass density before the jump, $\mu$ is the mean mass of the particles, and $\gamma$ is the adiabatic index. In such conditions, the thermal pressure after the jump writes
\begin{equation}
P = \frac{2}{(\gamma+1)} \rho_0 V_S^2,
\end{equation}
hence
\begin{equation} \label{Eq-postpress}
P = 5.1 \times 10^{6}\ {\rm K}\ \cc\ \left(\frac{\densini}{1\ \cc}\right)\ \left(\frac{V_S}{200\ \kms}\right)^2.
\end{equation}

After the jump, and if the magnetic pressure is small compared to the termal pressure, the gas cools down isobarically (green curve in Fig.~\ref{Fig-Trajectory-Main}). This isobaric equation of state comes from the fact that the shock is at steady-state and that the postshock thermal pressure must therefore be at equilibrium with the preshock ram pressure. As the gas cools down, the postshock density increases. Because the magnetic field is frozen in the ionized gas, the magnetic field strength increases accordingly.

The final point of the trajectory is obtained when the postshock gas reaches thermal equilibrium. If, on the one hand, the magnetic pressure remains negligible over the  postshock, the final density of the gas is given by the first intersection between the postshock thermal pressure (Eq. \ref{Eq-postpress}) and the thermal equilibrium curve (black curve in Fig.~\ref{Fig-Trajectory-Main}). If, on the other hand, the postshock magnetic pressure is no longer negligible compared to the postshock thermal pressure, the gas progressively shifts from an isobaric to an isochoric evolution (blue curve in Fig.~\ref{Fig-Trajectory-Main}). In this case, the final density is given by
\begin{equation} \label{Eq-densfin}
\densfin = 146\ \cc\ \left(\frac{\densini}{1\ \cc}\right)^{3/2} \left(\frac{V_S}{200\ \kms}\right)\ \left(\frac{B_0}{1\ \mu{\rm G}}\right)^{-1}.
\end{equation}

As already shown though theoretical works and numerical simulations (e.g., \citealt{Hennebelle1999a,Falle2020a}), all these properties imply that interstellar shocks  can induce a phase transition between the WNM and the CNM. This is illustrated in Fig.~\ref{Fig-Trajectory-Main}, which shows that the standard model naturally converts the initial warm neutral material into cold neutral material. From the above considerations, it is clear that phase transition occurs only if the thermal pressure is larger than the maximum pressure of the WNM and if the maximum density allowed by magnetic compression is larger than the minimum density of the CNM. For the gas-phase elemental abundances adopted in this work (see Sect.~\ref{Sect-chem}) and for $0.1 \leqslant G_0 \leqslant 10$, we estimate that the maximum pressure of the WNM scales as
\begin{equation}
P_{\rm WNM}^{\rm max} \sim 8 \times 10^{3}\ {\rm K}\ \cc\ G_0\,^{0.5}
,\end{equation}
and the minimum density of the CNM as
\begin{equation}
n_{\rm H,CNM}^{\rm min} \sim 8\ \cc\ G_0\,^{0.5}.
\end{equation}
These scaling relations combined with Eqs. \ref{Eq-postpress} and \ref{Eq-densfin} imply that interstellar shocks induce phase transition if
\begin{equation} \label{Eq-velcond1}
V_S \geqslant 8 \kms\ \left(\frac{\densini}{1\ \cc}\right)^{-1/2}\ G_0\,^{0.25}
,\end{equation}
and if
\begin{equation} \label{Eq-velcond2}
V_S \geqslant 11 \kms\ \left(\frac{\densini}{1\ \cc}\right)^{-3/2}\ \left(\frac{B_0}{1\ \mu{\rm G}}\right)\ G_0\,^{0.5}.
\end{equation}
The validity of these conditions is discussed in Sect. \ref{Sect-grid}.

\subsection{Emitted spectrum}

\begin{figure}[!h]
\begin{center}
\includegraphics[width=10.0cm,trim = 1.7cm 1.5cm 0.7cm 1.0cm, clip,angle=0]{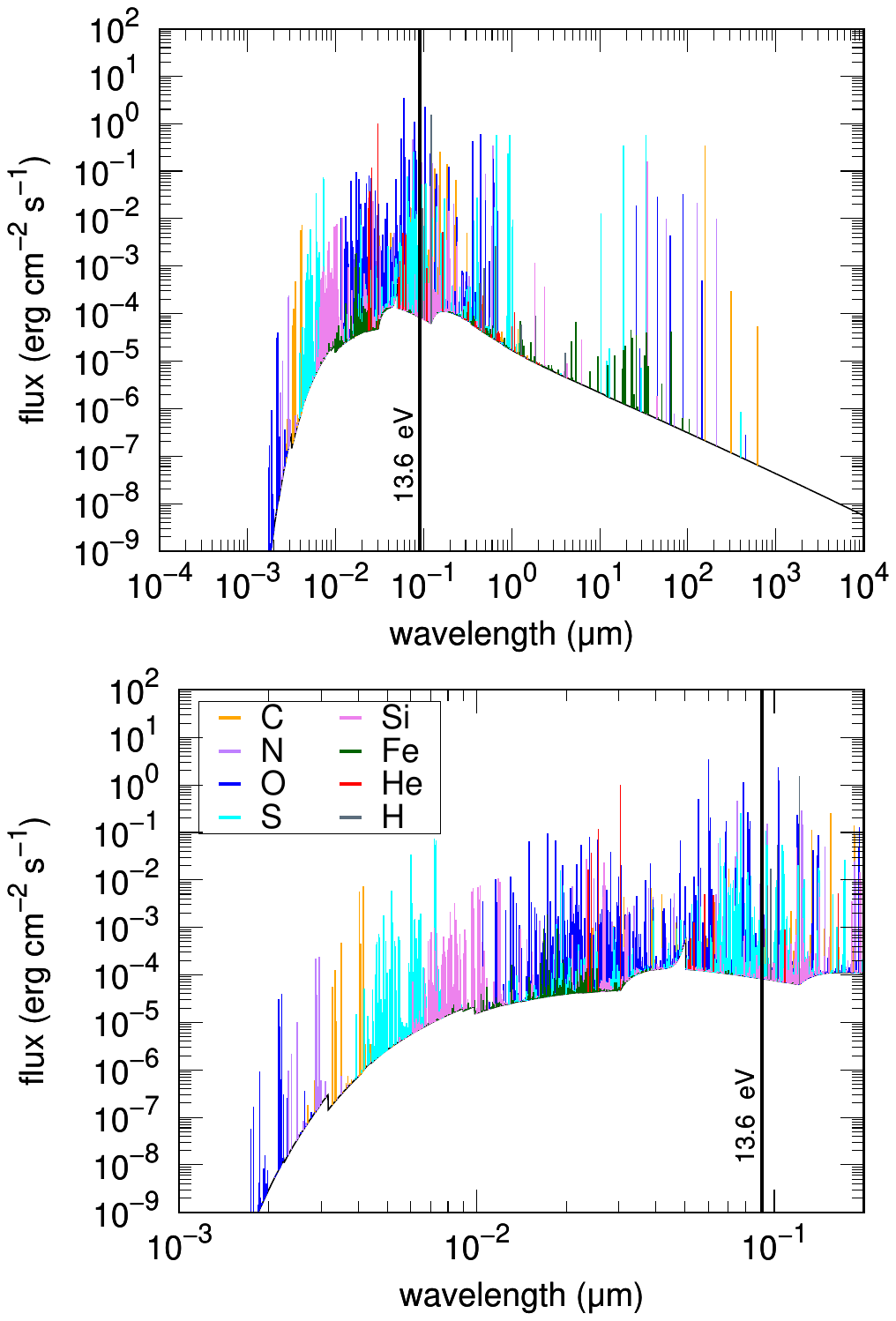}
\caption{Energy flux, computed as $\nu F_\nu^{\rm int}$, emitted toward the preshock at $z=0$ in the standard model as a function of the wavelength of the emitted photon. The black curve corresponds to the total continuum emission (including Bremsstrahlung, bound-free, and two-photon processes) and the colored spikes to line emissions. Spikes are colored according to the element responsible for the line emission regardless of its ionization state. The top panel displays the full wavelength range while the bottom panel is a zoom onto the range between 10 and 2000\,\,\AA. The black vertical line indicates the ionization threshold of hydrogen.}
\label{Fig-spectrum}
\end{center}
\end{figure}

As described in the previous sections, the radiation field generated by the shock has a fundamental impact on the thermochemical structure of the gas. In addition, since most of the initial kinetic energy is reprocessed into line and continuum emission \citep{Allen2008}, this radiation field is a unique signature of the shock itself. To illustrate the conversion process, we display in Fig.~\ref{Fig-spectrum} the energy flux, $\nu F_\nu^{\rm int}$, emitted toward the preshock, at the origin of the distances (defined as $z=0$, see Fig.~\ref{Fig-convergence}), by the standard model. Fig.~\ref{Fig-spectrum} summarizes the strengths and the relative contributions of the different emission processes, including the Bremsstrahlung, bound-free, and two-photon continuum emissions, and the emissions of resonant and non-resonant lines.

Figure~\ref{Fig-spectrum} displays the fluxes of all the 763970 lines contained in the CHIANTI database (see Sect. \ref{Sect-cooling}), regardless of their strength. For the standard model, we find that only $\sim 2200$ lines have an energy flux above 3\% of the continuum level. These dominant lines carry more than 90\% of the initial flux of mechanical energy. About 56\% of the line flux is emitted in the EUV and X-ray domain ($\lambda \leqslant 911$~\AA), 37\% in the UV range ($911 < \lambda \leqslant 2400$~\AA), and 7\% in the optical and the infrared ($\lambda > 2400$~\AA). All the elements display clear line features that extend from a few tens of Angstr\"oms to a few hundreds of micrometers. Yet, the energy budget shows that most of the photon emission is carried out by electronic lines of He$^+$, O$^{3+}$, O$^{4+}$, O$^{5+}$, and H, and to a lesser extent by the electronic lines of C$^+$, C$^{2+}$, C$^{3+}$, N$^{2+}$, N$^{3+}$, N$^{4+}$, S$^+$, and S$^{2+}$. Seemingly prominent in Fig.~\ref{Fig-spectrum}, both the Bremsstrahlung and two-photon continuum emissions only carry a few percent of the initial flux of mechanical energy.

\section{Grid of models of WNM shocks} \label{Sect-grid}

In the previous section, the Paris-Durham shock code was applied with the standard set of parameters in order to validate the recent updates and highlight the thermochemical state of the gas and the main processes at play in shocks propagating in the WNM. To test the robustness of the code and broaden its applications, we present here an exploration performed over a grid of models that encompass the typical physical conditions of the WNM expected in the solar neighborhood. The grid of models contains 475 runs with an initial preshock density varying between 0.1 and 2 \cc, a transverse magnetic field varying between 0.1 and 10 $\mu$G, and a shock velocity varying between 10 and 500 \kms.

\subsection{Critical velocities and expected shock types}

\begin{figure}[!h]
\begin{center}
\includegraphics[width=9.5cm,trim = 1.7cm 1.5cm 2.0cm 1.0cm, clip,angle=0]{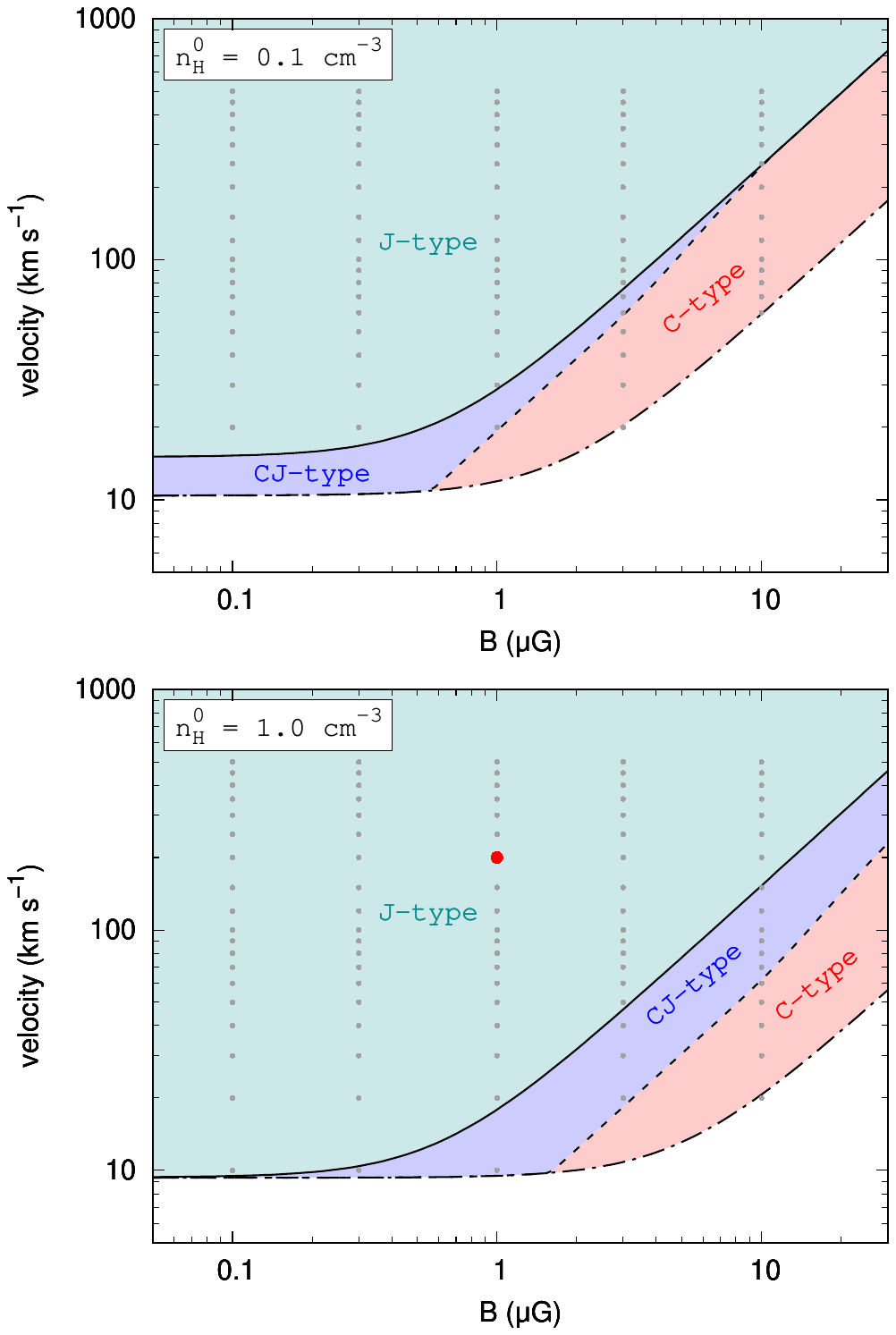}
\caption{Critical velocities of interstellar shocks for $\densini = 0.1$ \cc\ (top panel) and 1 \cc\ (bottom panel). The solid and dash-dotted lines show the magnetosonic speed computed in the ionized fluid (Eq. \ref{Eq-ims}) and the neutral fluid (Eq. \ref{Eq-nms}), respectively. The colored areas highlight the range of velocities over which stationary shocks are C-type (red), CJ-type (blue), and J-type (green) shocks. The transition between C-type and CJ-type shocks (dashed line) is estimated as the point where the maximum postshock thermal pressure is equal to the maximum magnetic pressure allowed by an adiabatic compression. The gray points indicate the values of the shock velocity and transverse magnetic field strength explored in the grid of models. The red dot highlights the position of the standard model.}
\label{Fig-Critical-Vel}
\end{center}
\end{figure}

The nature of an interstellar shock is the result of a subtle interplay between the strength of the magnetic field, the ionization degree, and the cooling rate (e.g.,  \citealt{Chernoff1987,Roberge1990,Godard2019}). The prototypical model presented in the previous section is a single fluid interstellar shock in which the ionized and the neutral fluids are fully coupled with each other. This kind of shock is referred to as a J-type shock because of the discontinuity in the temperature, velocity, and density profiles, which abruptly change at the scale of the particle mean free path (see Fig.~\ref{Fig-convergence} and Sect.~\ref{Sect-viscosity}). However, different physical conditions of the preshocked gas and different shock velocities may lead to a decoupling between the ionized and the neutral fluids and to the formation of a magnetic precursor. This process gives rise to other kinds of shock structures called  CJ-type and C-type shocks depending on whether the gas abruptly becomes subsonic at some point along the trajectory or not. The types of shocks obtained over the grid of models explored here are shown in Fig.~\ref{Fig-Critical-Vel}.

The division between the domains of existence of different types of shocks can be understood by comparing the shock velocity with the characteristic wave speeds of the interstellar gas. The magnetosonic speeds of a medium with a magnetic field strength $B$ are
\begin{equation} \label{Eq-nms}
c_{\rm nms} = \left(c_n^2 + B^2 / 4\pi \rho_n\right)^{1/2}
,\end{equation}
and
\begin{equation} \label{Eq-ims}
c_{\rm ims} = \left(c_i^2 + B^2 / 4\pi \rho_i\right)^{1/2}
,\end{equation}
where $c_n$ and $c_i$ are the sound speeds in the neutral and the ionized fluids, and $\rho_n$ and $\rho_i$ are the mass densities of the neutrals and the ions. In the limit of strong coupling between the ions and the neutrals, the minimum speed required for a shock to propagate is given by the neutrals magnetosonic speed (Eq. \ref{Eq-nms}). If the shock velocity is larger than the ions magnetosonic speed (Eq. \ref{Eq-ims}), the shock is a J-type shock. Below this limit, the shock is either a CJ-type or a C-type shock. Because the cooling rate of the WNM is inefficient to reduce the temperature gradients induced by the compression and the ion-neutral drift, the limit between these two kinds of shocks can be estimated as the point where the maximum postshock thermal pressure (given by Eq. \ref{Eq-postpress}) is equal to the maximum magnetic pressure allowed by an adiabatic compression. This limit is shown as a dashed line in Fig.~\ref{Fig-Critical-Vel} and is found to be in excellent agreement with the results of the Paris-Durham shock code, which shows that shocks below (respectively, above) this value are C-type (respectively, CJ-type).


\subsection{Phase trajectories}

\begin{figure*}[!h]
\begin{center}
\includegraphics[width=19.0cm,trim = 1.5cm 2.0cm 0.5cm 0.5cm, clip,angle=0]{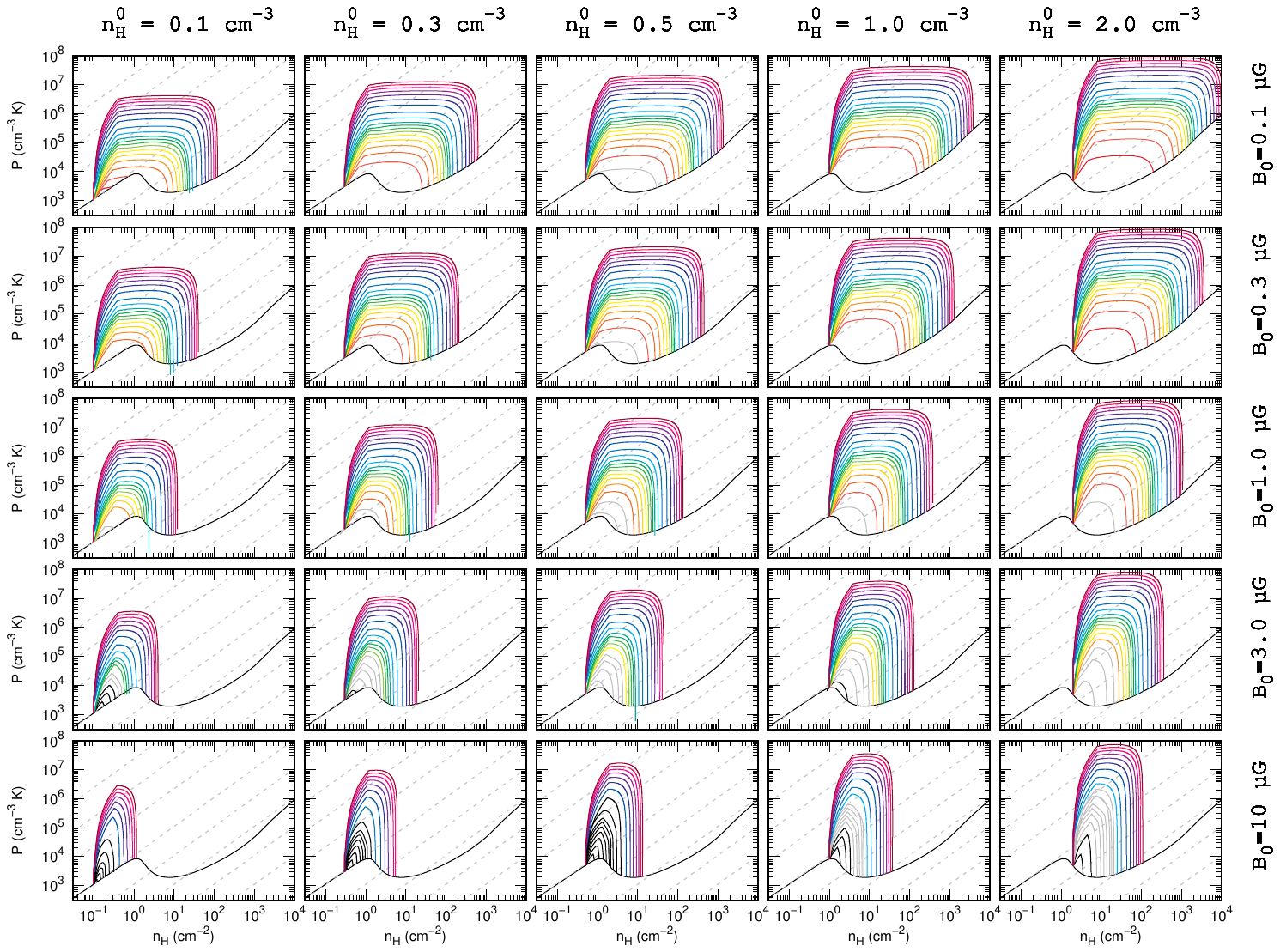}
\caption{Trajectories followed by interstellar shocks in a proton density versus thermal pressure diagram (see Fig.~\ref{Fig-Trajectory-Main}). Each panel corresponds to different values of the initial density $n_{\rm H}^0$ (top labels) and transverse magnetic field $B_0$ (right labels). The different curves correspond to the following values of the shock velocity ordered from top to bottom trajectories: 500, 450, 400, 350, 300, 250, 200, 150, 120, 100, 90, 80, 70, 60, 50, 40, 30, 20, and 10 \kms. Colored trajectories correspond to J-type shocks, gray trajectories to CJ-type shocks, and black trajectories to C-type shocks. The black curve common to all panels indicates the thermal equilibrium state of the diffuse gas as a function of the proton density obtained for $G_0=1$ (see Fig.~\ref{Fig-equilibrium}). Light gray lines are isothermal contours from 1 to $10^7$ K (from bottom right to top left). We note that the trajectories of low-velocity shocks are not always visible as they only slightly perturb the gas from its original state.}
\label{Fig-traj-grd}
\end{center}
\end{figure*}

The trajectories followed by all the models explored in the grid are shown in Fig.~\ref{Fig-traj-grd} in a proton density versus thermal pressure diagram. In many ways, this simple figure is the main outcome of this work. It provides, at a glance, an almost complete summary of all the models explored. It also demonstrates the robustness of the code for velocities between 10 and 500 \kms\ and the coherence of its predictions compared to the theoretical and analytical descriptions given in the previous sections.

For standard values of the magnetic field expected in the WNM ($B_0 \lesssim 1$ $\mu$G), most of the models are found to be J-type shocks, except at low velocities ($V_S \lesssim 20$ \kms) where the code converges toward stationary CJ-type solutions. This result is in agreement with the expectations given in Fig.~\ref{Fig-Critical-Vel} based on the sole consideration of characteristic wave speeds. In this parameter domain, the predicted trajectories are found to follow the description given in Sect. \ref{Sect-phase-standard}. In particular, interstellar shocks display successive adiabatic, isobaric, and isochoric evolutions with a maximum thermal pressure and a final compression that match the expected values given by Eqs. \ref{Eq-postpress} and \ref{Eq-densfin}. It follows that interstellar shocks are found to induce phase transition between the WNM and the CNM as long as the shock velocity verifies the conditions given in Eqs. \ref{Eq-velcond1} and \ref{Eq-velcond2}. These criteria are fulfilled by a vast majority of the shocks, except at low preshock density and low velocity where the postshock thermal pressure is below the maximum pressure of the WNM, and at low preshock density and large values of $B_0$ where the maximum density of the final state is below the minimum density of the CNM.

For larger values of the strength of the magnetic field ($B_0 \gtrsim 3$ $\mu$G), a substantial fraction of the models are found to be C-type shocks, which is in agreement with the expectation given in Fig.~\ref{Fig-Critical-Vel}. The strength of the preshock magnetic field has mostly two impacts on the shock structure. When the preshock magnetic pressure is comparable to the ram pressure, the magnetic field is strong enough to reduce or even remove the rise of the gas temperature. The initial growth of thermal pressure thus occurs with a lower slope than that predicted by the adiabatic hydrodynamic jump conditions. When the Alfv\'enic Mach number is $\lesssim 3$, the magnetic pressure in the postshock becomes strong enough to entirely suppress the isobaric cooling stage. It follows that the gas rapidly reaches an isochoric cooling evolution toward thermal equilibrium. Although many shocks in this domain of parameter occur through ion-neutral decoupling, or with reduced postshock thermal pressure, the criteria established to obtain phase transition are still valid. It is so because, for large values of the magnetic field, the sole criterion that dictates a possible transition between the WNM and the CNM is Eq. \ref{Eq-velcond2}, which is simply derived from the conversion of the initial ram pressure into magnetic pressure.

Interestingly, Fig.~\ref{Fig-traj-grd} shows that the final states of several shocks explored in the grid lie on the unstable branch of the thermal equilibrium state (black curve in Fig.~\ref{Fig-traj-grd}). Because the shocks are described as stationary plane-parallel structures, the model predicts that the postshock gas remains on this unstable branch indefinitely. At first sight, these solutions could appear to be unphysical \citep{Falle2020a}. However, when such shocks occur in nature, the gas effectively follows the trajectories displayed in Fig.~\ref{Fig-traj-grd}. The unstable nature of the final state just implies that the postshock gas then naturally splits up into a warm and a cold neutral components with mass fractions that depend on the thermal pressure of the surrounding gas. This final evolution doesn't taint the predictions of this work regarding the thermochemical evolution and the emission properties of the shock. 

Conversely, Fig.~\ref{Fig-traj-grd} shows that all the models at $\densini=2$~\cc\ start from an unstable thermal state. This is due to the fact that the preshock conditions are calculated assuming thermochemical equilibrium regardless of possible thermodynamical instabilities. These conditions might seem unrealistic if we consider that the diffuse medium is composed solely of WNM and CNM. This is not the case. Indeed, observations of the 21~cm line of atomic hydrogen show that the unstable phase, sometimes called the Lukewarm Neutral Medium (LNM, \citealt{Marchal2019a, Bellomi2020}), contains a substantial fraction ($\sim 20$\%) of the mass of the diffuse ISM (e.g., \citealt{Murray2018a,Marchal2019a}). Although its volume filling factor is small compared to that of the WNM, the LNM is still an important component of the diffuse matter where shocks could develop. The unstable initial conditions obtained at $\densini=2$~\cc\ can therefore be seen as an idealized modeling of shocks propagating in the LNM. 

\subsection{Preshock ionization state}

\begin{figure}[!h]
\begin{center}
\includegraphics[width=9.2cm,trim = 1.9cm 5.2cm 0.7cm 2.3cm, clip,angle=0]{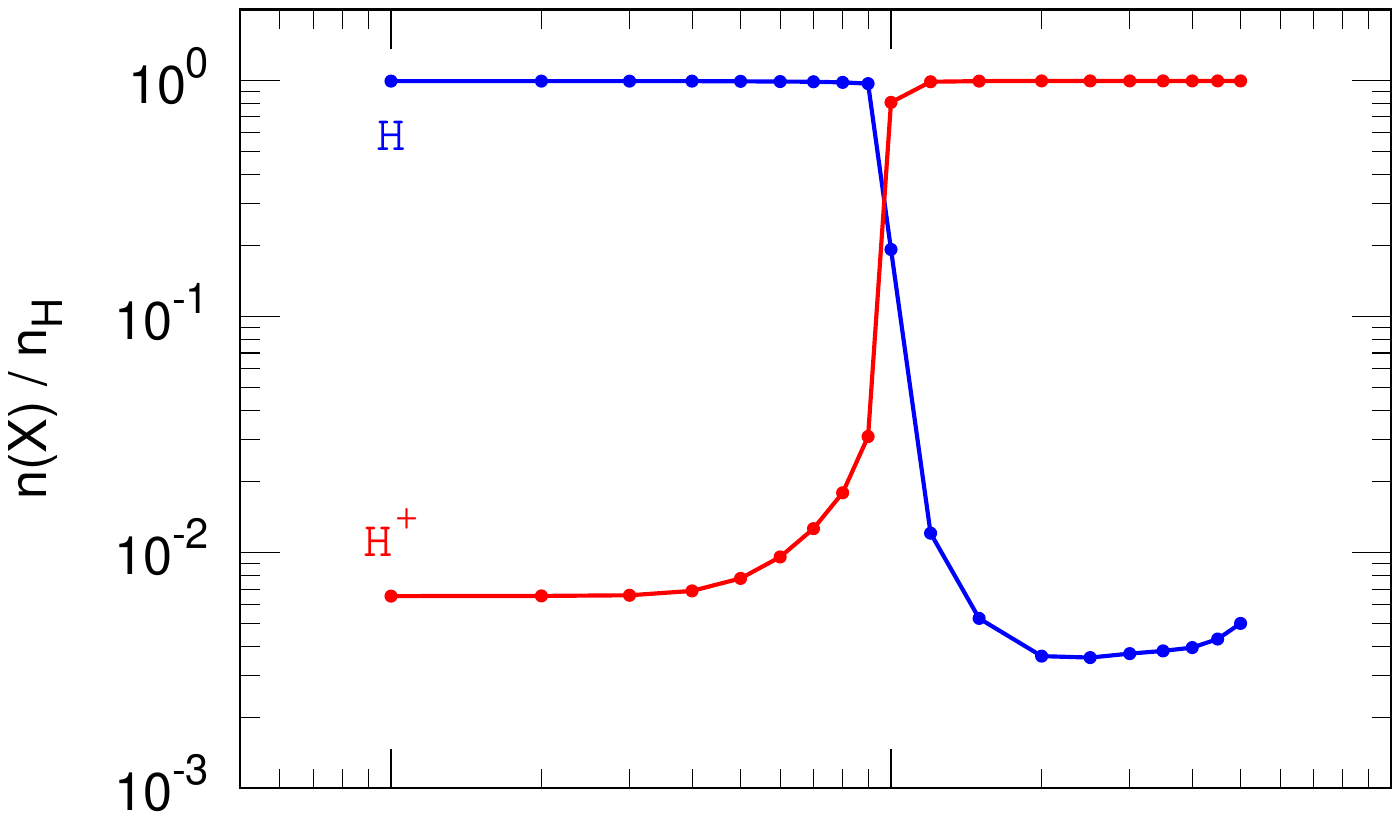}
\includegraphics[width=9.2cm,trim = 1.9cm 5.2cm 0.7cm 2.3cm, clip,angle=0]{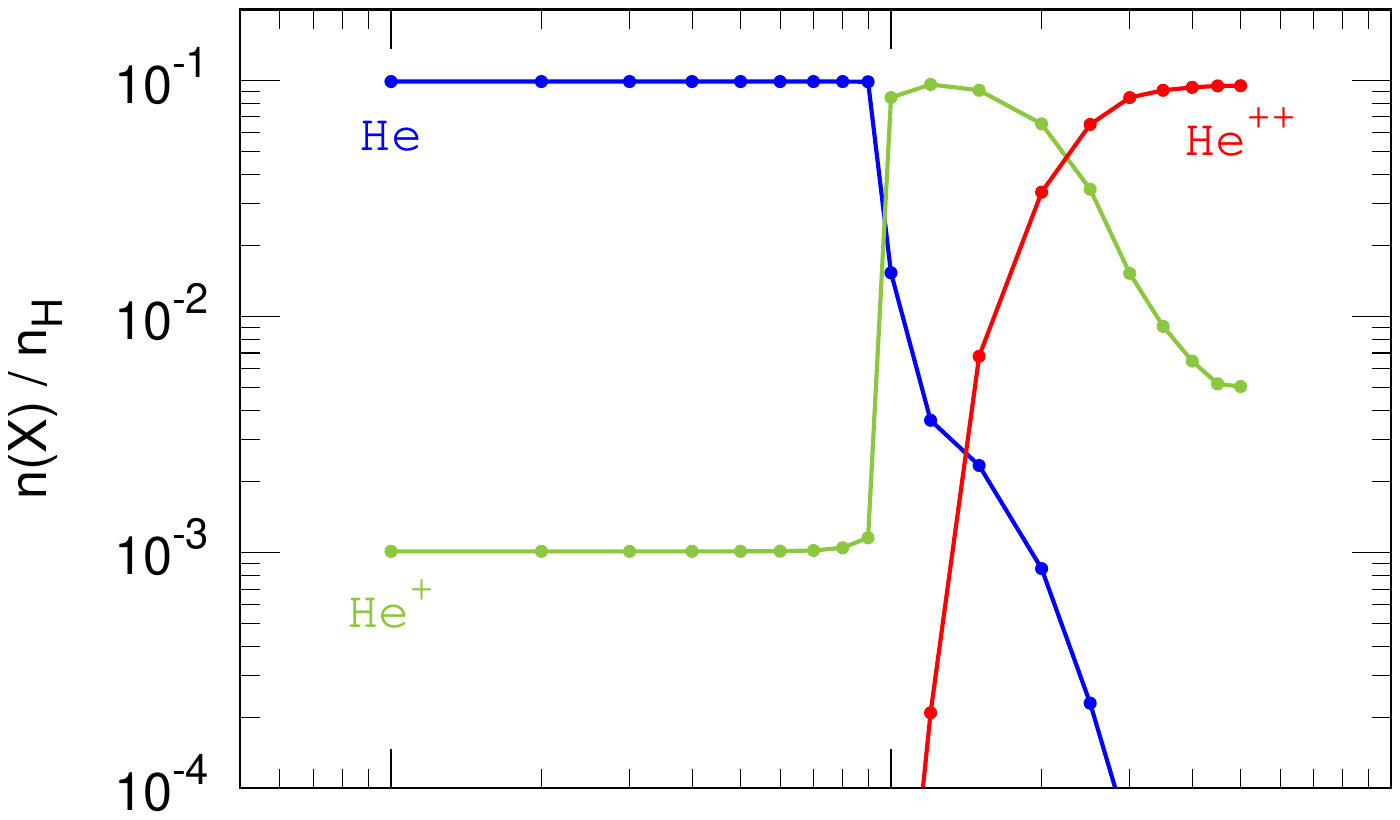}
\includegraphics[width=9.2cm,trim = 1.9cm 2.5cm 0.7cm 2.3cm, clip,angle=0]{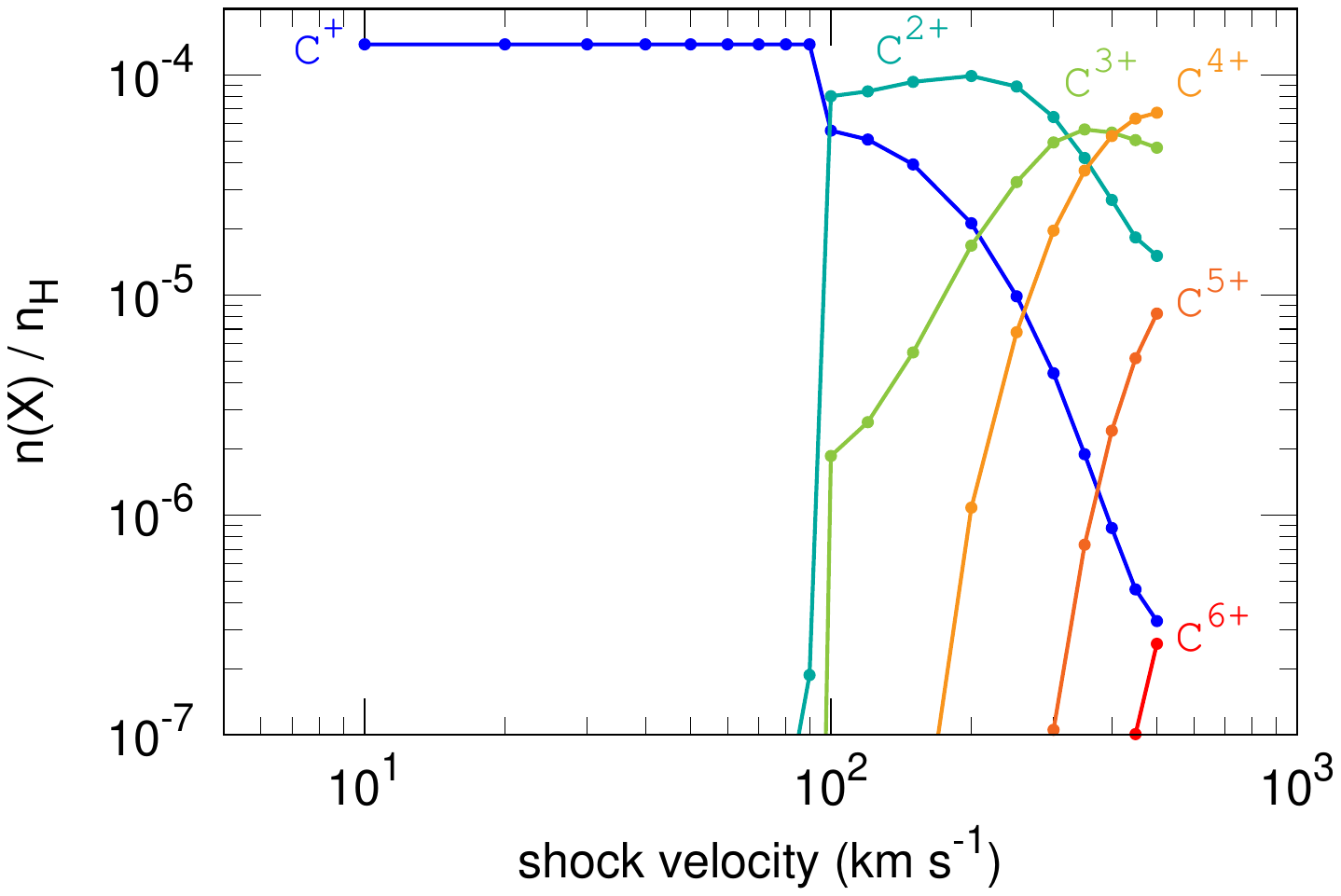}
\caption{Fractional abundances of H and \Hp\ (top panel), He, \Hep, and \Hepp\ (middle panel), and C$^+$, C$^{2+}$, C$^{3+}$, C$^{4+}$, C$^{5+}$, and C$^{6+}$ (bottom panel) in the gas entering the shock as functions of the shock velocity. Except for the velocity, all other parameters are set to their standard values (see Table~\ref{Tab-main}).}
\label{Fig-prsh-ionization}
\end{center}
\end{figure}

The thermochemical profiles of intermediate- and high-velocity shocks and the potential decoupling between the ionized and the neutral flows are driven, in part, by the impact of the radiation field generated by the shock on the thermochemical state of the preshock gas. To illustrate this aspect, we display in Fig.~\ref{Fig-prsh-ionization} the chemical composition of the radiative precursor (at $z=0$) in hydrogen-, helium-, and carbon-bearing compounds as a function of the velocity of the shock for $\densini=1$~\cc\ and $B_0=1$~$\mu$G. This figure shows that the ionization state of the preshock gas displays several transitions depending on the shock velocity, and therefore on the strength and hardness of the emitted radiation field.

At $V_S \sim 100$~\kms, the radiative precursor is found to abruptly switch to a medium dominated by H$^+$, He$^+$ and C$^{2+}$. The reason for this sharp transition is that a shock at 100~\kms\ not only collisionally ionizes neutral helium but is also strong enough to collisionally excite the 300~\AA\ lines of He$^+$. These lines, which dominate the EUV spectrum (see Fig.~\ref{Fig-spectra}), correspond to photon energies of $\sim 40$~eV much larger than the ionization potential of H ($\sim 13.6$~eV), He ($\sim 24.6$~eV), and C$^+$ ($\sim 24.4$~eV), yet still smaller than the ionization potential of C$^{2+}$ ($\sim 47.9$~eV). This behavior is in line with the Figs. 7, 8, and 9 of \citet{Allen2008} and in excellent agreement with the predictions of \citet{Hollenbach1989} who show that the high-velocity shocks fully ionize the radiative precursor for shock velocities $V_S \gtrsim 120$~\kms\ (Fig.~11 in their paper).

As the shock velocity increases, the radiative precursor displays progressive transitions toward higher ionization stages. For instance, we find that the preshock gas is mainly composed of He$^{2+}$ above $\sim 200$~\kms, and mainly composed of C$^{3+}$ and C$^{4+}$ above $\sim 300$~\kms. This smooth transitions toward higher ionization stages comes from the fact that shocks at larger and larger velocities progressively excite a forest of lines of atomic ions at larger and larger energies (see Fig.~\ref{Fig-spectra}), which contribute together to the ionization of the different elements.

\subsection{Reprocessing of kinetic energy}

\begin{figure*}
\begin{center}
\includegraphics[width=17.0cm,trim = 1.5cm 1.5cm 0.3cm 0.5cm, clip,angle=0]{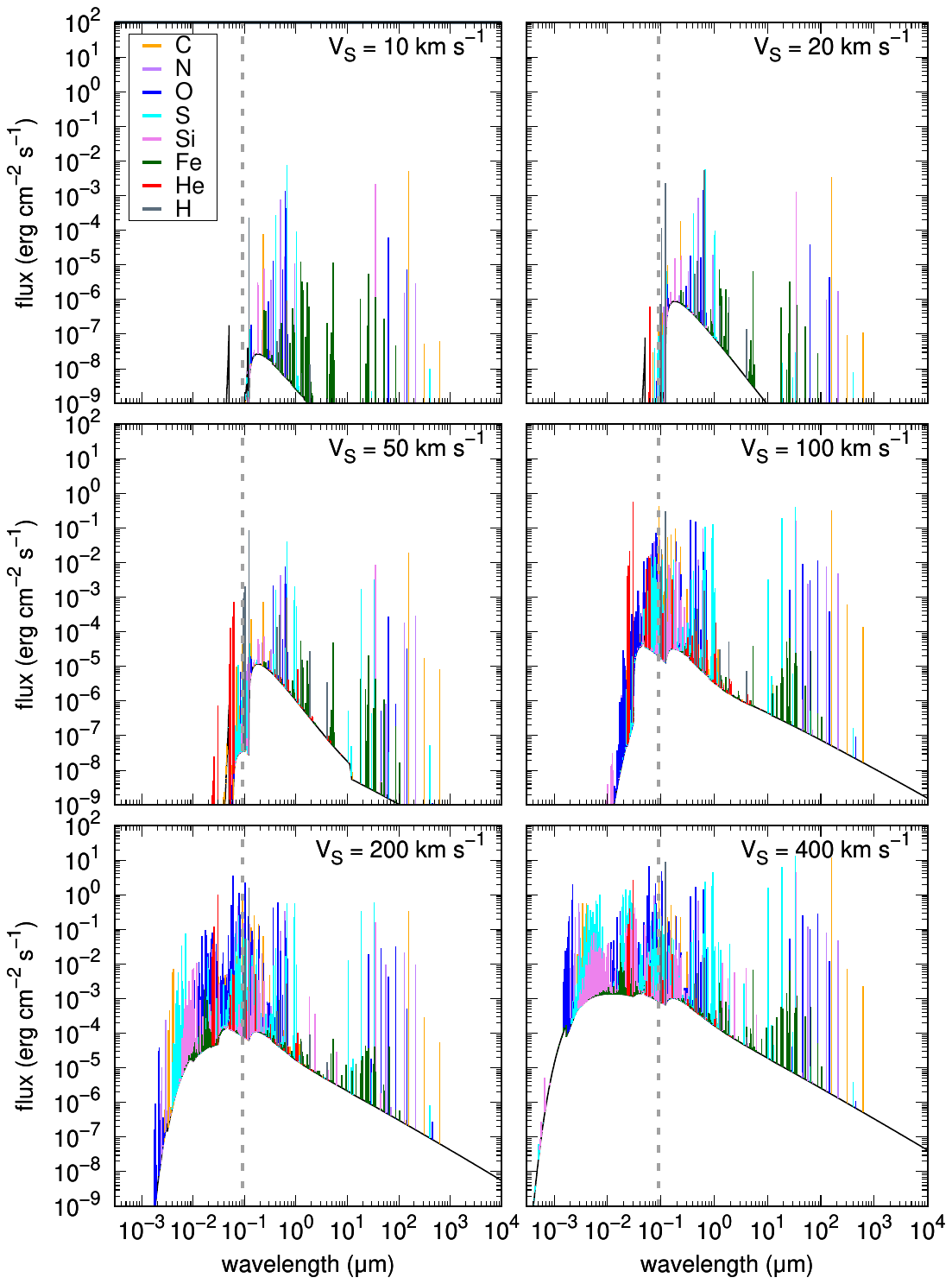}
\caption{Energy flux, computed as $\nu F_\nu^{\rm int}$, emitted toward the preshock at $z=0$ as a function of the wavelength of the emitted photon. The black curve corresponds to the total continuum emission (including Bremsstrahlung, bound-free, and two-photon processes) and the colored spikes to line emissions. Spikes are colored according to the element responsible for the line emission regardless of its ionization state. Predictions are shown for shocks at different velocities, ranging from 10 \kms\ (top left panel) to 400 \kms\ (bottom right panel). All other parameters are set to their standard values (see Table \ref{Tab-main}). The dashed gray vertical line indicates the ionization threshold of hydrogen.}
\label{Fig-spectra}
\end{center}
\end{figure*}

\begin{figure}[!h]
\begin{center}
\includegraphics[width=9.0cm,trim = 1.9cm 2.5cm 0.7cm 2.0cm, clip,angle=0]{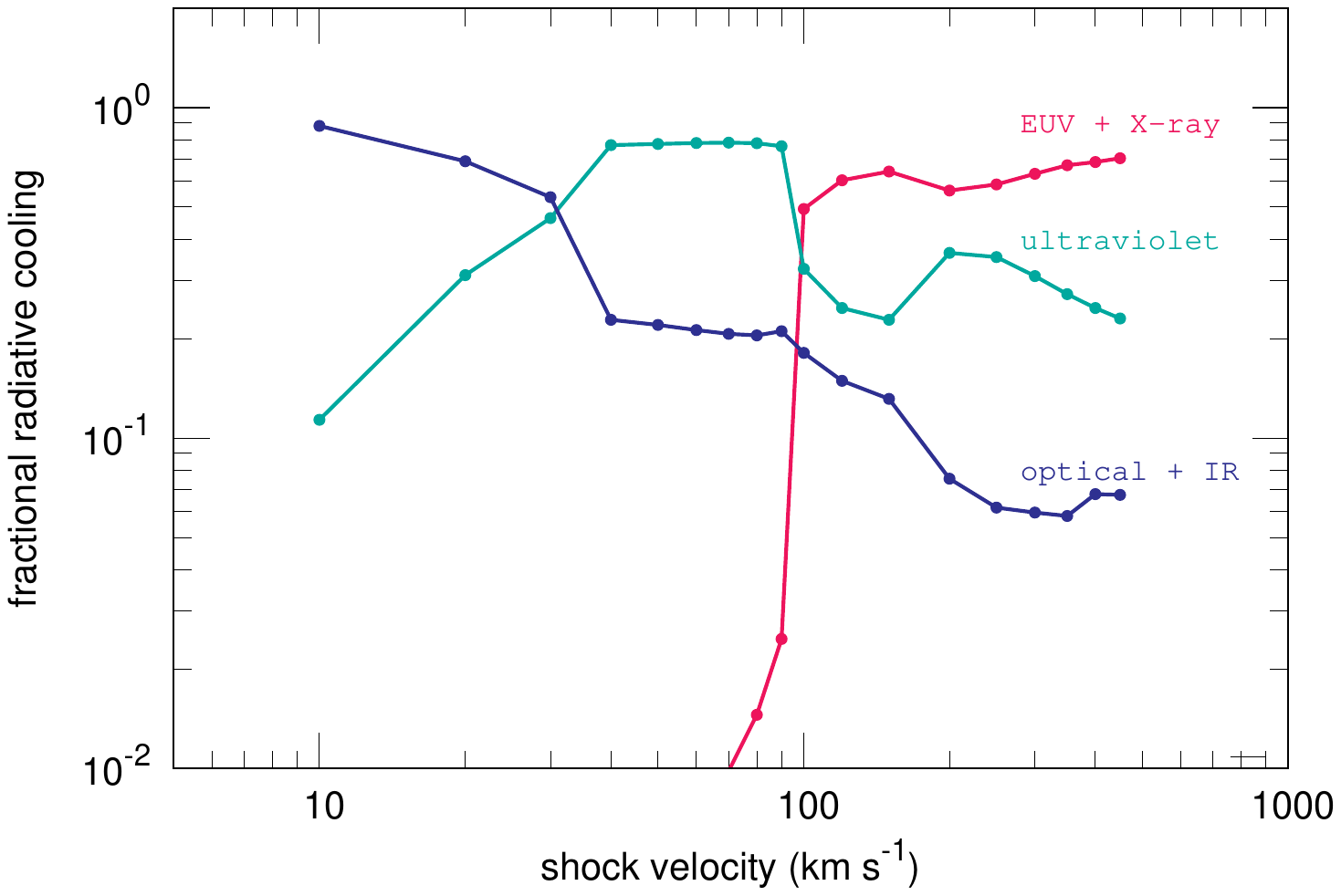}
\caption{Fraction of the radiation emitted by shocks in the X-ray and EUV domains (red curve), in the ultraviolet range (green curve), and in the optical and the infrared domains (blue curve) as functions of the shock velocity for $\densini = 1$~\cc\ and $B_0=1$~$\mu$G.}
\label{Fig-reprocessing}
\end{center}
\end{figure}

As an illustration of the reprocessing of the initial mechanical energy, we display in Fig.~\ref{Fig-spectra} the spectra emitted by six shocks with velocities of 10, 20, 50, 100, 200, and 400 \kms, and in Fig.~\ref{Fig-reprocessing} the fraction of the radiation (including lines and continuum) emitted over different domains of wavelength. For all the models displayed here, we find that most of the input flux of mechanical energy is reprocessed into line emission. This result extends almost over the entire grid of models, except for low Alfv\'enic Mach number where most of the energy goes in the compression of the initial magnetic field. Shocks at low velocity  ($V_S \leqslant 30$ \kms) mostly emit photons in the infrared and submillimeter domains. Shocks at intermediate velocities ($30 < V_S < 100$ \kms) emit most of their radiation in the UV and EUV range, while shocks at high velocity ($V_S \geqslant 100$ \kms) preferentially emit in the EUV and X-ray domains.

At low and intermediate velocities, the amount of ionizing radiation ($> 13.6$ eV) is found to be almost entirely carried by a few lines of He and He$^+$ (red spikes in Fig.~\ref{Fig-spectra}). This feature is responsible for the abrupt jump in the chemical state of the preshock medium obtained at $V_S \sim 100$ \kms\ (see Fig.~\ref{Fig-prsh-ionization}). Conversely, the ionizing radiation emitted by shocks at high velocity results from a combination of electronic lines of helium, carbon, oxygen, nitrogen, and sulfur in various stages of ionization. The gas-phase elemental abundances adopted here imply that the cooling is dominated by the different ionization stages of oxygen. However, figure~\ref{Fig-spectra} suggests that this would not necessarily be the case for a different set of elemental abundances.

The continuum flux is found to be largely dominated by two-photon emission of hydrogen and helium as long as $V_S \lesssim 200$ \kms. Above this value, most of the continuum photons originate from Bremsstrahlung emission. Even though the continuum radiation becomes more prominent at larger velocities, it remains subdominant compared to line cooling over the entire grid of models. Given the dependences of the cooling rates induced by collisional excitation and the free-free radiation on the gas temperature, we estimate that continuum radiation should dominate over line emission for shock velocities $V_S \gtrsim 1000$ \kms. This domain of parameter, which dangerously starts to fall in the realm of relativistic shocks, is far outside the scope of the present study.

One last remarkable aspect of Fig.~\ref{Fig-spectra} is the behavior of the infrared and submillimeter parts of the spectra as a function of the shock velocity. Indeed, the fluxes of the dominant infrared and submillimeter lines are modified from 10 to 400~\kms, but only slightly compared to the rise of the flux of mechanical energy ($6.4 \times 10^4$ over this range of velocities). This feature highlights the fact that these lines are emitted when the gas cools from $\sim 10^4$ to $\sim 70$ K. This evolution is common to all the shocks displayed here (see Fig.~\ref{Fig-traj-grd}). The resulting infrared and submillimeter spectrum therefore weakly depends on the shock velocity.

\subsection{Induced photo-ionization rates}

\begin{figure}[!h]
\begin{center}
\includegraphics[width=9.0cm,trim = 1.5cm 2.5cm 0.3cm 2.0cm, clip,angle=0]{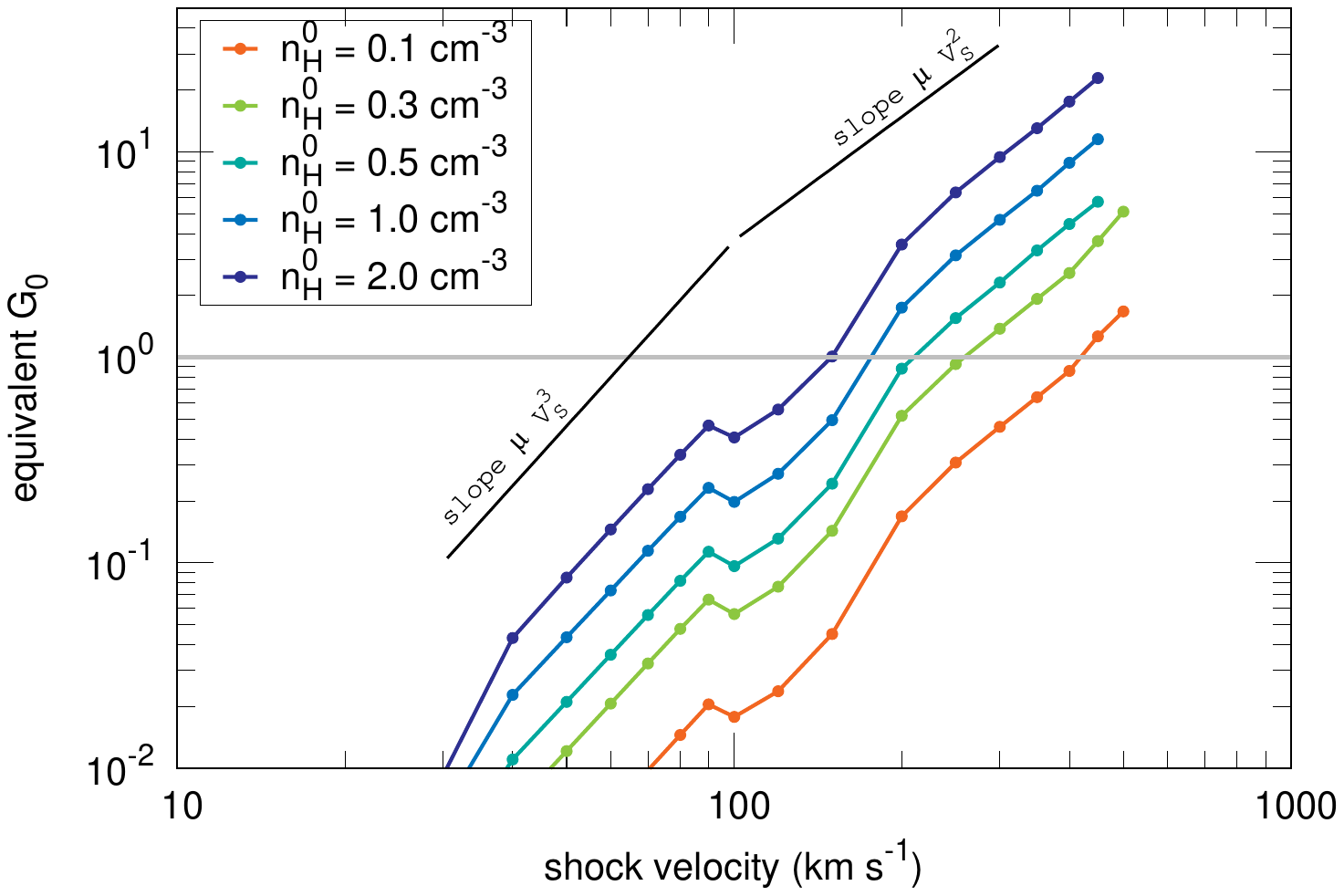}
\includegraphics[width=9.0cm,trim = 1.5cm 2.5cm 0.3cm 2.0cm, clip,angle=0]{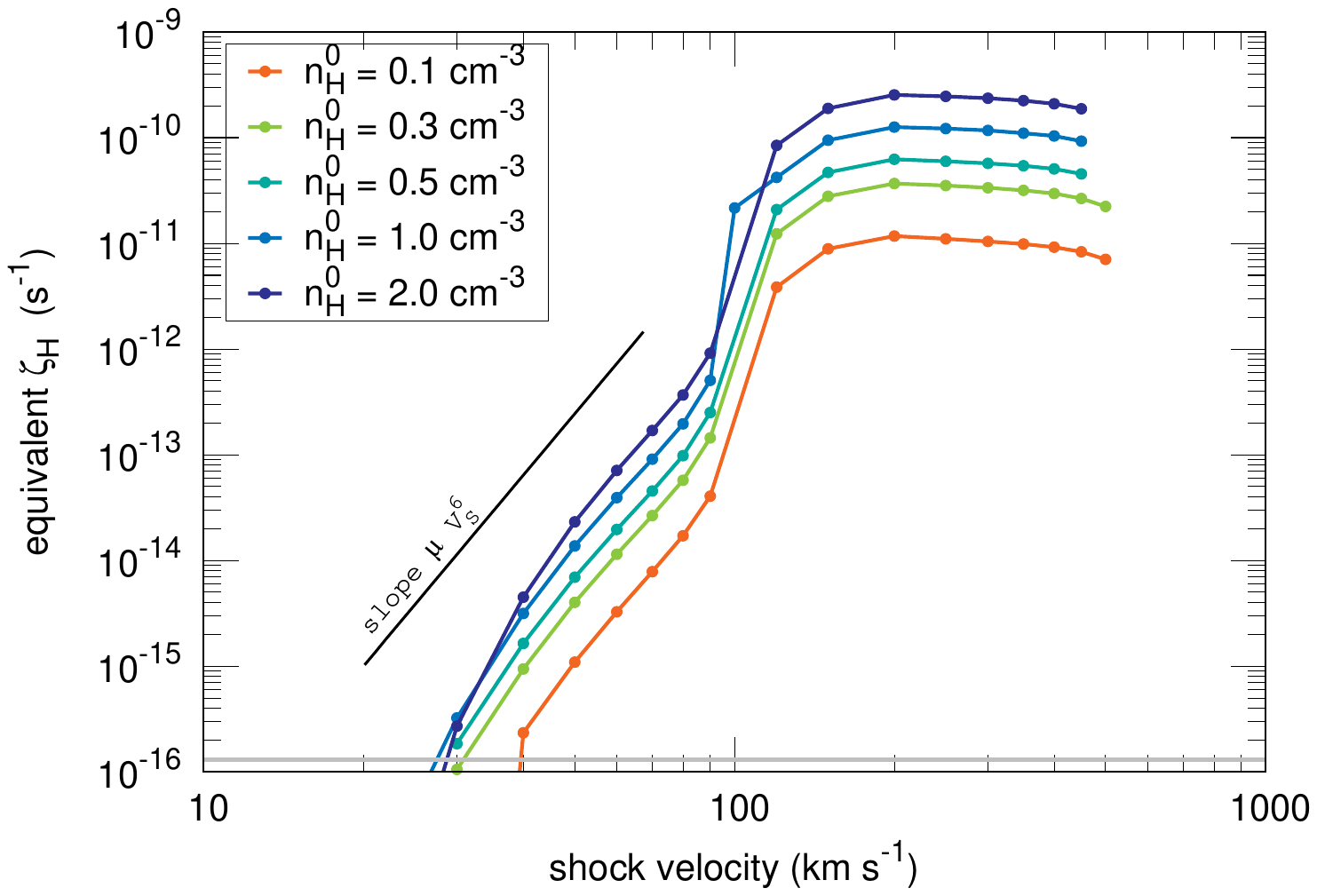}
\caption{Strength of the self-generated UV radiation field (top panel) and ionization rate of hydrogen induced by the self-generated EUV and X-ray radiation field (bottom panel) in the radiative precursor, at $z=0$. Both quantities are shown as functions of the shock velocity for different values of the preshock density and for $B_0=1$~$\mu$G. The strength of the UV radiation field is calculated from the integral of $I_\nu^{\rm int}$, over solid angle and frequency between 911 and 2400 \AA, normalized to the value obtained by Mathis \citep{Mathis1983}. The strength of the external UV radiation field used in this work ($G_0=1$) is shown as a gray line on the top panel. The cosmic ray ionization rate of atomic hydrogen used in this work ($\zeta_{\rm H} = \zeta_{{\rm H}_2} /2.3 = 1.3 \times 10^{-16}$ s$^{-1}$) is shown as a gray line on the bottom panel.}
\label{Fig-equivalent-ionization}
\end{center}
\end{figure}

\begin{figure}[!h]
\begin{center}
\includegraphics[width=9.0cm,trim = 1.9cm 2.0cm 0.7cm 1.0cm, clip,angle=0]{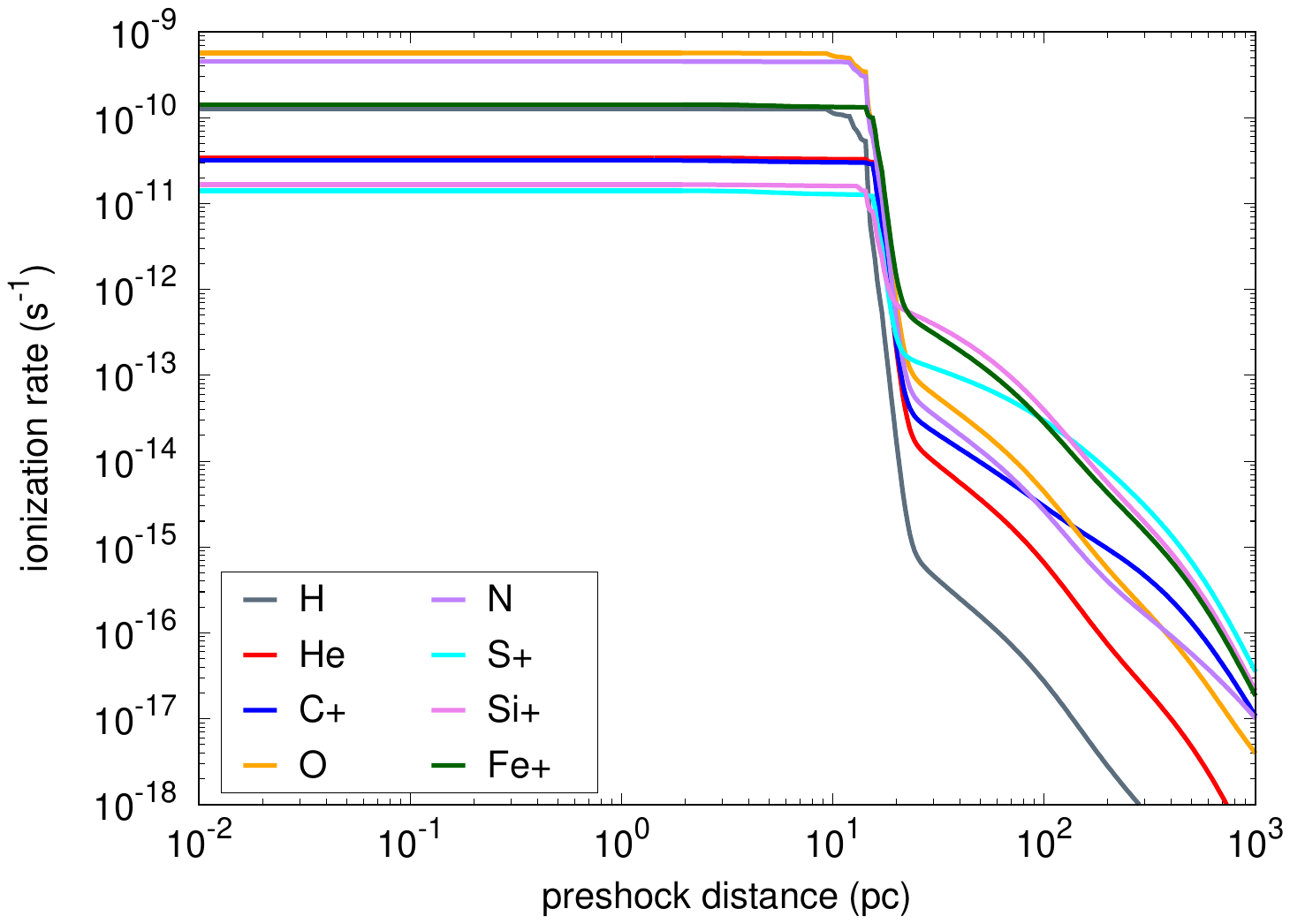}
\caption{Ionization rates of H, He, \Cp, O, N, \Sp, \Sip, and \Fep\ induced by the standard shock in the preshock gas as functions of the preshock distance. These species are selected because they are the main carriers of the elements included in the model in the WNM.}
\label{Fig-prsh-ionization-rate}
\end{center}
\end{figure}

Figure~\ref{Fig-equivalent-ionization} displays the strengths of the UV radiation field and the ionization rate of hydrogen resulting from the EUV and X-ray photons generated by the shocks for different velocities and preshock densities, and for $B_0=1$~$\mu$G. As already suggested by Figs. \ref{Fig-Critical-Vel} and \ref{Fig-traj-grd}, this result is found to be independent of the strength of the transverse magnetic field as long as $B_0 \leqslant 1$~$\mu$G, which correspond to the standard values expected in the WNM.

As intermediate-velocity shocks mostly emit UV photons, the strength of the UV radiation field generated by shocks is found to be proportional to the input flux of mechanical energy and to scale as $V_S^3$ as long as $V_S < 100$~\kms. Above this value, the self-generated UV radiation field still increases as a function of the shock velocity but with a shallower slope ($\propto V_S^2$), as most of the energy is radiated away in the X-ray domain. We find that the equivalent $G_0$ is below the fiducial value associated with the external UV radiation field ($G_0=1$) as long as $V_S < 200$~\kms\ $(0.5\ \cc/\densini)^{1/3}$. This indicates that only high-velocity shocks propagating in dense environments may have a substantial impact on the UV irradiation of the surrounding interstellar gas.

Intermediate-velocity shocks are found to generate a photoionization rate of hydrogen that scales as $V_S^6$ for $V_S < 100$~\kms. This rate suddenly jumps at $V_S \sim 100$~\kms\ because of the excitation of the electronic levels of He$^+$ and reaches a plateau for higher velocities, $\zeta_{\rm H} = 10^{-10}$~s$^{-1}$~$(\densini/1\ \cc)$. This plateau comes from the fact that high-velocity shocks radiate the input mechanical energy by hardening the X-ray radiation field rather than increasing the output flux of EUV photons (see Fig.~\ref{Fig-spectra}). Since the photoionization cross-section of hydrogen quickly drops as the photon energy increases ($\sigma_{\rm H}(\nu) \propto \nu^{-3}$, see Fig.~\ref{Fig-cross-sections}), the hardening of the X-ray radiation field has no impact on the ionization rate. All these behaviors only apply to atomic hydrogen. Ionization rates of other species in different ionization stages display different scalings as functions of the velocity of the shock.

The spatial extensions of the ionization rates of the main constituents of the WNM predicted by the standard model are shown in Fig.~\ref{Fig-prsh-ionization-rate}. While the induced ionization of atomic hydrogen roughly extends over the size of the radiative precursor ($\sim 10$ pc for the standard model), the ionization of other elements extends over much larger distances. This is due to the fact that these species are effectively ionized by photons at larger energies (see Fig.~\ref{Fig-cross-sections}), which propagate over higher distances in the preshock (see Fig.~\ref{Fig-alpha}).

\section{Discussion} \label{Sect-discuss}

\subsection{Viscosity and jump conditions} \label{Sect-viscosity}

The Paris-Durham shock code has the particularity of treating the viscous stresses in the evolution equation of the gas velocity for J-type shocks \citep{Lesaffre2013}. In practice, this is done by setting a constant viscous length $\lambda = 1 / \sigma \densini$, where $\sigma = 3 \times 10^{-15}$~cm$^{2}$ is the cross section of elastic collisions between \HH\ and \HH\ computed by \citet{Monchick1980a}. This prescription is at the root of the viscous jump shown, for instance, in Fig.~\ref{Fig-convergence} which has a typical size of $10^{-4}$~pc for $\densini = 1$ \cc. While the resulting evolutions of the thermodynamical state of the gas are found to be in perfect agreement with the jump conditions expected for adiabatic magnetohydrodynamic shocks, it is important to note, however, that this viscous jump is somehow artificial for the models explored in this paper.

First, the viscous length should be set from the cross section of elastic collision between hydrogen atoms and not molecular hydrogen. In addition, this viscous length is relevant only if the radiative precursor is not fully ionized. Indeed, when the radiative precursor is fully ionized, the adiabatic jump of interstellar shocks is known to be mediated by long range interactions and plasmoid instabilities leading to collisionless shock waves.

Since the adiabatic jump conditions are always fulfilled and have no impact on the subsequent cooling trajectory of the gas, regardless of the microphysics mechanism responsible for the jump, we deliberately decided to keep the prescription of the Paris-Durham shock code as it is. The only impact is that the length of the jump is somehow larger than it should be.

\subsection{Approximation of equilibrium for the level populations}

The postprocessing radiative-transfer algorithm (see Appendix \ref{Append-radtrans}) and the prescription adopted for the cooling induced by collisional excitation of ions (see Sect. \ref{Sect-cooling}) are built on the hypothesis that the populations of the levels of atoms and atomic ions are at equilibrium. To test this approximation, we developed two separated versions of the Paris-Durham shock code, one which solves the time dependent evolution of the populations of the first ten electronic levels of atomic hydrogen, and the other one the time dependent evolution of the three fine structure levels of atomic carbon. In both cases, the evolution of the populations of all the levels considered were found to be identical to those computed at equilibrium. It is so because the typical deexcitation timescales of electronic levels and of fine structure levels are orders of magnitude smaller than the cooling timescales. This statement also holds for levels that relax through the emission of a resonant photon which, because of photon trapping, have a local effective deexcitation coefficient several orders of magnitude below the spontaneous decay rate.

\subsection{Absorption of the external radiation field} \label{Sect-absorption}

We consider in this work that the shocked gas is irradiated by two radiation fields: the radiation field emitted by the shock itself, which extends at all wavelengths, and an external UV radiation field, which extends between 911~\AA\ and 2400~\AA. As described in Appendix~\ref{Append-radtrans-approx}, the internal radiation field is calculated taking into account all absorption processes including dust and PAHs. As stated in Sect.~\ref{Sect-param}, the external radiation field is considered constant throughout the entire shock trajectory.

It is important to note that the Paris-Durham shock code is capable of treating the absorption of the external radiation field by dust and PAHs and the self-shielding of \HH\ \citep{Godard2019}. We choose, however, to keep this field constant here to simplify the problem and to focus on the sole propagation of the internal radiation field. This is equivalent to assume that the shock is surrounded by external UV radiation sources, which is probably the case in the diffuse ISM. The validity of this approximation depends on the typical visual extinction of the shocks. The 1D visual extinction computed across shocks over the entire grid of models explored in this work is found to range between a few $10^{-3}$ and 0.3. These values indicate that dust absorption can indeed be neglected and has a negligible or weak impact on the emission of atomic species and on the immediate follow-up studies.

However, the range of visual extinction also suggest that self-shielding processes might be important. Indeed, as shown by \citet{Hollenbach1989}, the self-shielding of \HH\ leads to the reformation of molecular hydrogen in the postshock when the temperature falls below $10^4$~K. As \HH\ reforms, it efficiently cools the gas and triggers the formation of a wealth of molecules. The spectrum emitted by the shock not only contains strong rovibrational lines of \HH\ but also other molecular lines. This was recently illustrated by \citet{Lehmann2020} who showed that non-irradiated shocks at intermediate velocities induce substantive molecular emission.  The formation of molecules and the molecular tracers of intermediate- and high-velocity shocks propagating in the WNM, including the prediction of the full spectrum of molecular hydrogen, could be addressed in future works.

\subsection{X-ray emission induced by charge exchange}

While charge exchanges between singly ionized atoms and neutral species are included in the chemical network, we do not include the processes of charge exchange between strongly multi-ionized atoms and neutral species. Indeed, the recombination of strongly multi-ionized species is expected to be dominated by capture of free electrons rather than charge exchange with neutral species in stationary shocks. Moreover, the X-ray emission induced by this process has been shown to be negligible in shocks compared to the X-ray emission induced by collisional excitation by electron impact \citep{Wise1989a}. X-ray emission induced by charge exchange processes requires astrophysical environments where a neutral phase coexist with a highly ionized phase (e.g., \citealt{Lallement2004a}). Such mixing layers are out of the scope of this paper.

\subsection{Limitations of 1D models}

The model presented here is limited by its one-dimensional nature which constrains the final state of the postshock gas. Indeed, the postshock gas cannot escape in the x and y directions and is forced to be confined. It follows that the postshock trajectory predicted by 1D models necessarily becomes, at some point, unrealistic. This is true for shocks propagating in a single phase medium but also for shocks propagating in multiphase environments where the postshock gas is known to be subject to thermal instabilities (e.g., \citealt{Falle2020a}). As shown by \citet{Lesaffre2020} (see Fig.~21 in their paper), 1D models still provide robust predictions regarding the thermochemical state and the emission properties of shocks, providing that the postshock gas is not integrated indefinitely but up to a final point which, depending on the problem, can be chosen from the temperature or density profiles (e.g., \citealt{Wardle1999}) or energetic considerations (e.g., \citealt{Godard2019}).

Another limitation of the model is that it cannot treat shocks propagating in inhomogeneous media which are known to trigger various dynamical instabilities \citep{Falle2020a,Raymond2020a,Kupilas2021a,Markwick2021a}.

\section{Summary and conclusions}

In this paper, we present a new version of the Paris-Durham shock code, which extends its domain of application to self-irradiated shocks at intermediate ($30 < V_S < 100$~\kms) and high ($V_S\geqslant 100$~\kms) velocity. The code is updated to take into account the production of multi-ionized species and the cooling induced by free-free emission and by the collisional excitation of the electronic levels of any atom in any ionization stage. This last part is done through the inclusion of all the levels, radiative transitions, and  collisional rates contained in the CHIANTI database. The interactions of the photons produced by the shock with the radiative precursor and the shock itself are computed with a new, exact, and fast radiative-transfer algorithm for line emission based on the formalism of coupled escape probabilities. The chemical and thermodynamical structures of the radiative precursor and the shock are calculated self-consistently at steady-state with an iterative procedure that progressively accounts for the feedback of the self-generated radiation field on the thermal and chemical states of the gas.

The resulting model is explored over a large grid of parameters including low-, intermediate-, and high-velocity shocks propagating in poorly and highly magnetized ($0.1 \leqslant B_0 \leqslant 10$~$\mu$G) diffuse environments ($\densini \leqslant 2$ \cc). As far as we know, this is the first time that multifluid shocks and shocks irradiated by themselves and by an external UV radiation field are studied over this range of velocities and densities. The exploration reveals a diversity of stationary solutions including C-type, CJ-type, and J-type shocks whose existence depends on the shock velocity compared to the speeds of the magnetosonic waves in the radiative precursor.

As the gas is shocked, it undergoes a well-defined trajectory in a proton density versus thermal pressure diagram. Depending on the physical conditions of the preshock gas and on the shock velocity, shocks can induce phase transition, that is the conversion of the initial WNM material into CNM material. For standard conditions of the WNM ($\densini \geqslant 0.5$~\cc, $B_0 \leqslant 1$~$\mu$G), the criteria required to obtain phase transition (Eqs. \ref{Eq-velcond1} and \ref{Eq-velcond2}) are met by most of the models, indicating that interstellar shocks may play a fundamental role in the exchange of mass and energy between the warm and cold phases of the diffuse interstellar gas.

We find that most of the flux of mechanical energy is reprocessed into line emission over the entire grid of models and that the resulting spectrum hardens as the shock velocity increases. While the spectra of shocks propagating at low velocity ($V_S \leqslant 30$~\kms) is dominated by the emission of infrared photons, high-velocity shocks ($V_S \geqslant 100$~\kms) mostly emit EUV and X-ray photons through the deexcitation of the electronic levels of He$^+$, and of different ionization stages of oxygen, carbon, nitrogen, and sulfur. These photons efficiently heat and modify the chemical state of the radiative precursor. In particular, the strength of the ionizing radiation field is found to be sufficient to fully ionize H, He, and C$^+$ for $V_S \gtrsim 100$~\kms, fully ionize He$^+$ for $V_S \gtrsim 200$~\kms, and fully ionize C$^{2+}$ for $V_S \gtrsim 300$~\kms.

The thermal pressure reached by the shocked gas and the subsequent cooling trajectories lead to specific thermal and chemical properties and specific emission spectra whose signatures may already be present in ancillary observations of the diffuse interstellar matter. A detailed modeling of these phenomena therefore provides new clues for the interpretation of observations and the understanding of several fundamental mechanisms, including the injection and dissipation of mechanical energy in the ISM and the processes by which the gas is converted from the WNM to the CNM. Follow-up papers are currently underway to interpret ancillary data of atoms and ions observed in emission and absorption in the solar neighborhood. The production of molecules and molecular lines ---which naturally arises during the phase transition process and the cooling of the postshock gas toward the CNM--- will be addressed in future works.

\begin{acknowledgements}

We are very grateful to the referee for their thorough reading and their comments. The research leading to these results has received fundings from the European Research  Council, under the European Community's Seventh framework Programme, through the Advanced Grant MIST (FP7/2017-2022, No 742719). The grid of models used in this work has been run on the computing cluster Totoro of the ERC MIST, administered by MesoPSL. The models were developed using the atomic data currently available on the CHIANTI database. CHIANTI is a collaborative project involving George Mason University, the University of Michigan (USA), University of Cambridge (UK) and NASA Goddard Space Flight Center (USA). We would also like to acknowledge the support from the Programme National ``Physique et Chimie du Milieu Interstellaire'' (PCMI) of CNRS/INSU with INC/INP co-funded by CEA and CNES. 

\end{acknowledgements}

\bibliographystyle{aa} 
\bibliography{mybib}

\begin{thebibliography}{101}
\expandafter\ifx\csname natexlab\endcsname\relax\def\natexlab#1{#1}\fi

\bibitem[{{Abdel-Naby} {et~al.}(2012){Abdel-Naby}, {Nikoli{\'c}}, {Gorczyca},
  {Korista}, \& {Badnell}}]{Abdel-Naby2012}
{Abdel-Naby}, S.~A., {Nikoli{\'c}}, D., {Gorczyca}, T.~W., {Korista}, K.~T., \&
  {Badnell}, N.~R. 2012, \aap, 537, A40

\bibitem[{{Allen} {et~al.}(2008){Allen}, {Groves}, {Dopita}, {Sutherland}, \&
  {Kewley}}]{Allen2008}
{Allen}, M.~G., {Groves}, B.~A., {Dopita}, M.~A., {Sutherland}, R.~S., \&
  {Kewley}, L.~J. 2008, \apjs, 178, 20

\bibitem[{{Altun} {et~al.}(2007){Altun}, {Yumak}, {Yavuz}, {Badnell}, {Loch},
  \& {Pindzola}}]{Altun2007}
{Altun}, Z., {Yumak}, A., {Yavuz}, I., {et~al.} 2007, \aap, 474, 1051

\bibitem[{{Appleton} {et~al.}(2023){Appleton}, {Guillard}, {Emonts},
  {Boulanger}, {Togi}, {Reach}, {Alatalo}, {Cluver}, {Diaz Santos}, {Duc},
  {Gallagher}, {Ogle}, {O'Sullivan}, {Voggel}, \& {Xu}}]{Appleton2023a}
{Appleton}, P.~N., {Guillard}, P., {Emonts}, B., {et~al.} 2023, \apj, 951, 104

\bibitem[{{Athay} \& {Skumanich}(1971)}]{Athay1971a}
{Athay}, R.~G. \& {Skumanich}, A. 1971, \apj, 170, 605

\bibitem[{{Audit} \& {Hennebelle}(2005)}]{Audit2005a}
{Audit}, E. \& {Hennebelle}, P. 2005, \aap, 433, 1

\bibitem[{{Badnell}(2006)}]{Badnell2006}
{Badnell}, N.~R. 2006, \apjs, 167, 334

\bibitem[{{Badnell} {et~al.}(2003){Badnell}, {O'Mullane}, {Summers}, {Altun},
  {Bautista}, {Colgan}, {Gorczyca}, {Mitnik}, {Pindzola}, \&
  {Zatsarinny}}]{Badnell2003}
{Badnell}, N.~R., {O'Mullane}, M.~G., {Summers}, H.~P., {et~al.} 2003, \aap,
  406, 1151

\bibitem[{{Bellomi} {et~al.}(2020){Bellomi}, {Godard}, {Hennebelle},
  {Valdivia}, {Pineau des For{\^e}ts}, {Lesaffre}, \&
  {P{\'e}rault}}]{Bellomi2020}
{Bellomi}, E., {Godard}, B., {Hennebelle}, P., {et~al.} 2020, \aap, 643, A36

\bibitem[{{Best} {et~al.}(2000){Best}, {R{\"o}ttgering}, \&
  {Longair}}]{Best2000a}
{Best}, P.~N., {R{\"o}ttgering}, H.~J.~A., \& {Longair}, M.~S. 2000, \mnras,
  311, 23

\bibitem[{{Chernoff}(1987)}]{Chernoff1987}
{Chernoff}, D.~F. 1987, \apj, 312, 143

\bibitem[{{Crutcher} {et~al.}(2010){Crutcher}, {Wandelt}, {Heiles},
  {Falgarone}, \& {Troland}}]{Crutcher2010}
{Crutcher}, R.~M., {Wandelt}, B., {Heiles}, C., {Falgarone}, E., \& {Troland},
  T.~H. 2010, \apj, 725, 466

\bibitem[{{Del Zanna} {et~al.}(2021){Del Zanna}, {Dere}, {Young}, \&
  {Landi}}]{Del-Zanna2021}
{Del Zanna}, G., {Dere}, K.~P., {Young}, P.~R., \& {Landi}, E. 2021, \apj, 909,
  38

\bibitem[{{Dere} {et~al.}(2019){Dere}, {Del Zanna}, {Young}, {Landi}, \&
  {Sutherland}}]{Dere2019}
{Dere}, K.~P., {Del Zanna}, G., {Young}, P.~R., {Landi}, E., \& {Sutherland},
  R.~S. 2019, \apjs, 241, 22

\bibitem[{{Dere} {et~al.}(1997){Dere}, {Landi}, {Mason}, {Monsignori Fossi}, \&
  {Young}}]{Dere1997}
{Dere}, K.~P., {Landi}, E., {Mason}, H.~E., {Monsignori Fossi}, B.~C., \&
  {Young}, P.~R. 1997, \aaps, 125, 149

\bibitem[{{Dopita}(1978)}]{Dopita1978a}
{Dopita}, M.~A. 1978, \apjs, 37, 117

\bibitem[{{Dopita} \& {Sutherland}(1995)}]{Dopita1995}
{Dopita}, M.~A. \& {Sutherland}, R.~S. 1995, \apj, 455, 468

\bibitem[{{Dopita} \& {Sutherland}(1996)}]{Dopita1996}
{Dopita}, M.~A. \& {Sutherland}, R.~S. 1996, \apjs, 102, 161

\bibitem[{{Dopita} \& {Sutherland}(2003)}]{Dopita2003}
{Dopita}, M.~A. \& {Sutherland}, R.~S. 2003, Astrophysics of the Diffuse
  Universe (Springer, Berlin, Heidelberg)

\bibitem[{{Dopita} \& {Sutherland}(2017)}]{Dopita2017a}
{Dopita}, M.~A. \& {Sutherland}, R.~S. 2017, \apjs, 229, 35

\bibitem[{{Dopita} {et~al.}(2013){Dopita}, {Sutherland}, {Nicholls}, {Kewley},
  \& {Vogt}}]{Dopita2013a}
{Dopita}, M.~A., {Sutherland}, R.~S., {Nicholls}, D.~C., {Kewley}, L.~J., \&
  {Vogt}, F. P.~A. 2013, \apjs, 208, 10

\bibitem[{{Draine}(2011)}]{Draine2011}
{Draine}, B.~T. 2011, {Physics of the Interstellar and Intergalactic Medium}
  (Princeton University Press)

\bibitem[{{Draine} \& {Katz}(1986)}]{Draine1986a}
{Draine}, B.~T. \& {Katz}, N. 1986, \apj, 310, 392

\bibitem[{{Draine} \& {Lee}(1984)}]{Draine1984}
{Draine}, B.~T. \& {Lee}, H.~M. 1984, \apj, 285, 89

\bibitem[{{Draine} \& {Li}(2007)}]{Draine2007}
{Draine}, B.~T. \& {Li}, A. 2007, \apj, 657, 810

\bibitem[{{Draine} \& {Sutin}(1987)}]{Draine1987}
{Draine}, B.~T. \& {Sutin}, B. 1987, \apj, 320, 803

\bibitem[{Dullemond(2012)}]{Dullemond_book}
Dullemond. 2012, Radiative Transfer in Astrophysics, Theory, Numerical Methods
  and Applications

\bibitem[{{Elitzur} \& {Asensio Ramos}(2006)}]{Elitzur2006a}
{Elitzur}, M. \& {Asensio Ramos}, A. 2006, \mnras, 365, 779

\bibitem[{{Elmegreen} \& {Scalo}(2004)}]{Elmegreen2004}
{Elmegreen}, B.~G. \& {Scalo}, J. 2004, \araa, 42, 211

\bibitem[{{Falle} {et~al.}(2020){Falle}, {Wareing}, \& {Pittard}}]{Falle2020a}
{Falle}, S.~A.~E.~G., {Wareing}, C.~J., \& {Pittard}, J.~M. 2020, \mnras, 492,
  4484

\bibitem[{{Field}(1965)}]{Field1965a}
{Field}, G.~B. 1965, \apj, 142, 531

\bibitem[{{Flower}(2010)}]{Flower2010}
{Flower}, D. 2010, in Lecture Notes in Physics, Berlin Springer Verlag, Vol.
  793, Lecture Notes in Physics, Berlin Springer Verlag, ed. P.~J.~V. {Garcia}
  \& J.~M. {Ferreira}, 161

\bibitem[{{Flower} \& {Pineau des For\^ets}(1998)}]{Flower1998c}
{Flower}, D.~R. \& {Pineau des For\^ets}, G. 1998, \mnras, 297, 1182

\bibitem[{{Flower} \& {Pineau des For{\^e}ts}(2003)}]{Flower2003}
{Flower}, D.~R. \& {Pineau des For{\^e}ts}, G. 2003, \mnras, 343, 390

\bibitem[{{Flower} {et~al.}(1985){Flower}, {Pineau des For\^ets}, \&
  {Hartquist}}]{Flower1985}
{Flower}, D.~R., {Pineau des For\^ets}, G., \& {Hartquist}, T.~W. 1985, \mnras,
  216, 775

\bibitem[{{Frick} {et~al.}(2001){Frick}, {Stepanov}, {Shukurov}, \&
  {Sokoloff}}]{Frick2001a}
{Frick}, P., {Stepanov}, R., {Shukurov}, A., \& {Sokoloff}, D. 2001, \mnras,
  325, 649

\bibitem[{{Godard} {et~al.}(2024){Godard}, {Pineau Des For{\^e}ts}, {La Porte},
  \& {Merlin-Weck}}]{Godard2024b}
{Godard}, B., {Pineau Des For{\^e}ts}, G., {La Porte}, J., \& {Merlin-Weck}, M.
  2024, arXiv e-prints, arXiv:2406.19719

\bibitem[{{Godard} {et~al.}(2019){Godard}, {Pineau des For{\^e}ts}, {Lesaffre},
  {Lehmann}, {Gusdorf}, \& {Falgarone}}]{Godard2019}
{Godard}, B., {Pineau des For{\^e}ts}, G., {Lesaffre}, P., {et~al.} 2019, \aap,
  622, A100

\bibitem[{{Guillard} {et~al.}(2009){Guillard}, {Boulanger}, {Pineau des
  For{\^e}ts}, \& {Appleton}}]{Guillard2009}
{Guillard}, P., {Boulanger}, F., {Pineau des For{\^e}ts}, G., \& {Appleton},
  P.~N. 2009, \aap, 502, 515

\bibitem[{{Guillet} {et~al.}(2009){Guillet}, {Jones}, \& {Pineau des
  For{\^e}ts}}]{Guillet2009}
{Guillet}, V., {Jones}, A.~P., \& {Pineau des For{\^e}ts}, G. 2009, \aap, 497,
  145

\bibitem[{{Heiles} \& {Troland}(2003)}]{Heiles2003}
{Heiles}, C. \& {Troland}, T.~H. 2003, \apj, 586, 1067

\bibitem[{{Hennebelle} \& {P{\'e}rault}(1999)}]{Hennebelle1999a}
{Hennebelle}, P. \& {P{\'e}rault}, M. 1999, \aap, 351, 309

\bibitem[{{Hennebelle} \& {P{\'e}rault}(2000)}]{Hennebelle2000a}
{Hennebelle}, P. \& {P{\'e}rault}, M. 2000, \aap, 359, 1124

\bibitem[{{Hindmarsh}(1983)}]{Hindmarsh1983}
{Hindmarsh}, A.~C. 1983, {Scientific Computing} (Amsterdam: North-Holland)

\bibitem[{{Hollenbach} \& {McKee}(1979)}]{Hollenbach1979}
{Hollenbach}, D. \& {McKee}, C.~F. 1979, \apjs, 41, 555

\bibitem[{{Hollenbach} \& {McKee}(1989)}]{Hollenbach1989}
{Hollenbach}, D. \& {McKee}, C.~F. 1989, \apj, 342, 306

\bibitem[{{Indriolo} {et~al.}(2007){Indriolo}, {Geballe}, {Oka}, \&
  {McCall}}]{Indriolo2007}
{Indriolo}, N., {Geballe}, T.~R., {Oka}, T., \& {McCall}, B.~J. 2007, \apj,
  671, 1736

\bibitem[{{Indriolo} {et~al.}(2015){Indriolo}, {Neufeld}, {Gerin}, {Schilke},
  {Benz}, {Winkel}, {Menten}, {Chambers}, {Black}, {Bruderer}, {Falgarone},
  {Godard}, {Goicoechea}, {Gupta}, {Lis}, {Ossenkopf}, {Persson},
  {Sonnentrucker}, {van der Tak}, {van Dishoeck}, {Wolfire}, \&
  {Wyrowski}}]{Indriolo2015}
{Indriolo}, N., {Neufeld}, D.~A., {Gerin}, M., {et~al.} 2015, \apj, 800, 40

\bibitem[{{Inoue} \& {Inutsuka}(2008)}]{Inoue2008a}
{Inoue}, T. \& {Inutsuka}, S.-i. 2008, \apj, 687, 303

\bibitem[{{Jenkins} \& {Tripp}(2011)}]{Jenkins2011}
{Jenkins}, E.~B. \& {Tripp}, T.~M. 2011, \apj, 734, 65

\bibitem[{{Jones} {et~al.}(1996){Jones}, {Tielens}, \&
  {Hollenbach}}]{Jones1996}
{Jones}, A.~P., {Tielens}, A.~G.~G.~M., \& {Hollenbach}, D.~J. 1996, \apj, 469,
  740

\bibitem[{{Kaastra} \& {Mewe}(1993)}]{Kaastra1993}
{Kaastra}, J.~S. \& {Mewe}, R. 1993, \aaps, 97, 443

\bibitem[{{Karzas} \& {Latter}(1961)}]{Karzas1961a}
{Karzas}, W.~J. \& {Latter}, R. 1961, \apjs, 6, 167

\bibitem[{{Kopsacheili} {et~al.}(2020){Kopsacheili}, {Zezas}, \&
  {Leonidaki}}]{Kopsacheili2020a}
{Kopsacheili}, M., {Zezas}, A., \& {Leonidaki}, I. 2020, \mnras, 491, 889

\bibitem[{{Kristensen} {et~al.}(2023){Kristensen}, {Godard}, {Guillard},
  {Gusdorf}, \& {Pineau des For{\^e}ts}}]{Kristensen2023a}
{Kristensen}, L.~E., {Godard}, B., {Guillard}, P., {Gusdorf}, A., \& {Pineau
  des For{\^e}ts}, G. 2023, \aap, 675, A86

\bibitem[{{Kupilas} {et~al.}(2021){Kupilas}, {Wareing}, {Pittard}, \&
  {Falle}}]{Kupilas2021a}
{Kupilas}, M.~M., {Wareing}, C.~J., {Pittard}, J.~M., \& {Falle}, S.~A.~E.~G.
  2021, \mnras, 501, 3137

\bibitem[{{Lallement}(2004)}]{Lallement2004a}
{Lallement}, R. 2004, \aap, 422, 391

\bibitem[{{Landini} \& {Monsignori Fossi}(1990)}]{Landini1990}
{Landini}, M. \& {Monsignori Fossi}, B.~C. 1990, \aaps, 82, 229

\bibitem[{{Laor} \& {Draine}(1993)}]{Laor1993}
{Laor}, A. \& {Draine}, B.~T. 1993, \apj, 402, 441

\bibitem[{{Lehmann} {et~al.}(2016){Lehmann}, {Federrath}, \&
  {Wardle}}]{Lehmann2016}
{Lehmann}, A., {Federrath}, C., \& {Wardle}, M. 2016, \mnras, 463, 1026

\bibitem[{{Lehmann} {et~al.}(2020){Lehmann}, {Godard}, {Pineau des For{\^e}ts},
  \& {Falgarone}}]{Lehmann2020}
{Lehmann}, A., {Godard}, B., {Pineau des For{\^e}ts}, G., \& {Falgarone}, E.
  2020, \aap, 643, A101

\bibitem[{{Lehmann} {et~al.}(2022){Lehmann}, {Godard}, {Pineau des For{\^e}ts},
  {Vidal-Garc{\'\i}a}, \& {Falgarone}}]{Lehmann2022}
{Lehmann}, A., {Godard}, B., {Pineau des For{\^e}ts}, G., {Vidal-Garc{\'\i}a},
  A., \& {Falgarone}, E. 2022, \aap, 658, A165

\bibitem[{{Lennon} {et~al.}(1988){Lennon}, {Bell}, {Gilbody}, {Hughes},
  {Kingston}, {Murray}, \& {Smith}}]{Lennon1988}
{Lennon}, M.~A., {Bell}, K.~L., {Gilbody}, H.~B., {et~al.} 1988, Journal of
  Physical and Chemical Reference Data, 17, 1285

\bibitem[{{Lesaffre} {et~al.}(2013){Lesaffre}, {Pineau des For{\^e}ts},
  {Godard}, {Guillard}, {Boulanger}, \& {Falgarone}}]{Lesaffre2013}
{Lesaffre}, P., {Pineau des For{\^e}ts}, G., {Godard}, B., {et~al.} 2013, \aap,
  550, A106

\bibitem[{{Lesaffre} {et~al.}(2020){Lesaffre}, {Todorov}, {Levrier},
  {Valdivia}, {Dzyurkevich}, {Godard}, {Tram}, {Gusdorf}, {Lehmann}, \&
  {Falgarone}}]{Lesaffre2020}
{Lesaffre}, P., {Todorov}, P., {Levrier}, F., {et~al.} 2020, \mnras, 495, 816

\bibitem[{{Marchal} {et~al.}(2019){Marchal}, {Miville-Desch{\^e}nes}, {Orieux},
  {Gac}, {Soussen}, {Lesot}, {d'Allonnes}, \& {Salom{\'e}}}]{Marchal2019a}
{Marchal}, A., {Miville-Desch{\^e}nes}, M.-A., {Orieux}, F., {et~al.} 2019,
  \aap, 626, A101

\bibitem[{{Markwick} {et~al.}(2021){Markwick}, {Frank}, {Carroll-Nellenback},
  {Liu}, {Blackman}, {Lebedev}, \& {Hartigan}}]{Markwick2021a}
{Markwick}, R.~N., {Frank}, A., {Carroll-Nellenback}, J., {et~al.} 2021,
  \mnras, 508, 2266

\bibitem[{{Mathis} {et~al.}(1983){Mathis}, {Mezger}, \& {Panagia}}]{Mathis1983}
{Mathis}, J.~S., {Mezger}, P.~G., \& {Panagia}, N. 1983, \aap, 128, 212

\bibitem[{{Mathis} {et~al.}(1977){Mathis}, {Rumpl}, \&
  {Nordsieck}}]{Mathis1977}
{Mathis}, J.~S., {Rumpl}, W., \& {Nordsieck}, K.~H. 1977, \apj, 217, 425

\bibitem[{{McCall} {et~al.}(2003){McCall}, {Huneycutt}, {Saykally}, {Geballe},
  {Djuric}, {Dunn}, {Semaniak}, {Novotny}, {Al-Khalili}, {Ehlerding},
  {Hellberg}, {Kalhori}, {Neau}, {Thomas}, {{\"O}sterdahl}, \&
  {Larsson}}]{McCall2003}
{McCall}, B.~J., {Huneycutt}, A.~J., {Saykally}, R.~J., {et~al.} 2003, \nat,
  422, 500

\bibitem[{{Monchick} \& {Schaefer}(1980)}]{Monchick1980a}
{Monchick}, L. \& {Schaefer}, J. 1980, \jcp, 73, 6153

\bibitem[{{Monteiro} {et~al.}(1988){Monteiro}, {Flower}, {Pineau des For\^ets},
  \& {Roueff}}]{Monteiro1988}
{Monteiro}, T.~S., {Flower}, D.~R., {Pineau des For\^ets}, G., \& {Roueff}, E.
  1988, \mnras, 234, 863

\bibitem[{{Murray} {et~al.}(2018){Murray}, {Stanimirovi{\'c}}, {Goss},
  {Heiles}, {Dickey}, {Babler}, \& {Kim}}]{Murray2018a}
{Murray}, C.~E., {Stanimirovi{\'c}}, S., {Goss}, W.~M., {et~al.} 2018, \apjs,
  238, 14

\bibitem[{{Neufeld} \& {Wolfire}(2017)}]{Neufeld2017}
{Neufeld}, D.~A. \& {Wolfire}, M.~G. 2017, \apj, 845, 163

\bibitem[{{Olson} {et~al.}(1986){Olson}, {Auer}, \& {Buchler}}]{Olson1986a}
{Olson}, G.~L., {Auer}, L.~H., \& {Buchler}, J.~R. 1986, \jqsrt, 35, 431

\bibitem[{{Pan} {et~al.}(2016){Pan}, {Padoan}, {Haugb{\o}lle}, \&
  {Nordlund}}]{Pan2016a}
{Pan}, L., {Padoan}, P., {Haugb{\o}lle}, T., \& {Nordlund}, {\r{A}}. 2016,
  \apj, 825, 30

\bibitem[{{Porter} {et~al.}(2002){Porter}, {Pouquet}, \&
  {Woodward}}]{Porter2002}
{Porter}, D., {Pouquet}, A., \& {Woodward}, P. 2002, \pre, 66, 026301

\bibitem[{{Priestley} {et~al.}(2022){Priestley}, {Chawner}, {Barlow}, {De
  Looze}, {Gomez}, \& {Matsuura}}]{Priestley2022a}
{Priestley}, F.~D., {Chawner}, H., {Barlow}, M.~J., {et~al.} 2022, \mnras, 516,
  2314

\bibitem[{{Raymond}(1979)}]{Raymond1979}
{Raymond}, J.~C. 1979, \apjs, 39, 1

\bibitem[{{Raymond} {et~al.}(2020){Raymond}, {Slavin}, {Blair}, {Chilingarian},
  {Burkhart}, \& {Sankrit}}]{Raymond2020a}
{Raymond}, J.~C., {Slavin}, J.~D., {Blair}, W.~P., {et~al.} 2020, \apj, 903, 2

\bibitem[{{Richard} {et~al.}(2022){Richard}, {Lesaffre}, {Falgarone}, \&
  {Lehmann}}]{Richard2022a}
{Richard}, T., {Lesaffre}, P., {Falgarone}, E., \& {Lehmann}, A. 2022, \aap,
  664, A193

\bibitem[{{Roberge} \& {Draine}(1990)}]{Roberge1990}
{Roberge}, W.~G. \& {Draine}, B.~T. 1990, \apj, 350, 700

\bibitem[{{Rybicki} \& {Hummer}(1991)}]{Rybicki1991a}
{Rybicki}, G.~B. \& {Hummer}, D.~G. 1991, \aap, 245, 171

\bibitem[{{Sarkar} {et~al.}(2021{\natexlab{a}}){Sarkar}, {Gnat}, \&
  {Sternberg}}]{Sarkar2021b}
{Sarkar}, K.~C., {Gnat}, O., \& {Sternberg}, A. 2021{\natexlab{a}}, \mnras,
  504, 583

\bibitem[{{Sarkar} {et~al.}(2021{\natexlab{b}}){Sarkar}, {Sternberg}, \&
  {Gnat}}]{Sarkar2021a}
{Sarkar}, K.~C., {Sternberg}, A., \& {Gnat}, O. 2021{\natexlab{b}}, \mnras,
  503, 5807

\bibitem[{{Sarkar} {et~al.}(2022){Sarkar}, {Sternberg}, \&
  {Gnat}}]{Sarkar2022a}
{Sarkar}, K.~C., {Sternberg}, A., \& {Gnat}, O. 2022, \apj, 940, 44

\bibitem[{{Seifried} {et~al.}(2011){Seifried}, {Schmidt}, \&
  {Niemeyer}}]{Seifried2011}
{Seifried}, D., {Schmidt}, W., \& {Niemeyer}, J.~C. 2011, \aap, 526, A14

\bibitem[{{Shull} \& {McKee}(1979)}]{Shull1979}
{Shull}, J.~M. \& {McKee}, C.~F. 1979, \apj, 227, 131

\bibitem[{{Stone} {et~al.}(1998){Stone}, {Ostriker}, \& {Gammie}}]{Stone1998a}
{Stone}, J.~M., {Ostriker}, E.~C., \& {Gammie}, C.~F. 1998, \apjl, 508, L99

\bibitem[{{Sutherland} {et~al.}(2003){Sutherland}, {Bicknell}, \&
  {Dopita}}]{Sutherland2003a}
{Sutherland}, R.~S., {Bicknell}, G.~V., \& {Dopita}, M.~A. 2003, \apj, 591, 238

\bibitem[{{Sutherland} \& {Dopita}(2017)}]{Sutherland2017}
{Sutherland}, R.~S. \& {Dopita}, M.~A. 2017, \apjs, 229, 34

\bibitem[{{van Hoof} {et~al.}(2014){van Hoof}, {Williams}, {Volk}, {Chatzikos},
  {Ferland}, {Lykins}, {Porter}, \& {Wang}}]{van-Hoof2014a}
{van Hoof}, P.~A.~M., {Williams}, R.~J.~R., {Volk}, K., {et~al.} 2014, \mnras,
  444, 420

\bibitem[{{van Loo} {et~al.}(2010){van Loo}, {Falle}, \&
  {Hartquist}}]{van-Loo2010a}
{van Loo}, S., {Falle}, S.~A.~E.~G., \& {Hartquist}, T.~W. 2010, \mnras, 406,
  1260

\bibitem[{{Vazquez-Semadeni} {et~al.}(1996){Vazquez-Semadeni}, {Passot}, \&
  {Pouquet}}]{Vazquez-Semadeni1996a}
{Vazquez-Semadeni}, E., {Passot}, T., \& {Pouquet}, A. 1996, \apj, 473, 881

\bibitem[{{Verner} \& {Yakovlev}(1995)}]{Verner1995}
{Verner}, D.~A. \& {Yakovlev}, D.~G. 1995, \aaps, 109, 125

\bibitem[{{Vestuto} {et~al.}(2003){Vestuto}, {Ostriker}, \&
  {Stone}}]{Vestuto2003}
{Vestuto}, J.~G., {Ostriker}, E.~C., \& {Stone}, J.~M. 2003, \apj, 590, 858

\bibitem[{{Wardle}(1999)}]{Wardle1999}
{Wardle}, M. 1999, \apjl, 525, L101

\bibitem[{{Wise} \& {Sarazin}(1989)}]{Wise1989a}
{Wise}, M.~W. \& {Sarazin}, C.~L. 1989, \apj, 345, 384

\bibitem[{{Wolfire} {et~al.}(2003){Wolfire}, {McKee}, {Hollenbach}, \&
  {Tielens}}]{Wolfire2003}
{Wolfire}, M.~G., {McKee}, C.~F., {Hollenbach}, D., \& {Tielens}, A.~G.~G.~M.
  2003, \apj, 587, 278

\bibitem[{Zari {et~al.}(2018)Zari, Hashemi, Brown, Jardine, \&
  de~Zeeuw}]{Zari2018}
Zari, E., Hashemi, H., Brown, A. G.~A., Jardine, K., \& de~Zeeuw, P.~T. 2018,
  \aap, 620, A172

\bibitem[{{Zhu} {et~al.}(2023){Zhu}, {Kewley}, \& {Sutherland}}]{Zhu2023a}
{Zhu}, P., {Kewley}, L.~J., \& {Sutherland}, R.~S. 2023, \apj, 954, 175

\end{thebibliography}

\appendix

\section{Radiative-transfer algorithm} \label{Append-radtrans}

\begin{figure}[!h]
\begin{center}
\includegraphics[width=9.0cm,trim = 5cm 3cm 5cm 5cm, clip,angle=0]{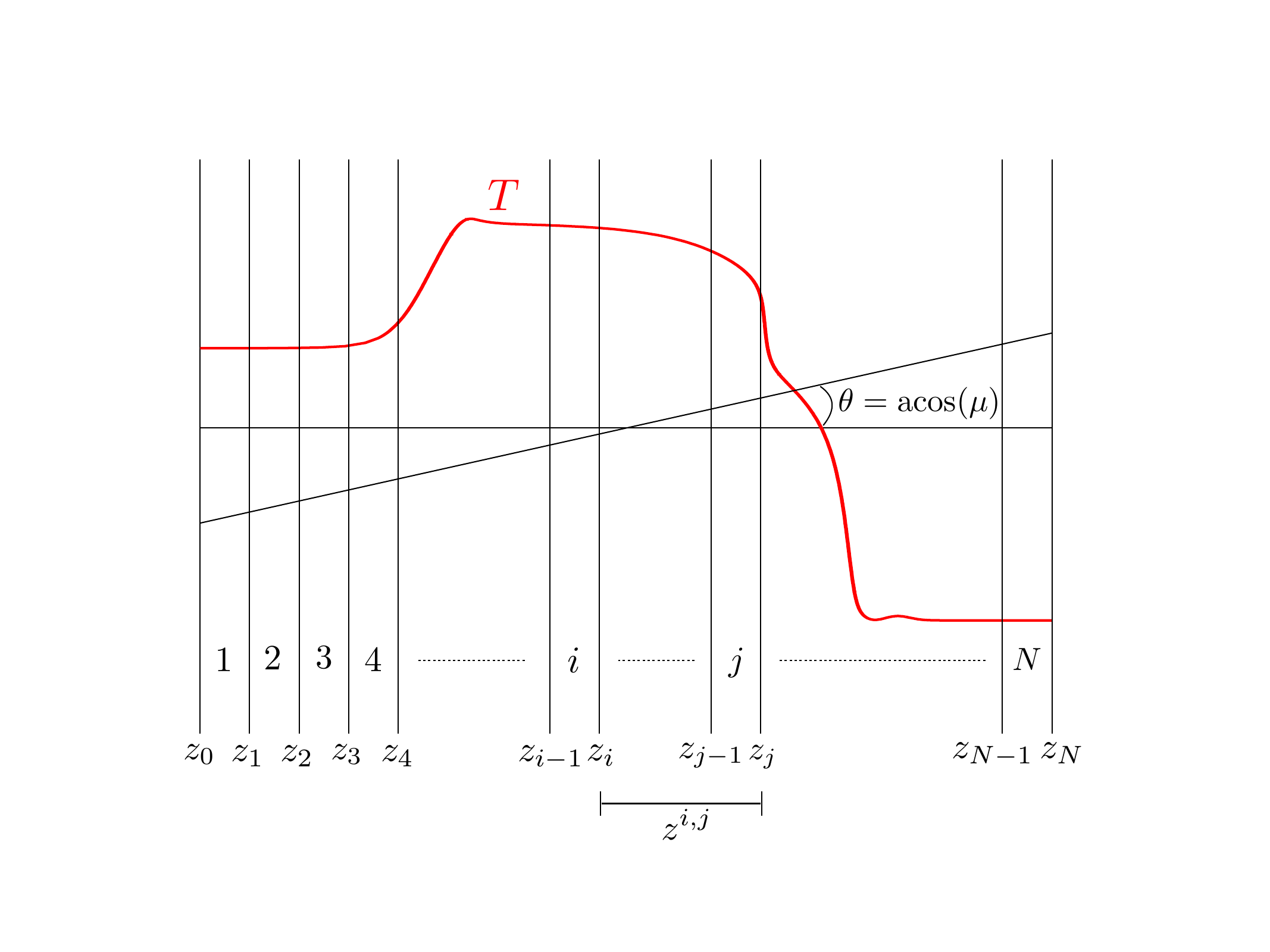}
\caption{Schematic view of the subdivision of a plane-parallel shock into zones. The red curve displays a typical temperature profile induced by an interstellar shock wave in arbitrary units as a function of the distance in logscale. A shock profile is divided into $N$ zones. Each zone is labelled with an index $i$ and is supposed to extend between the boundaries $z_{i-1}$ and $z_i$ which are not homogeneously separated (as the scheme would lead to believe). The classical definitions of the angle $\theta$ and its cosine $\mu$ are used to account for the angular dependance of the radiation field in a plane-parallel geometry.}
\label{Fig-zones}
\end{center}
\end{figure}

The iterative procedure used to compute the thermochemical structure of self-irradiated shocks relies on the proper calculation of the propagation of photons emitted by the shock and their interactions with the shocked and preshocked material. To perform this step we developed an exact radiative transfer code based on the formalism of coupled escape probability (CEP) for line emission \citep{Elitzur2006a}.

\subsection{Choice of the method} \label{Append-radtrans-choice}

One of the main issue of radiative transfer is the treatment of the chicken-or-egg dilemma which arises from the fact that the radiation field propagating in a given direction in a cloud depends on the state of matter everywhere which itself depends on the radiation field propagating in every direction \citep{Dullemond_book}. An exact and proven solution to this problem is the Accelerated Lambda Iteration (ALI) formalism in which the radiation field is computed through the iterative application of an operator $\Lambda$ onto the source function (e.g., \citealt{Olson1986a}). Because ALI algorithms and their developments for multilevel systems (MALI, \citealt{Rybicki1991a}) can be numerically expansive, \citet{Elitzur2006a} proposed a new and exact numerical method for solving the radiative transfer for line emission. In this method, designated as coupled escape probability, the effect of the radiation field is encoded into nonlocal couplings between the level populations at different positions in the cloud. At steady-state, the level populations are thus fully described by a set of non linear algebraic equations. \citet{Elitzur2006a} showed that solving this system provides an exact solution to the radiative transfer without having to compute the radiation field itself, and for a computational cost substantially smaller than those of MALI algorithms.

In their original paper, \citet{Elitzur2006a} developed the CEP for a plane-parallel geometry assuming a fixed line profile and applied it to standard problems with constant physical conditions. To solve the radiative transfer in plane-parallel interstellar shocks we propose here a modified version of the algorithm that takes into account spatial variations of the line profile and a new conditioning of the nonlocal coupling terms. To further reduce the computation cost, we finally apply additional simplifications that hold for diffuse material.

\subsection{Formal solution for an emission line} \label{Append-radtrans-formal}

Let's consider the radiative transfer within a line $u-l$ of a chemical species, where $u$ designates the upper energy state and $l$ the lower energy state of the line. In a plane-parallel geometry, the specific intensity at frequency $\nu$ along a ray, defined by its angle $\theta = {\rm acos}(\mu)$ with respect to the normal direction (see Fig.~\ref{Fig-zones}), obeys the radiative transfer equation
\begin{equation}\label{Eq-radtrans}
\mu \frac{dI_\nu(\mu,z)}{dz} = j_\nu(\mu,z) - \kappa_\nu(\mu,z) I_\nu(\mu,z),
\end{equation}
The emissivity $j_\nu$ and absorption coefficient $\kappa_\nu$ write
\begin{equation}
j_\nu(\mu,z) = \frac{h\nu_{ul}}{4\pi} A_{ul} n_u(z) \phi_\nu(\mu,z)
\end{equation}
and
\begin{equation}
\kappa_\nu(\mu,z) = \frac{h\nu_{ul}}{4\pi} \left[ B_{lu} n_l(z) - B_{ul} n_u(z) \right] \phi_\nu(\mu,z),
\end{equation}
where $A_{ul}$, $B_{ul}$, and $B_{lu}$ are the Einstein coefficients of the line, $n_u(z)$ and $n_l(z)$ are the local densities of the upper and lower energy states of the species under consideration, and 
\begin{equation}
\phi_\nu(\mu,z) = \frac{1}{\sqrt{2\pi} \sigma_\nu(z)} {\rm exp} \left[-\frac{1}{2}\left(\frac{\nu-\nu_0(\mu,z)}{\sigma_\nu(z)}\right)^2\right]
\end{equation}
is the line profile. For the sake of simplicity, we consider here a Gaussian line profile centered at the local rest frequency
\begin{equation}
\nu_0(\mu,z) = \nu_{ul}\left(1 + \frac{\upsilon_z(z) \mu}{c} \right)
\end{equation}
and with a thermal broadening (in frequency space)
\begin{equation}
\sigma_\nu(z) = \sqrt{\frac{kT(z)}{m}} \frac{\nu_{ul}}{c}
\end{equation}
where $\upsilon_z(z)$ and $T(z)$ are the local velocity and temperature of the fluid and $m$ is the mass of the species under consideration. Assuming complete frequency redistribution, the source function of the line writes
\begin{equation}
S(z) = \frac{\varepsilon(z)}{\gamma(z)}  = \frac{A_{ul}n_u(z)}{B_{lu} n_l(z) - B_{ul} n_u(z)},
\end{equation}
where
\begin{equation}
\varepsilon(z) = \frac{h\nu_{ul}}{4\pi} A_{ul} n_u(z)
\end{equation}
and
\begin{equation}
\gamma(z) = \frac{h\nu_{ul}}{4\pi} \left[ B_{lu} n_l(z) - B_{ul} n_u(z) \right]
\end{equation}
are the integrals over frequency of the emissivity and the absorption coefficient, respectively.

With these notations, and assuming that the cloud is not irradiated by an external radiation source, the radiative transfer equation (\ref{Eq-radtrans}) has the following formal solution
\begin{equation} \label{Eq-forsol}
\begin{split}
I_\nu(\mu,z) & = \int_0^z {\rm e}^{\displaystyle -\int_y^z \frac{\gamma(x) \phi_\nu(\mu,x)}{\mu} dx} \frac{\varepsilon(y)\phi_\nu(\mu,y)}{\mu}dy \quad {\rm for}\,\,\mu > 0 \\
I_\nu(\mu,z) & = \int_{z_N}^z {\rm e}^{\displaystyle -\int_y^z \frac{\gamma(x) \phi_\nu(\mu,x)}{\mu} dx} \frac{\varepsilon(y)\phi_\nu(\mu,y)}{\mu}dy \quad {\rm for}\,\,\mu < 0 \end{split},
\end{equation}
where $z_N$ is the final distance of the cloud (see Fig.~\ref{Fig-zones}). Now, let's define the net radiative bracket of the line \citep{Athay1971a}
\begin{equation}
p(z) = 1 - \frac{\overline{J}(z)}{S(z)},
\end{equation}
where
\begin{equation}
\overline{J}(z) = \frac{1}{2} \int d\nu \int_{-1}^1 \phi_\nu(\mu,z) I_\nu(\mu,z) d\mu
\end{equation}
is the profile integrated averaged intensity. The formal solution of the radiative-transfer equation (Eq. \ref{Eq-forsol}) implies that
\begin{equation} \label{Eq-bracket}
\begin{split}
p(z) & = 1 - \frac{1}{2 S(z)} \int d\nu \int_{0}^1 \phi_\nu(\mu,z) d\mu\\ 
& \times \int_0^{z_N} {\rm e}^{\displaystyle - \left| \int_y^z \frac{\gamma(x) \phi_\nu(\mu,x)}{\mu} dx \right|} \frac{\varepsilon(y)\phi_\nu(\mu,y)}{\mu}dy.
\end{split}
\end{equation}
This equation is equivalent to Eq. 7 of \citet{Elitzur2006a} except that the integration is performed over $z$ rather than the opacity and that the line profile is supposed to have an explicit dependance on $\mu$ and $z$. This expression of the net radiative bracket is at the root of the CEP approach. On the one hand, $p(z)$ solely depends on the densities $n_u(z)$ and $n_l(z)$ across the cloud. On the other hand, the equations of evolution $n_u(z)$ and $n_l(z)$ solely depends on $p(z)$ (see Sect. \ref{Append-radtrans-multilev}). The net radiative bracket therefore contains all the nonlocal terms that couple the level populations at different positions in the cloud. In the following, we therefore develop the expression of $p(z)$ over a discretized grid of positions (Sect. \ref{Append-radtrans-discrete}) and then explicit the system of equations we need to solve for a multilevel species (Sect. \ref{Append-radtrans-multilev}).

\subsection{Net radiative bracket over a discretized grid} \label{Append-radtrans-discrete}

As described in Fig.~\ref{Fig-zones} and following \citet{Elitzur2006a}, the plane-parallel cloud (an interstellar shock in our case) is divided into $N$ zones. The boundaries of the zones are designated by $z_i$. The $i^{\rm th}$ zone is defined as the zone occupying the range of distance $z_{i-1} < z \leqslant z_{i}$. The distance between two boundaries $z_i$ and $z_j$ is designated as
\begin{equation}
z^{i,j} = |z_i - z_j|.
\end{equation}
All the physical and chemical conditions within a zone $i$ are assumed to be constant and are calculated as averaged of the actual profile between $z_{i-1}$ and $z_i$.

With these assumptions, the net radiative bracket for a specific line $u-l$ in the zone $i$ writes
\begin{equation}
p^i = \frac{1}{z^{i-1,i}} \int_{z_{i-1}}^{z_i} p(z) dz.
\end{equation}
The combination of this equation with Eq. \ref{Eq-bracket} gives
\begin{equation} \label{Eq-bracket2}
\begin{split}
p^i & = 1 - \frac{1}{2S^i z^{i-1,i}} \sum_{j=1}^N \varepsilon^j \int_{z_{i-1}}^{z_i} dy' \int_{z_{j-1}}^{z_j} dy \int d\nu \\
& \times \int_0^1 d\mu\,\, \phi_\nu^i(\mu) {\rm e}^{\displaystyle - \left| \int_y^{y'} \frac{\gamma(x) \phi_\nu(\mu,x)}{\mu} dx \right|} \frac{\phi_\nu^j(\mu)}{\mu},
\end{split}
\end{equation}
where all the superscripts refer to physical quantities averaged over the corresponding zone. This last equation can be separated into
\begin{equation} \label{Eq-bracket3}
p^i = \beta^{i,i} + \sum_{\substack{j = 1 \\ j \neq i}}^N \beta^{i,j} 
\end{equation}
where $\beta^{i,i}$ corresponds to the contribution of the zone $i$ to $p^i$ and $\beta^{i,j}$ corresponds to the contribution of the zone $j$ to $p^i$.

Let's first consider $\beta^{i,i}$. From Eqs. \ref{Eq-bracket2} \& \ref{Eq-bracket3}, we have
\begin{equation}
\begin{split}
\beta^{i,i} & = 1 - \frac{\varepsilon^i}{2S^i z^{i-1,i}} \int_{z_{i-1}}^{z_i} dy' \int_{z_{i-1}}^{z_i} dy \int d\nu \\
& \times \int_0^1 d\mu\,\, \frac{(\phi_\nu^i(\mu))^2}{\mu} {\rm e}^{\displaystyle - \left| \int_y^{y'} \frac{\gamma(x) \phi_\nu(\mu,x)}{\mu} dx \right|}.
\end{split}
\end{equation}
Performing the integration over $y$ (treating successively the cases $y \leqslant y'$ and $y>y'$) and then over $y'$ leads to the following expression
\begin{equation}
\beta^{i,i} = \frac{1}{z^{i-1,i}\gamma^i} \int d\nu \left[\frac{1}{2} - \int_0^1 \mu d\mu\,\, {\rm e}^{\displaystyle - \frac{\gamma^i \phi_\nu^i(\mu) z^{i-1,i}}{\mu}} \right].
\end{equation}
Assuming that the line profile is independent of $\mu$, this expression can be simplified further into
\begin{equation}\label{Eq-betaon-fin}
\beta^{i,i} = \frac{1}{z^{i-1,i}\gamma^i} \int d\nu \left[\frac{1}{2} - E_3(\gamma^i \phi_\nu^i z^{i-1,i}) \right],
\end{equation}
where $E_3$ is the third exponential integral. This final equation for $\beta^{i,i}$ is completely equivalent to Eq. 22 of \citet{Elitzur2006a}.

Let's now consider the nonlocal coupling term $\beta^{i,j}$. From Eqs. \ref{Eq-bracket2} \& \ref{Eq-bracket3}, we have
\begin{equation}
\begin{split}
\beta^{i,j} & = - \frac{\varepsilon^j}{2S^i z^{i-1,i}} \int_{z_{i-1}}^{z_i} dy' \int_{z_{j-1}}^{z_j} dy \int d\nu \\
& \times \int_0^1 d\mu\,\, \frac{\phi_\nu^i(\mu) \phi_\nu^j(\mu)}{\mu} {\rm e}^{\displaystyle - \left| \int_y^{y'} \frac{\gamma(x) \phi_\nu(\mu,x)}{\mu} dx \right|}.
\end{split}
\end{equation}
Performing the successive integrations over $y$ and $y'$ for $j < i$ leads to
\begin{equation} \label{Eq-betaoff1}
\beta^{i,j} = \frac{S^j}{2S^i \gamma^i z^{i-1,i}} \int d\nu \int_0^1 d\mu\,\, \mu \left[ \alpha^{i,j} - \alpha^{i-1,j} - \alpha^{i,j-1} + \alpha^{i-1,j-1} \right]
\end{equation}
where
\begin{equation}
\alpha^{i,j} = {\rm e}^{\displaystyle - \int_{z_j}^{z_i} \frac{\gamma(x) \phi_\nu(\mu,x)}{\mu}dx}.
\end{equation}
To simplify this expression further, we make the approximation that the differences in Eq. \ref{Eq-betaoff1} can be replaced by derivatives, that is
\begin{equation} \label{Eq-diffalpha1}
\begin{split}
\alpha^{i,j} - \alpha^{i,j-1} & \approx z^{j-1,j} \frac{d}{dz}\left[ {\rm e}^{\displaystyle - \int_{z}^{z_i} \frac{\gamma(x) \phi_\nu(\mu,x)}{\mu}dx} \right] \left( \frac{z_{j-1}+z_{j}}{2} \right) \\
& \approx z^{j-1,j} \frac{\gamma^i \phi_\nu^j(\mu)}{\mu} {\rm e}^{\displaystyle - \int_{\frac{z_{j-1}+z_{j}}{2}}^{z_i} \frac{\gamma(x) \phi_\nu(\mu,x)}{\mu}dx}
\end{split}
\end{equation}
and
\begin{equation} \label{Eq-diffalpha2}
\begin{split}
\alpha^{i-1,j} - \alpha^{i-1,j-1} & \approx z^{j-1,j} \frac{d}{dz}\left[ {\rm e}^{\displaystyle - \int_{z}^{z_{i-1}} \frac{\gamma(x) \phi_\nu(\mu,x)}{\mu}dx} \right] \left(\frac{z_{j-1}+z_{j}}{2} \right) \\
& \approx z^{j-1,j} \frac{\gamma^i \phi_\nu^j(\mu)}{\mu} {\rm e}^{\displaystyle - \int_{\frac{z_{j-1}+z_{j}}{2}}^{z_{i-1}} \frac{\gamma(x) \phi_\nu(\mu,x)}{\mu}dx}.
\end{split}
\end{equation}
Performing the same approximation on the difference between Eqs. \ref{Eq-diffalpha1} and \ref{Eq-diffalpha2} and inserting the result in Eq. \ref{Eq-betaoff1}, we obtain
\begin{equation} \label{Eq-betaoff2}
\begin{split}
\beta^{i,j} & = -\frac{S^j \gamma^j z^{j-1,j}}{2S^i} \int d\nu \int_0^1 d\mu\,\, \frac{\phi_\nu^i(\mu) \phi_\nu^j(\mu)}{\mu} \\
& \times {\rm e}^{\displaystyle - \int_{\frac{z_{j-1}+z_{j}}{2}}^{\frac{z_{i-1}+z_{i}}{2}} \frac{\gamma(x) \phi_\nu(\mu,x)}{\mu}dx}.
\end{split}
\end{equation}
We note that this equation is valid for $j<i$. The same reasoning applied to the case $j>i$ virtually leads to the same expression except that the boundaries in the final integral are obviously inverted. This last equation has a deep physical meaning. It basically states that the amplitude of the coupling induced by zone $j$ on zone $i$ is proportional to the source function and the opacity\footnote{the opacity of a zone $i$ is proportional to $\gamma^i z^{i-1,i}$} of zone $j$, to the overlap of the line profiles of zones $i$ and $j$, and to the probability of nonabsorption of the photons between zone $i$ and $j$ which is given by the exponential term. Assuming that the line profile is independent of $\mu$, Eq. \ref{Eq-betaoff2} can be written in the more concise form, valid for any couple $i$ and $j$,
\begin{equation} \label{Eq-betaoff-fin}
\beta^{i,j} = -\frac{S^j \gamma^j z^{j-1,j}}{2S^i} \int d\nu \phi_\nu^i \phi_\nu^j E_1\left(\left|\int_{\frac{z_{j-1}+z_{j}}{2}}^{\frac{z_{i-1}+z_{i}}{2}} \gamma(x) \phi_\nu(x) dx \right| \right)
\end{equation}
where $E_1$ is the first exponential integral. It should be noted that this equation is very different from the expression of $\beta^{i,j}$ given by \citet{Elitzur2006a}. This comes from the facts that we take into account the spatial variation of the line profile and that we replace the differences in Eq. \ref{Eq-betaoff1} by derivatives. Beside the physical insight described above, such an expression has the numerical advantage that $\beta^{i,j}$ and the induced nonlocal coupling term are necessarily negative. It therefore provides a well behaved conditioning of the numerical problem.

\subsection{Multilevel system and fluxes} \label{Append-radtrans-multilev}

Let's consider a chemical species with $L$ energy levels. In any zone $i$, the density of the species, $n^i$, naturally verifies
\begin{equation}\label{Eq-cons-pop}
n^i = \sum_{k=1}^L n_k^i = n^i \sum_{k=1}^L \mathrm{x}_k^i 
\end{equation}
where $n_k^i$ and $\mathrm{x}_k^i$ are the density and fractional population of the species in the level k and zone $i$. Taking into account excitations and de-excitations by collisions with electrons, spontaneous and stimulated emissions, and absorption processes, the fractional population of level $k$ in zone $i$ obeys the following evolution equation
\begin{equation}\label{Eq-system}
\begin{split}
\frac{d \mathrm{x}_k^i}{dt} & = \sum_{u=k+1}^L \mathrm{x}_u^i A_{uk} p_{uk}^i + C_{uk}^i n_e^i \left(\mathrm{x}_u^i - \frac{g_u}{g_k}\mathrm{x}_k^i {\rm e}^{-E_{uk}/kT^i} \right)\\
& - \sum_{l=1}^{k-1} \mathrm{x}_k^i A_{kl} p_{kl}^i + C_{kl}^i n_e^i \left(\mathrm{x}_k^i - \frac{g_k}{g_l}\mathrm{x}_l^i {\rm e}^{-E_{kl}/kT^i} \right),
\end{split}
\end{equation}
where $p_{ul}^i$ is the net radiative bracket of the transition $u-l$, $n_e^i$ is the density of electrons, $g_k$ is the degeneracy of level $k$, and $C_{ul}^i$ is the de-excitation rate (in cm$^{3}$ s$^{-1}$) from level $u$ to level $l$ by collision with electrons. At steady-state, the above equations form a large set of non linear algebraic equations. Because all the nonlocal couplings are contained in $p_{ul}^i$, solving this system provides an exact solution to the radiative transfer. The computation of the intensity and the flux of any line is then straightforward. From the formal solution of the radiative transfer (Eq. \ref{Eq-forsol}), the specific intensity of a line $u-l$ at boundary $z_j$ writes
\begin{equation}
I_{\nu,ul}(\mu,z_j) = \sum_{i=1}^{j}     S_{ul}^i \left[\alpha_{ul}^{j,i} - \alpha_{ul}^{j,i-1}\right] \quad {\rm for}\,\,\mu > 0
\end{equation}
and
\begin{equation}
I_{\nu,ul}(\mu,z_j) = - \sum_{i=j+1}^{z_N} S_{ul}^i \left[\alpha_{ul}^{j,i} - \alpha_{ul}^{j,i-1}\right] \quad {\rm for}\,\,\mu < 0.
\end{equation}
Again, replacing differences by derivatives (as we did from Eq. \ref{Eq-betaoff1} to Eq. \ref{Eq-betaoff2}), we obtain
\begin{equation}
\begin{split}
I_{\nu,ul}(\mu,z_j) & = \sum_{i=1}^{j} S_{ul}^i z^{i-1,i} \frac{\gamma_{ul}^i \phi_{\nu,ul}^i(\mu)}{\mu} \\
& \times {\rm e}^{\displaystyle - \int_{\frac{z_{i-1}+z_{i}}{2}}^{z_{j}} \frac{\gamma_{ul}(x) \phi_{\nu,ul}(\mu,x)}{\mu}dx} \quad {\rm for}\,\,\mu > 0
\end{split}
\end{equation}
and
\begin{equation}
\begin{split}
I_{\nu,ul}(\mu,z_j) & = -\sum_{i=j+1}^{N} S_{ul}^i z^{i-1,i} \frac{\gamma_{ul}^i \phi_{\nu,ul}^i(\mu)}{\mu} \\
& \times {\rm e}^{\displaystyle - \int_{\frac{z_{i-1}+z_{i}}{2}}^{z_j} \frac{\gamma_{ul}(x) \phi_{\nu,ul}(\mu,x)}{\mu}dx} \quad {\rm for}\,\,\mu < 0.
\end{split}
\end{equation}
If the line profile is independent of $\mu$, the flux densities in the forward and backward directions at boundary $z_j$, defined as
\begin{equation}
F_{\nu,ul}^+(z_j) = 2\pi \int_0^1 I_{\nu,ul}(\mu,z_j) \mu d\mu
\end{equation}
and 
\begin{equation}
F_{\nu,ul}^-(z_j) = 2\pi \int_{-1}^0 I_{\nu,ul}(\mu,z_j) \mu d\mu,
\end{equation}
simply write
\begin{equation} \label{Eq-flux-p}
F_{\nu,ul}^+(z_j) = 2 \pi \sum_{i=1}^{j} S_{ul}^i z^{i-1,i} \gamma_{ul}^i \phi_{\nu,ul}^i E_2 \left( \int_{\frac{z_{i-1}+z_{i}}{2}}^{z_j} \gamma_{ul}(x) \phi_{\nu,ul}(x) dx \right)
\end{equation}
and
\begin{equation} \label{Eq-flux-m}
F_{\nu,ul}^-(z_j) = 2 \pi \sum_{i=j+1}^{N} S_{ul}^i z^{i-1,i} \gamma_{ul}^i  \phi_{\nu,ul}^i E_2 \left( \int_{z_j}^{\frac{z_{i-1}+z_{i}}{2}} \gamma_{ul}(x) \phi_{\nu,ul}(x) dx \right),
\end{equation}
where $E_2$ is the second exponential integral. These previous equations for the intensities and the flux densities are different from Eqs. 19 and 20 of \citet{Elitzur2006a}. This is, again, because we approximate differences by derivatives and because we include a position dependent line profile. We stress that for an optically thin line, the second exponential integral is close to 1. In this case, it is remarkable to note that, while the specific intensity diverges for small values of $\mu$, the forward and backward flux densities are well behaved and give exactly the value we would obtain by integrating the emissivity along the $\mu = 1$ direction and multiplying the result by $2\pi$.

A final remark should be made regarding the balance of energy. All the equations above have been derived considering that the sole source of energy is thermal. The level populations calculated over the entire cloud therefore necessarily fullfil the energy conservation equation
\begin{equation} \label{Eq-energy-cons}
\begin{split}
\sum_{i=1}^{N} \sum_{u=1}^{L} \sum_{l=1}^{u-1} E_{ul} z^{i-1,i} C_{ul}^i n_e^i  \left(n_l^i \frac{g_u}{g_l} {\rm e}^{-E_{ul}/kT^i} - n_u^i\right) = \\
\sum_{u=1}^{L}\sum_{l=1}^{u-1} \int_\nu \left( F_{\nu,ul}^-(0) + F_{\nu,ul}^+(z_N) \right) d\nu
\end{split}
\end{equation}
where the left hand side corresponds to the total cooling flux induced by collisions, and the right hand side corresponds to the total line flux escaping the cloud. This equation provides a fundamental sanity check of the results obtained from the numerical scheme used to solved the multilevel system.

\subsection{Approximations for diffuse interstellar shocks} \label{Append-radtrans-approx}

\begin{figure}[!h]
\begin{center}
\includegraphics[width=10.0cm,trim = 1.7cm 6cm 0.7cm 1.0cm, clip,angle=0]{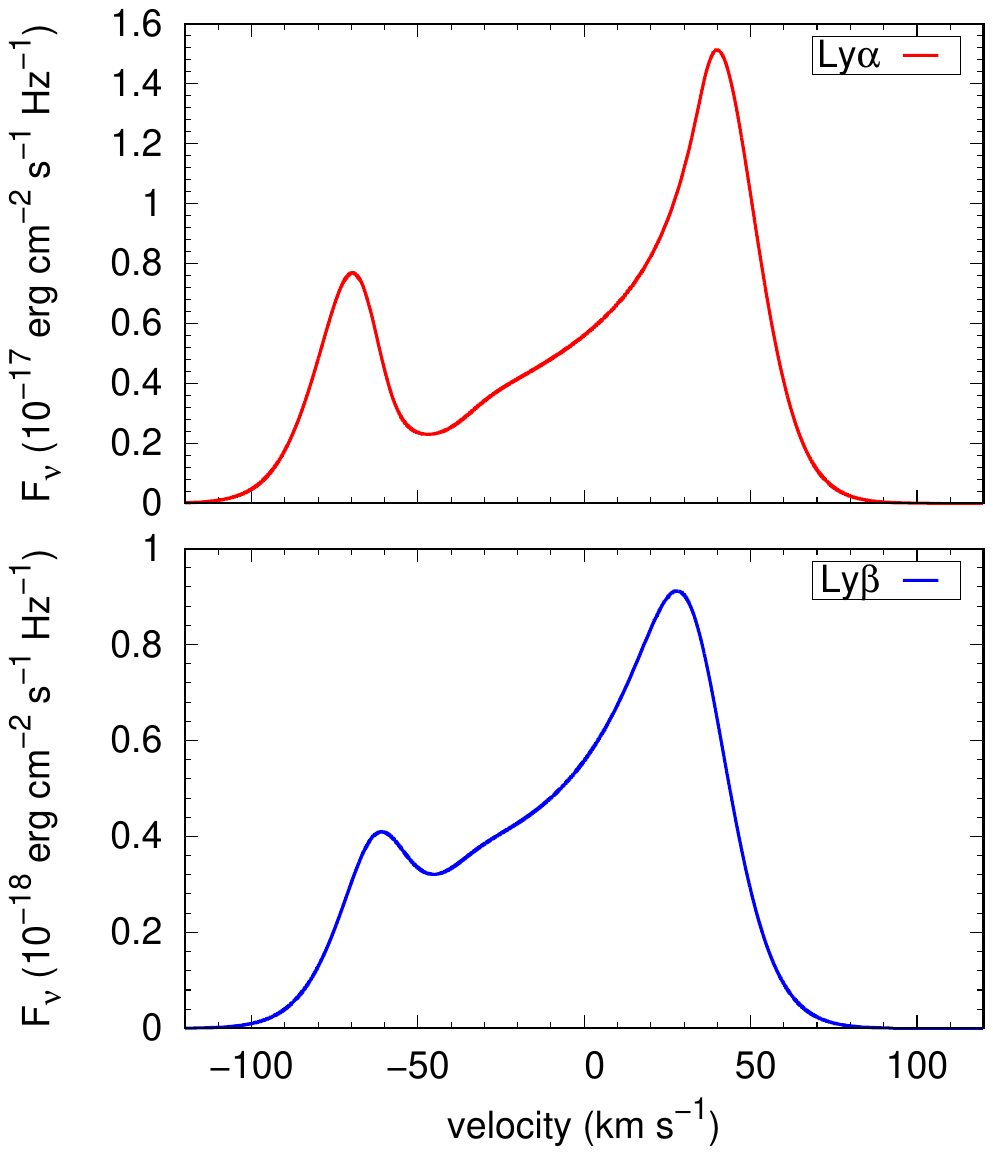}
\caption{Flux densities of the Ly$\alpha$ (top panel) and Ly$\beta$ (bottom panel) lines of atomic hydrogen obtained by applying the exact radiative-transfer algorithm of Appendix \ref{Append-radtrans} on an interstellar shock. The flux densities are computed in the backward direction, that is in the direction of the preshock material, at the position $z=0$. The model displayed corresponds to a shock at 50 \kms\ propagating in a medium with a density of 1 \cc\ and a magnetic field strength of 1 $\mu$G, and irradiated by an external UV radiation field with a scaling factor $G_0=1$ (in Mathis units).}
\label{Fig-H-profile}
\end{center}
\end{figure}

The general methodology presented in the previous sections is used to postprocess the radiative transfer in interstellar shocks within all the lines of any atom in any ionization state. As described above, the method relies on the subdivision of the plane-parallel structure into zones. In the case of interstellar shocks, the number of zones $N$ and their boundaries $z_i$ are chosen so that the density of the species under consideration, $n^i$, the increment of total column density, $n_{\rm H}^i z^{i-1,i}$, the density of electrons, $n_e^i$, the temperature of the gas, $T^i$, and its velocity, $\upsilon_z^i$, vary by less than 10\% from one zone to the next.

In practice the nonlinear system given by Eqs. \ref{Eq-bracket3}, \ref{Eq-betaon-fin}, \ref{Eq-betaoff-fin}, \ref{Eq-cons-pop}, and \ref{Eq-system} could be solved, at steady-state, using a Newton-Raphson algorithm. However, an additional simplification can be made by noticing that in the diffuse interstellar medium, only the ground electronic state of atoms and atomic ions are substantially populated, regardless of the gas temperature. It implies that the fractional population of the ground electronic state can be set to one and that only the dipole-allowed electronic lines connected to this level should be considered as resonant and treated with a detailed calculation of their net radiative bracket. All the other transitions, including the two-photon emission processes, can be considered as optically thin with a net radiative bracket $p_{ul}^i = 1$ for any zone $i$. With this approximation, the original system of $N \times L$ nonlinear equations is transformed into $L$ systems of $N$ linear equations that can be solved sequentially, starting from the highest level, $L$, and going downward.

The method described in the previous sections has been developed neglecting the impacts of the external radiation field and the shock continuum emission (through Bremsstrahlung and recombination radiation) on the radiative pumping of excited states. As discussed by \citet{Elitzur2006a} (in their Appendix A), these effects are not particularly difficult to include and simply imply to add pumping rates from the ground electronic states in Eq. \ref{Eq-system}. In this paper, we deliberately ignore this aspect in order to check that the method ensures the conservation of energy (Eq. \ref{Eq-energy-cons}). The impact of external and continuum radiations onto the level excitations will be included in future works if the method proves to be effective in a wide variety of environments.

It may appear surprising that all the equations are written without including the absorption induced by photoionization and dust and PAH extinctions. Solving this problem self-consistently is, in fact, straightforward and requires to include the opacity of each process computed in the $\mu=1$ direction, over the zone $i$ in the exponential integral of Eq. \ref{Eq-betaon-fin}, and between zones $i$ and $j$ in the exponential integrals of Eqs. \ref{Eq-betaoff-fin}, \ref{Eq-flux-p}, and \ref{Eq-flux-m}. Despite the simplicity of the task, we choose here to adopt the following approach: the level populations and the line fluxes are first calculated without these processes over the entire shock profile ; the line fluxes are then computed, in the shock and the preshock media, including gas and dust absorptions. This choice is done, again, to check that the method effectively ensures the conservation of energy (Eq. \ref{Eq-energy-cons}). The self-consistent treatment will be adopted in future works once the method has proven successful over a larger set of parameters than that explored here. We stress, however, that this self-consistent treatment should not impact the results of this paper. It is so because the absorption lengths of gas and dust are not homogeneous but display extremely sharp transitions in the shock and the preshock materials (see Appendix \ref{Append-photo}, Fig.~\ref{Fig-alpha}).

\subsection{Example of line profiles}

As an example of application of the postprocessing radiative-transfer algorithm, we display in Fig.~\ref{Fig-H-profile} the flux densities of the Ly$\alpha$ and Ly$\beta$ lines of atomic hydrogen in the backward direction (i.e., escaping toward the preshocked gas) obtained for a shock propagating at 50 \kms\ in a medium with a density of 1 \cc\ and a magnetic field strength of 1 $\mu$G. As expected in self-irradiated shocks, both the Ly$\alpha$ and Ly$\beta$ lines display complex and asymmetric profiles. Because of the large opacity of atomic hydrogen across the shock, Ly$\alpha$ and Ly$\beta$ photons can only escape in the wings of the line. Moreover, the flux density of the Ly$\beta$ line is found to be about 20 times smaller than that of the Ly$\alpha$ line. This is the signature of the conversion of Ly$\beta$ photons into H$\alpha$ and Ly$\alpha$ photons which occur in environments with large opacities. All these features are in agreement with the line profiles we recently computed using a MALI algorithm for shocks at similar velocities, yet at significantly larger densities (see Fig.~6 of \citealt{Lehmann2020}).

\section{Photon-matter interactions} \label{Append-photo}

\begin{figure}[!h]
\begin{center}
\includegraphics[width=10.0cm,trim = 1.7cm 1.5cm 0.7cm 1.0cm, clip,angle=0]{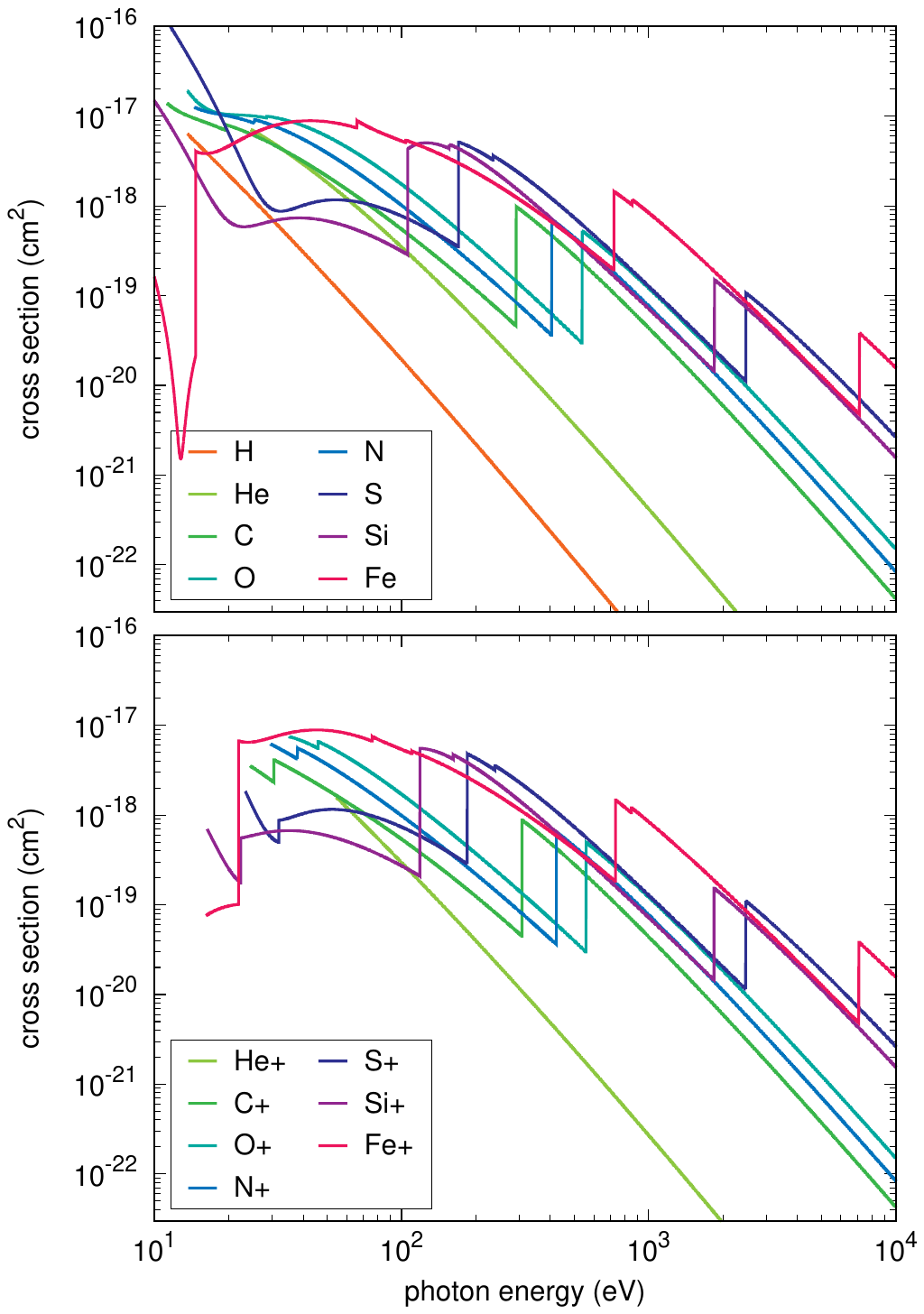}
\caption{Photoionization cross sections of neutral atoms (top panel) and their first ionization stage (bottom panel) as functions of the photon energy, derived from the analytical formula of \citet{Verner1995} (Eq.\,\,1 and Table\,\,1 in their paper). The discontinuities are due to jumps of the cross sections when the photon energy becomes larger than the different thresholds required to extract an electron from internal subshells of the species.}
\label{Fig-cross-sections}
\end{center}
\end{figure}

\begin{figure*}[!h]
\begin{center}
\includegraphics[width=9.0cm,trim = 0cm 1.5cm 28cm 0.5cm, clip,angle=0]{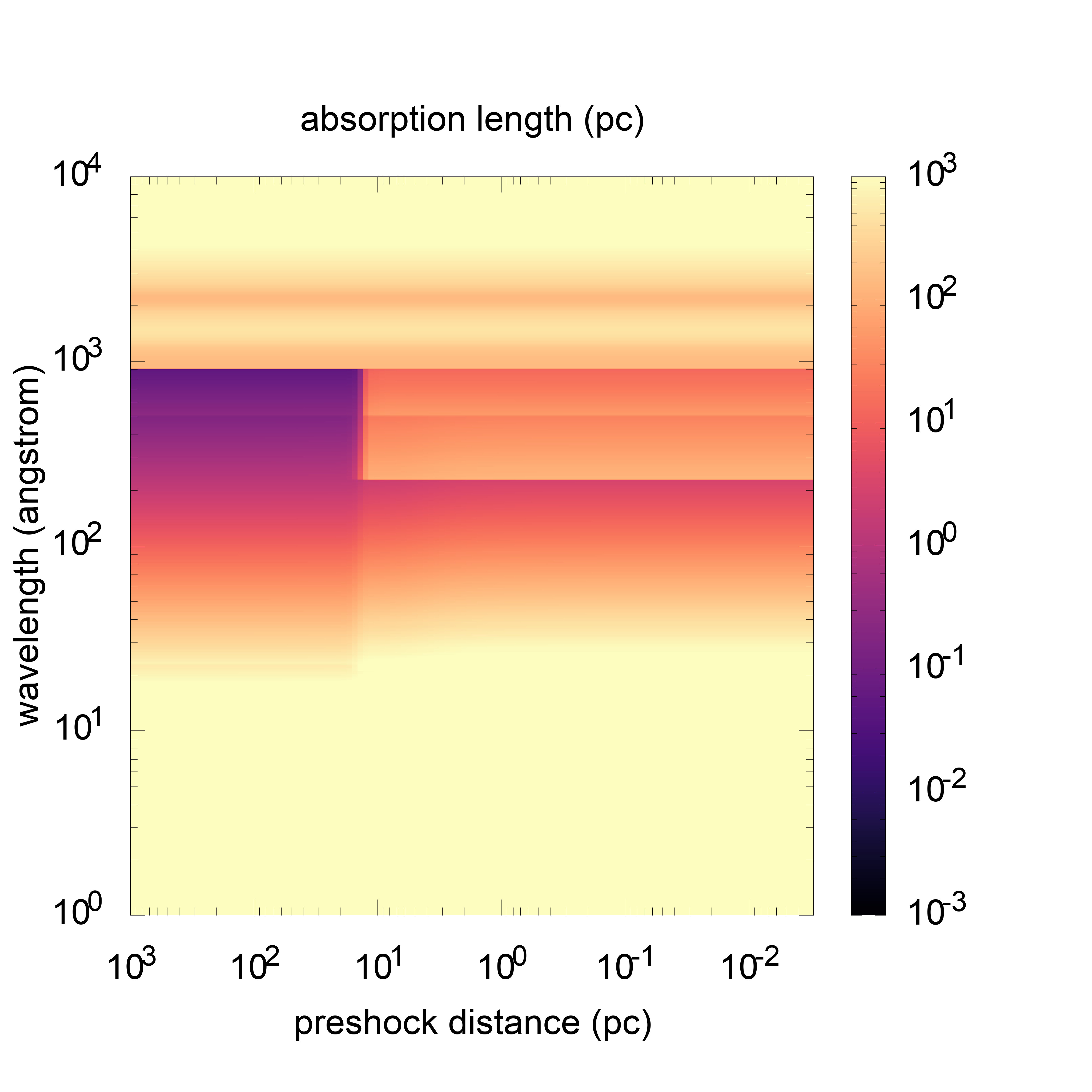}
\includegraphics[width=9.0cm,trim = 18.0cm 1.5cm 10cm 0.5cm, clip,angle=0]{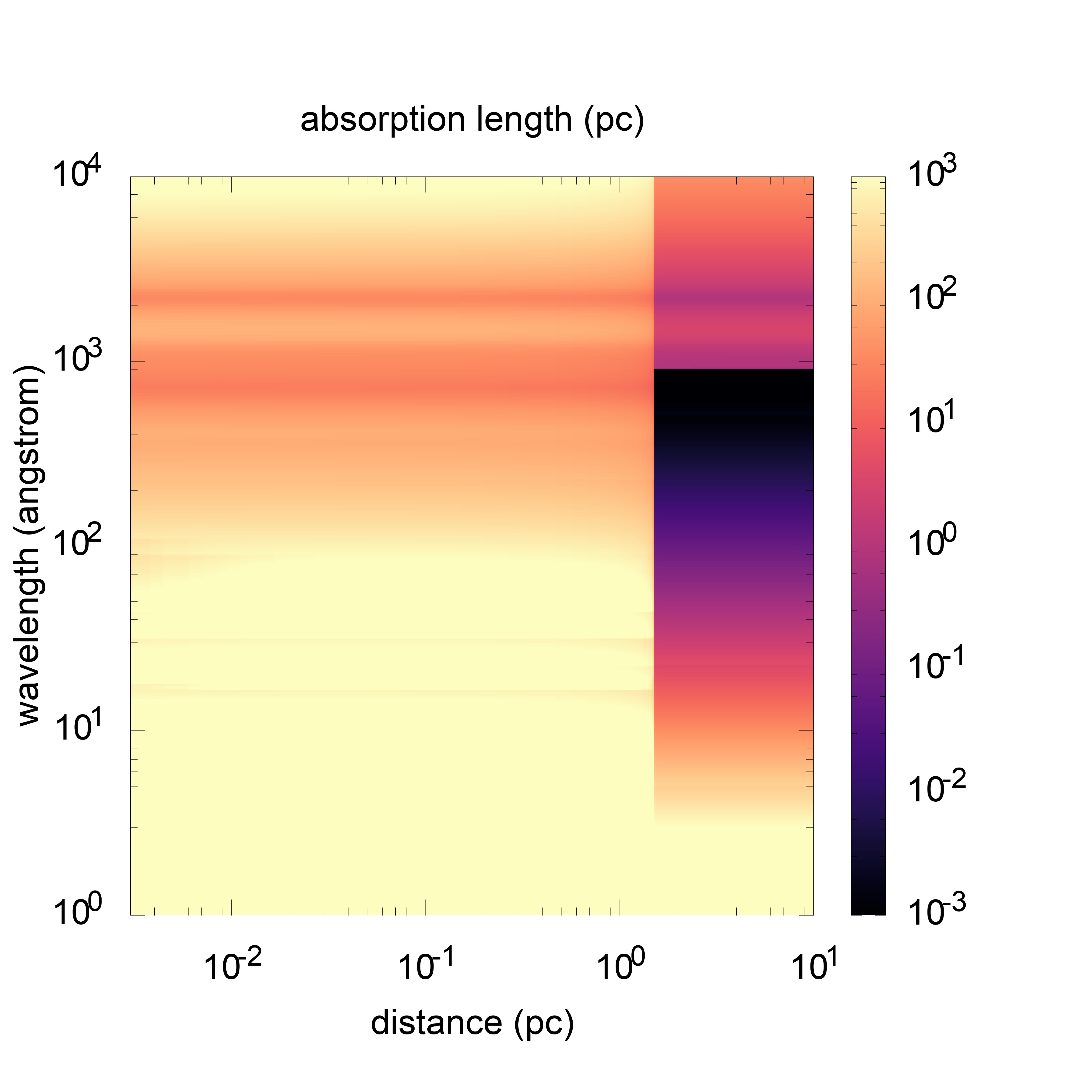}
\caption{Local absorption length of photons obtained in the final converged iteration of the standard shock model. The absorption length, given in pc with the color scale, is shown as a function of the photon energy and the distance in the preshock gas (left panel) and in the shock (right panel). Above 911\,\,\AA, photons are preferentially absorbed by dust and PAHs. Below this limit, photons are preferentially absorbed by atoms, in particular H, He, and \Hep. The walls seen by the photons below 911\,\,\AA\ correspond to the regions in the preshock and the postshock gas where neutral hydrogren is reformed.}
\label{Fig-alpha}
\end{center}
\end{figure*}

The propagation of photons in the shock and the preshock media is controlled by their interactions with the interstellar matter and the cross sections of the associated processes. The main interactions of UV, EUV, and X-ray photons in atomic environments are the absorption by PAHs and dust and the photoionization of atoms and atomic ions. These processes not only drive the physical and chemical states of the shocked and the preshocked material but also determine the propagation scale of the radiation. Following our previous work \citep{Godard2019}, the extinction by PAHs and dust is modeled using the absorption coefficients of spherical graphite grains derived by \citet{Draine1984} and \citet{Laor1993}. All PAHs are assumed to have a constant size of 6.4 \AA. The size of dust particles is computed from the mean square radius of grains taking into account the erosion of grain cores (see Appendix A of \citealt{Godard2019}). The photoionization of gas phase species is modeled using the prescription of \citet{Verner1995} who performed analytical fits of the partial ionization cross sections of atoms and atomic ions for all subshells $nl$ of the associated ground electronic state. The total ionization cross section of any species is the sum of these partial ionization cross sections. As an illustration, we display in Fig.~\ref{Fig-cross-sections} the total photoionization cross sections of the several elements, in their neutral form and in their first ionization stage, as functions of the photon energy. In addition to the expected asymptotic behavior (in $\nu^{-3.5}$ at large $\nu$; e.g., \citealt{Dopita2003}), this figure highlights the sudden jumps of the cross sections at specific energy thresholds. These jumps are due to the openings of new ionization channels corresponding to the removal of one electron (and possibly several Auger's electrons) from the inner subshells of the species.

The impacts of PAHs, dust, and gas phase particles on the absorption of the radiation field is shown in Fig.~\ref{Fig-alpha} which displays the local absorption length of photons in the preshocked and shocked materials as a function of the photon wavelength, computed with the standard shock model. In high-velocity shocks, photons with wavelength below 911~\AA\ are subject to two walls of absorption, one in the preshock, and one in the postshock. This feature is the signature of the chemical composition of the gas. Between these walls, hydrogen and helium are mostly ionized and EUV and X-ray photons are mostly absorbed by ionized metals (in the preshock), and PAHs and dust (in the shock). Beyond these walls, the gas is mostly neutral and high energy photons are quickly absorbed by both H and He. In contrast, photons with wavelength above 911 \AA\ are subjected to a single wall of absorption in the postshock. This comes from the fact that UV photons are preferentially absorbed by dust and PAHs. Since the density is roughly constant in the preshock medium, so is the absorption length. The wall seen in the postshock medium corresponds to the position where the gas cools from a few $10^5$~K to a few tens of K (see Fig.~\ref{Fig-convergence}) and is induced by the isobaric increase of the density up to the point where the gas is supported by the magnetic pressure.

\end{document}